\documentclass[prd,aps,showpacs,tightenlines,notitlepage,nofootinbib,superscriptaddress,groupedaddress]{revtex4-1}
\usepackage{mathrsfs}
\usepackage{amsmath}
\usepackage{amsfonts}
\usepackage{array}
\usepackage{verbatim}
\usepackage{epsfig}
\usepackage{graphicx} 
\usepackage[usenames, dvipsnames]{color}
\usepackage{marginnote}
\usepackage{slashed}
\usepackage{bm}

\newcounter{mynote}% a new counter for use in margin notes

\begin{document}
\title{The role of the chiral anomaly in polarized deeply inelastic scattering I: Finding the triangle graph inside the box diagram in Bjorken and Regge asymptotics}
\author{Andrey Tarasov$^{1,2}$}
\author{Raju Venugopalan$^3$}

\affiliation{$^1$Department of Physics, The Ohio State University, Columbus, OH 43210, USA\\
$^2$Joint BNL-SBU Center for Frontiers in Nuclear Science (CFNS) at Stony Brook University, Stony Brook, NY 11794, USA\\
$^3$Physics Department, Brookhaven National Laboratory,
Bldg. 510A, Upton, NY 11973, U.S.A.}

\begin{abstract}
We revisit the role of the chiral ``triangle" anomaly in deeply inelastic scattering (DIS) of electrons off polarized protons employing a powerful worldline formalism. We demonstrate how the triangle anomaly appears at high energies in the DIS box diagram for the polarized proton structure function 
$g_1(x_B,Q^2)$ in both the Bjorken limit of large $Q^2$ and in the Regge limit of small $x_B$. We show that the operator product expansion is not required to extract the anomaly in either asymptotics though it is sufficient in the Bjorken limit. Likewise, the infrared pole in the anomaly arises in both limits. 
The leading contribution to $g_1$, in both Bjorken and Regge asymptotics, is therefore given by the expectation value of the topological charge density, generalizing a result previously argued by Jaffe and Manohar to hold for the first moment of $g_1$. In follow-up work, we will show how our results  motivate the derivation of a helicity-dependent effective action  incorporating the physics of the anomaly at small $x_B$ and shall discuss the QCD evolution of $g_1(x_B,Q^2)$ in this framework.  

\end{abstract}

%\date{\today}

\maketitle
 
%\tableofcontents

\section{Introduction}

It has long been realized that deeply inelastic scattering (DIS) off polarized protons probes the physics of the chiral anomaly in QCD~\cite{Kodaira:1979pa} though its precise role has been the subject of some debate~\cite{Altarelli:1988nr,Carlitz:1988ab,Jaffe:1989jz}. The purpose of this work is revisit and cast new light on the chiral anomaly with a view to better understand the interplay between parton dynamics and the topology of the QCD vacuum in the helicity structure of the proton at high energies.

In this paper, we will focus on the triangle graph~\cite{Adler:1969gk,Adler:1969er,Bell:1969ts,Kogut:1974kt} whereby the anomaly manifests itself in the coupling of the isosinglet axial vector current to the topological charge density in the polarized proton. A careful treatment of the triangle graph is essential to a first principles understanding of polarized DIS. For instance, as we shall discuss, the off-forward matrix element for the polarized $g_1$ structure function contains an infrared pole that appears to diverge in the forward limit~\cite{Jaffe:1989jz}. It is well known that the triangle graph is embedded in the usual box diagram for polarized DIS in the Bjorken limit of large squared momentum transfer $Q^2$. Our analysis, performed in a worldline formalism  particularly suited to discussions of the anomaly~\cite{Polyakov:1987ez,AlvarezGaume:1983ig,Strassler:1992zr,DHoker:1995aat,DHoker:1995uyv,Haack:1998uy,McKeon:1998et,Schubert:2001he,Bastianelli:2003bg}, will show however that the usual operator product expansion (OPE) formalism is not necessary for this result though it is sufficient. 

Our novel result is that the triangle graph appears identically in the box diagram for high energy polarized DIS in the $x_{B}\ll 1$ Regge asymptotics~\cite{Jaffe:1996zw} of the Bjorken variable $x_{B}$. While there have been  qualitative discussions~\cite{Bass:1997zz,Bass:2004xa,Wakamatsu:2019ain} of the triangle anomaly at small $x_{B}$, a quantitative discussion has been lacking thus far. Our aim is to redress this lack especially in view  of polarized DIS experiments  at the Electron Ion Collider (EIC) in the near future that will access very small values of $x_{\rm B}$ for the first time~\cite{Accardi:2012qut,Aschenauer:2017jsk,Aschenauer:2020pdk}. 

Albeit our focus here is on the triangle anomaly, follow-up papers (Papers II \& III) in preparation will discuss further the fundamental issues underlying the non-perturbative regularization of the infrared pole of the anomaly. Some of these issues were discussed previously 
 by Shore and Veneziano~\cite{Shore:1990zu,Shore:1991dv}, and by Shore, Narison and Veneziano~\cite{Narison:1994hv,Narison:1998aq}--for a nice review, see \cite{Shore:2007yn}. Specifically, in Paper II, we will motivate in the worldline formalism an effective action for Regge asymptotics that is  consistent with anomalous chiral Ward identities~\cite{Wess:1971yu}. In Paper III, we will  discuss the energy evolution of helicity dependent distributions in this framework. We note that there is a considerable body of work on perturbative resummations of the large logarithms in $x_B$ that drive the energy evolution of helicity dependent distributions in Regge asymptotics~\cite{Kirschner:1983di,Bartels:1995iu,Bartels:1996wc,Kovchegov:2015pbl,Kovchegov:2016weo,Kovchegov:2017jxc,Chirilli:2018kkw,Boussarie:2019icw,Cougoulic:2019aja,Kovchegov:2020hgb,Cougoulic:2020tbc}.

To proceed further, we will recap briefly the discussion of the chiral anomaly in polarized DIS and some of the subtle issues in its interpretation. Polarized inclusive deeply inelastic scattering (DIS) is defined as the process 
\begin{eqnarray}
l(l) + N(P, S)\to l(l') + X\,,
\end{eqnarray}
where the lepton $l$ interacts with the polarized target hadron $N$ via the exchange of a virtual photon $\gamma^\ast$ with momentum $q = l - l'$. Here the target is characterized by its momentum vector $P = (P^+, M^2/2P^+, 0_\perp)$ and the covariant spin vector satisfies $S^2 = -1$.

The hadron tensor in DIS is the matrix element of the product of electromagnetic currents~\cite{Blumlein:2012bf},
\begin{eqnarray}
W^{\mu\nu}(q, P, S) = \frac{1}{2\pi} \int d^4x \,e^{iqx} \langle P,S| j^\mu(x) j^\nu(0)|P,S\rangle\,,
\label{hT}
\end{eqnarray}
where $j^\mu = \sum_f e_f \bar{\Psi}_f\gamma^\mu \Psi_f$ is bilinear in the quark and antiquark field operators and  $e_f$ denotes the electric charge of a quark of flavor $f$.
It can be split into symmetric and antisymmetric parts as
\begin{eqnarray}
W^{\mu\nu}(q, P, S) = \bar{W}^{\mu\nu}(q, P) + i \tilde{W}^{\mu\nu}(q, P, S)\,.
\label{SAT}
\end{eqnarray}
Since our interest in this paper is on spin effects in DIS,  our focus will be on the antisymmetric part of Eq.~(\ref{SAT}), which can be expressed in terms of spin dependent structure functions~\cite{Anselmino:1994gn} as
\begin{eqnarray}
\tilde{W}_{\mu\nu}(q, P, S)= \frac{2M_N}{P\cdot q}\epsilon_{\mu\nu\alpha\beta} q^\alpha\Big\{ S^\beta g_1(x_B, Q^2) + \Big[S^\beta - \frac{(S\cdot q)P^\beta}{P\cdot q}\Big]g_2(x_B, Q^2)\Big\}\,,
\label{WA}
\end{eqnarray}
where the virtuality of the incoming virtual photon $Q^2 = -q^2 > 0$,  the Bjorken variable $x_B = Q^2/(2P\cdot q)$, $M_N$ is the proton mass and the totally antisymmetric Levi-Civita tensor $\epsilon_{\mu\nu\alpha\beta}$  is defined with $\epsilon_{0123} = -1$. It is convenient to consider a longitudinally polarized target with the covariant spin vector $S^\mu(\lambda) \simeq \frac{2{\tilde \lambda}_P}{M_N}P^\mu$, where ${\tilde \lambda}_P = \pm \frac{1}{2}$ is the helicity. In this case, the $g_2$ structure function does not contribute.

In the parton model, at leading twist, this expression simplifies to read~\cite{Leader:2001gr}
\begin{equation}
\label{eq:parton-g1}
g_1(x_B, Q^2) = \frac{1}{2} \sum_f e_f^2 \left( \Delta q_f(x_B,Q^2) +\Delta {\bar q}_f(x_B,Q^2)\right)\,,
\end{equation}
where the polarized parton distribution function (pdf)
\begin{equation}
\Delta q_f(x_B,Q^2) = \frac{1}{4\pi} \int dy^- \,e^{-iy^- x_B \,P^+}\,\langle P,S| {\bar \Psi_f}(0,y^-,0_\perp) \gamma^+ \gamma_5 \Psi_f(0)|P,S\rangle\,.
\end{equation}
Here $P^+$ is the large light cone component of the momentum of the hadron. In light front quantization, $\Delta q_f(x_B,Q^2)$ has the physical interpretation of the difference in the density of left and right 
handed quarks of a given quark flavor. Likewise, $\Delta {\bar q}_f(x_B,Q^2)$ denotes the difference in the density of left and right handed anti-quarks of the given flavor. 

The first moment of Eq.~(\ref{eq:parton-g1}) can be expressed~\cite{Ellis:1973kp}, assuming flavor $SU(3)$, as
\begin{equation}
\int_0^1 dx_B\, g_1(x_B,Q^2) = \frac{1}{18} \left(3 F + D +
2 \,\Sigma(Q^2)\right) \,.
\label{g1fm}
\end{equation}
Here $F$ and $D$ in the combinations $F+D$ and $3 F-D$ are proportional respectively to the isotriplet axial vector current and the octet axial vector current. The former is nothing but $g_A$ the 
nucleon's axial vector coupling and is determined quite precisely from $\beta$-decay experiments. Likewise, the latter is well known from hyperon decay experiments. Their running is very weak and they can be treated 
for all relevant purposes as constants. 

The object of interest in this equation is the net light quark helicity  $\Sigma(Q^2)$ defined as the flavor singlet sum:
\begin{equation}
\Sigma(Q^2) = \sum_f \int_0^1 dx_B\,\left( \Delta q_f(x_B,Q^2) +\Delta {\bar q}_f(x_B,Q^2)\right) \,,
\end{equation}
and one can write, to leading twist accuracy,
\begin{equation}
\label{eq:Jmu5}
S^\mu \Sigma(Q^2) = \frac{1}{M_N}\sum_f \langle P,S| {\bar \Psi}_f\gamma^\mu \gamma_5 \Psi_f |P,S\rangle \equiv \frac{1}{M_N}\langle P,S| J^\mu_5 (0) |P,S\rangle \,,
\end{equation}
where $J^\mu_5$ is the flavor isosinglet axial vector current in QCD. $\Sigma(Q^2)$ contributes to the spin sum rule for the proton and its value was  first extracted in pioneering experiments by the European Muon Collaboration (EMC)~\cite{Ashman:1989ig,Ashman:1987hv}; best current estimates from COMPASS~\citep{Alekseev:2010ub} give $2 \,\Sigma(Q^2) = 0.32 \pm  0.03({\rm stat.}) \pm 0.03({\rm syst.})$ at $Q^2= 3$ GeV$^2$, which is in good agreement with the extraction by HERMES~\cite{Airapetian:2007mh} at $Q^2=5$ GeV$^2$ of $2\, \Sigma(Q^2) = 0.330 \pm 0.011({\rm th.}) \pm 0.025({\rm exp.}) \pm 0.028({\rm evol.})$. 
This is significantly below the ``naive" quark model expectation~\cite{Jaffe:1989jz} of $2\,\Sigma(Q^2) = 0.6\pm 0.12$, which would result from the Ellis-Jaffe sum rule equating the isosinglet and octet axial vector currents; for more detailed discussions, see \cite{Aidala:2012mv,Deur:2018roz}. This is of course the famous ``spin crisis" of the proton -- and has lead to a large body~\cite{Kuhn:2008sy} of theoretical and experimental work since. 

The role of the anomaly becomes relevant for this discussion because the isosinglet axial vector current in Eq.~(\ref{eq:Jmu5}) is not conserved, satisfying the anomaly equation
\begin{equation}
\label{eq:anomaly}
\partial^\mu J_\mu^5(x) = \frac{n_f\alpha_s}{2\pi}~{\rm Tr} \left(F_{\mu\nu}(x)\tilde{F}^{\mu\nu}(x)\right) \,,
\end{equation}
where $F_{\mu\nu}$ is the QCD field strength tensor, its dual ${\tilde F}^{\mu\nu} = \frac{1}{2}\epsilon^{\mu\nu\rho\sigma} F_{\rho\sigma}$,  $n_f$ is the number of light quark flavors and $\alpha_s=\frac{g^2}{4\pi}$, where $g$ is the QCD coupling. 
One may however rewrite Eq.~(\ref{eq:Jmu5}) in terms of a conserved current as 
\begin{equation}
S^\mu \Sigma(Q^2) = S^\mu {\tilde \Sigma}(Q^2) + 2\, n_f \frac{1}{M_N}\,\langle P,S| K^\mu|P,S\rangle \,,
\label{eq:real-Sigma}
\end{equation}
where $S^\mu {\tilde \Sigma}(Q^2) = \frac{1}{M_N}\langle P,S| {\tilde J}_\mu^5 |P,S\rangle$. Here ${\tilde J}_\mu^5 = J_\mu^5 - 2 n_f K_\mu$ is a conserved current since the anomaly satisfies the equation 
\begin{equation}
\partial_\mu J^\mu_5 = 2\, n_f \,\partial_\mu K^\mu \,,
\end{equation}
with the Chern-Simons current $K^\mu$ defined to be 
\begin{equation}
\label{eq:C-S}
K_\mu = \frac{\alpha_S}{4 \pi} \epsilon_{\mu\nu\rho\sigma} \left[  A_a^\nu \left( \partial^\rho A_a^\sigma -\frac{1}{3} g f_{abc} A_b^\rho A_c^\sigma \right)  \right]\,.
\end{equation}

One possible explanation for the small value of $\Delta \Sigma$, advanced early \cite{Altarelli:1988nr,Carlitz:1988ab,Lampe:1998eu} after the EMC discovery, is that if first moment of $g_1$ were providing information on ${\tilde \Sigma}$ rather than $\Sigma$, that would provide a potential resolution of the spin crisis with the framework of the parton model itself. More specifically, it was argued on the basis of the gauge structure of $K^\mu$ that one could write 
\begin{equation}
 {\tilde \Sigma}(Q^2) = \Sigma(Q^2) - \frac{n_f\alpha_S}{2\pi}  \Delta G \,,
\end{equation}
where $\Delta G$ is the gluon helicity pdf. If $\Delta G$ is large, this would provide a natural explanation of the spin crisis. However as pointed out by Jaffe and Manohar, this identification is 
intrinsically problematic because while the gluon helicity pdf $\Delta G$ is manifestly gauge invariant, the same cannot be said of the Chern-Simons current. 
The latter is not gauge invariant under large gauge transformations $U$, which give,
\begin{equation}
K_\mu \rightarrow K_\mu + i\frac{g}{8 \pi^2}\epsilon_{\mu\nu\alpha\beta}\partial^\nu \left((U^\dagger \partial^\alpha U) A^\beta\right) + \frac{1}{24\pi^2} \epsilon_{\mu\nu\alpha\beta}\left[(U^\dagger \partial^\nu U) (U^\dagger \partial^\alpha U) (U^\dagger \partial^\beta U)\right]\,.
\end{equation}

The resolution of the problem, as discussed by Jaffe and Manohar~\cite{Jaffe:1989jz} (see also \cite{Efremov:1989sn}), lies in how one takes limits when $U_A(1)$ is broken. This is because the breaking of this symmetry lifts an apparent pole in the forward scattering amplitude. Indeed this is the fundamental reason why the $\eta^\prime$-meson gets a mass (distinct from the pseudoscalar octet) in QCD~\cite{Witten:1979vv,Veneziano:1979ec,Bass:2004xa,Shore:2007yn}. We will now spell out the argument as sketched in \cite{Jaffe:1989jz}. For our convenience, and that of the reader familiar with their paper, we will use their notations. 

We begin by first considering the r.h.s of Eq.~(\ref{eq:Jmu5}) and writing its off-forward counterpart as
\begin{equation}
\label{eq:off-forward-Jmu5}
\frac{1}{M_N}\langle P^\prime,S| J^\mu_5 (0)|P,S\rangle = \Sigma(Q^2,t)\,S^\mu + h(Q^2,t)\,l\cdot S \,l^\mu \,,
\end{equation}
where $l^\mu = {P^\prime}^\mu - P^\mu$ is the momentum transfer between the outgoing and incoming proton and $t=l^2$. Here $\Sigma(Q^2,t)$ and $h(Q^2,t)$ can be interpreted as form factors that represent respectively the coupling of the isosinglet axial vector charge and the isosinglet pseudoscalar charge to the proton at finite momentum transfer. In particular, the former represents the triangle diagram of the anomaly 
\begin{equation}
i l\cdot S\, {\kappa}(Q^2,t) = \frac{1}{M_N}\langle P^\prime,S| \frac{\alpha_s n_f}{2\pi}~{\rm Tr} \left(F_{\mu\nu}\tilde{F}^{\mu\nu}\right)(0)|P,S\rangle \,,
\end{equation}
which, as suggested by the r.h.s, represents the coupling of topological charge to the nucleon. The other form factor $h(t)$ represents the isosinglet pseudoscalar form factor, given by the coupling of the $\eta^\prime$ meson to the nucleon~\cite{Veneziano:1989ei,Hatsuda:1989bi,Efremov:1989sn}. 

Then from the anomaly equation (Eq.~(\ref{eq:anomaly})), and from Eq.~(\ref{eq:off-forward-Jmu5}), one obtains,
\begin{equation}
\kappa(t) = \Sigma(Q^2,t) + t \,h(Q^2,t) \, .
\end{equation}
Further, since the $\eta^\prime$ is massive, $h(t)$ has no pole, which gives $\Sigma(Q^2,0)=\kappa(Q^2,0)$. However, as stated in \cite{Jaffe:1989jz}, the triangle graph gives $S^\mu\,\Sigma(Q^2,t) \propto -i \frac{\alpha_s}{2\pi}\, \frac{l^\mu}{l^2}\, {\rm  Tr}\,(F{\tilde F})$. One cannot therefore naively take the forward limit. 

More specifically, the statement that ``$h(t)$ has no pole" and the observation that the triangle graph has an infrared pole are intimately connected and it is the interplay between the two that leads to a finite result. 
Indeed, as noted in \cite{Jaffe:1989jz}, the limit of momentum zero transfer must then be understood by writing the r.h.s of Eq.~(\ref{eq:off-forward-Jmu5}) as 
\begin{equation}
S^\mu\, \Sigma(Q^2,t) + l\cdot S\, l^\mu\, h(Q^2,t) \longrightarrow \frac{l\cdot S\,l^\mu}{t} {\kappa}(Q^2,t) + \left(S^\mu -\frac{l\cdot S\, l^\mu}{t}\right) \lambda(Q^2,t) \,.
\end{equation}
This decomposition separates the triangle graph from other contributions. For the forward matrix element of $J_\mu^5$ to appear as a smooth limit of $l^\mu\rightarrow 0$ (because the pole must be lifted by the mass of the $\eta^\prime$)  the 
triangle contribution  must cancel  a similar contribution extracted from the pseudoscalar coupling, as suggested by the second term above. One way to think about this is that there is a mixing of the contributions from the topological charge and an isosinglet component of a pseudo-Goldstone nonet which can be separated out in this manner~\cite{Diakonov:1981nv,Shore:1990zu,Ji:1990fj,Shore:1991dv,Efremov:1990nj,Dvali:2005an,Dvali:2005ws,Kharzeev:2015xsa}. For this to hold, one must of course 
have $\lambda(0) = \kappa(0)$.  As we will discuss in paper II, this follows from the Wess-Zumino-Witten 
term~\cite{Wess:1971yu,Witten:1983tw,Leutwyler:1996sa} for the $\eta^\prime$. 

What survives then on the r.h.s as $l^\mu\rightarrow 0$ is  $S^\mu \lambda(0)\equiv S^\mu \kappa(0)$. This gives the result
\begin{equation}
 \Sigma(Q^2) = \frac{n_f\,\alpha_s}{2\pi\,M_N} 
 \lim_{l_\mu \to 0}\langle P^\prime,S|\frac{ 1 }{i l\cdot s  } {\rm Tr} \left(F\tilde{F}\right)(0) |P,S\rangle\,\,.
\label{eq:sigma-anomaly}
\end{equation}
While the matrix element of ${\rm Tr} \left(F\tilde{F}\right)(0)$ is naively zero in the forward limit, the matrix element as defined above is finite when one combines the contribution from the density matrix $|P^\prime\rangle\langle P|$ and the triangle operator. This is often done in careful perturbative QCD computations by introducing a mass term or like infrared regulator which cancels between the two to give the finite result~\cite{Manohar:1990jx,Bodwin:1990fk,Vogelsang:1990ug}.  However they do not arrive at the expression in Eq.~(\ref{eq:sigma-anomaly}) because they do not further impose the constraints required by the soft dynamics\footnote{For early discussions of these in the context of $U_A(1)$ breaking by instantons, following the seminal work of t'Hooft~\cite{tHooft:1976snw,tHooft:1986ooh}, we refer the reader to Refs.~\cite{Forte:1989qq,Forte:1990xb,Dorokhov:1993ym,Qian:2015wyq}. We note that while instantons provide an attractive dynamical mechanism, this interpretation of the phenomenon is by no means unique.} of $U_A(1)$ breaking in QCD. If they are not imposed, anomaly matching cannot occur and there  will remain an unrequited pole from a pseudoscalar coupling to the triangle graph~\cite{Preskill:1990fr}. For elegant reviews of the role of the anomaly and the  Wess-Zumino-Witten term for the $\eta^\prime$ in chiral effective lagrangians for the pseudoscalar nonet, we refer the reader to \cite{Leutwyler:1997yr,HerreraSiklody:1996pm}. Lattice computations of $\Sigma(Q^2)$ implementing anomalous Ward identities are discussed in \cite{Liang:2018pis} and references therein. 

As we noted previously, our purpose here is to go beyond the discussion of the triangle anomaly in $\Sigma(Q^2)$ {\it a la} Eq.~(\ref{eq:Jmu5}) and to discuss its role in $g_1(x,Q^2)$ itself in both Bjorken and Regge asymptotics. We obtain the striking result that the formal structure of our results is identical in the two asymptotic limits of perturbative QCD. Our quantitative results for the latter in particular are novel.  As the discussion above suggests,  they have strong implications for our understanding spin diffusion at small $x_B$; these  
 will be discussed at length in Papers II \& III. 

The outline of the paper is as follows. In section~\ref{sec:world-line-hadron-tensor}, we will extend the worldline formalism developed for unpolarized DIS by us previously~\cite{Tarasov:2019rfp} to the case of polarized DIS. We will first write down the most general expression for the 
box diagram corresponding to $g_1(x_B,Q^2)$. We will then consider the Bjorken asymptotics of $Q^2\rightarrow\infty$ in section~\ref{sec:Bj} and demonstrate explicitly in Sec.~\ref{subsec:Bj-worldline} how the triangle anomaly appears and  provides the leading contribution in this asymptotics. Our result is given in Eq.~(\ref{eq:g1-Bj-final}). A discussion of this result and its implications is provided in Sec.~\ref{subsec:Bj-discussion} for readers who may not be interested in the details of the worldline derivation. Our result here is of interest firslyt because most treatments in the literature are of $\Sigma(Q^2)$ rather than $g_1(x_B,Q^2)$ itself. Moreover, unlike these discussions, we do not make use of the OPE. Our worldline framework allows us to classify graphs into those that contain the anomaly structure, and those that do not, with the latter being suppressed in the Bjorken limit. Such a  classification may be of value in the computations of other DIS observables. We also comment on the consistency of our results with an analysis of the perturbative evolution of $\Sigma (Q^2)$ and $\Delta G(Q^2)$ to high loop accuracy. 

In section~\ref{sec:Regge}, we show that the anomaly provides the leading contribution in the Regge asymptotics of $x_B\rightarrow 0$. The worldline derivation in Sec.~\ref{subsec:Rj-worldline} is very similar to that of the previous section.  Though our final result (Eq.~(\ref{resultRj})) is formally identical to that in the Bjorken limit, subtle differences in the two derivations indicate that the computation of the matrix element of the anomaly will differ both  qualitatively and quantitatively in the two limits. This is discussed in Sec.~\ref{subsec:Rj-discussion}.

A final section summarizes our results and briefly discusses their implications for the computation of $g_1(x_B,Q^2)$. As noted earlier, this computation in the Regge limit will be discussed at length in Papers II\,\& III. Appendix A discusses details of the computation of the box diagram of polarized DIS in the worldline formalism. Appendix B provides detailed expressions for coefficient functions encountered in intermediate steps of the computation. The  computation of the triangle graph in this formalism is discussed in Appendix C.

\section{Worldline representation of antisymmetric part of the hadron tensor in polarized DIS}
\label{sec:world-line-hadron-tensor}
To compute $g_1(x,Q^2)$, we will require (see Eq.~(\ref{WA})) the antisymmetric part $\tilde{W}^{\mu\nu}$ of the hadron tensor in the worldline representation of DIS introduced in \cite{Tarasov:2019rfp}. One first reexpresses the hadron tensor in Eq.~(\ref{hT}) in terms of the second derivative of the effective action $\Gamma[A]$ with respect to the electromagnetic field $a_\mu(x)$  corresponding to the insertion of  incoming virtual photons $\gamma^\ast$ at two spacetime points: 
\begin{eqnarray}
W^{\mu\nu}(q, P, S) = \frac{1}{\pi e^2}{\rm Im}\  \int d^4x~ e^{iqx}\langle P,S| \frac{\delta^2 \Gamma[a, A]}{\delta a_\mu(\frac{x}{2})\delta a_\nu(-\frac{x}{2})} |P,S\rangle\,.
\label{TprodEffex}
\end{eqnarray}
Here $A$ denotes the gluon background field of the target. 

It is sufficient for our discussion of the triangle graph to work with the one loop QED+QCD representation\footnote{For a discussion of higher loop contributions to the effective action, we refer the reader to \cite{Fliegner:1997ra,Pawlowski:2008xh,Magnea:2013lna}. } of the worldline  effective action \cite{Schubert:2001he},
\begin{eqnarray}
&&\Gamma_{QCD}[a, A] = -\frac{1}{2} \int^T_0 \frac{dT}{T} {\rm Tr_c} \int \mathcal{D}x \int \mathcal{D} \psi \exp\Big\{-\int^T_0 d\tau \Big(\frac{1}{4} \dot{x}^2 + \frac{1}{2}\psi_\mu\dot{\psi}^\mu + ig\dot{x}^\mu (A_\mu + a_\mu) - ig \psi^\mu \psi^\nu F_{\mu\nu}(A+a)\Big)\Big\}\,,
\nonumber\\
\label{MLag}
\end{eqnarray}
which is characterized by 0+1 dimensional worldline trajectories of  Boson ($x^\mu(\tau)$) and Grassmann ($\psi^\mu(\tau)$) variables coupled to background electromagnetic ($a_\mu$) and gluon ($A_\mu$) fields. Note that the boson functional integral has periodic (P) boundary conditions while the Grassmann functional integral has anti-periodic (AP) boundary conditions. 

It is convenient to rewrite the operator in Eq. (\ref{TprodEffex}) in terms of the Fourier transformation of the effective action,
\begin{eqnarray}
{\tilde{\Gamma}}^{\mu\nu}[k_1, k_3] \equiv \int d^4z_1 d^4z_3 \frac{\delta^2 \Gamma[a, A]}{\delta a_\mu(z_1)\delta a_\nu(z_3)}|_{a=0} \,e^{ik_1 z_1} e^{ik_3 z_3}\,,
\label{secder}
\end{eqnarray}
where $k_1$ and $k_3$ denote the incoming photon four-momenta; separating out the antisymmetric part of Eq. (\ref{TprodEffex}), we obtain\footnote{Note that Eq. (\ref{WModFu}) is written in Euclidean space-time with signature $\eta = (1,1,1,1)$. In our calculation, we perform an analytical continuation to Minkowski space-time with $g = (1, -1, -1, -1)$ by the replacement $\eta_{\mu\nu}\to - g_{\mu\nu}$, see Ref. \cite{Schubert:2001he}.}
\begin{eqnarray}
i\tilde{W}^{\mu\nu}(q, P, S) = \frac{ 1 }{ 2 \pi e^2}\,{\rm Im}\ \int d^4x \,e^{-iqx} 
 \int \frac{d^4k_1}{(2\pi)^4} \int \frac{d^4k_3}{(2\pi)^4} e^{-ik_1 \frac{x}{2}} e^{ik_3 \frac{x}{2} }\langle P,S| {\tilde{\Gamma}}^{\mu\nu}_A[k_1, k_3] |P,S\rangle\,,
\label{WModFu}
\end{eqnarray}
where ${\tilde{\Gamma}}^{\mu\nu}_A[k_1, k_3] \equiv {\tilde{\Gamma}}^{\mu\nu}[k_1, k_3] - (\mu\leftrightarrow\nu)$. 

The hadron tensor in Eq. (\ref{TprodEffex}) is taken in the forward limit when $k_1=-k_3=-q$. However as discussed in the introduction, to obtain the infrared pole of the anomaly, one needs to calculate the off-forward matrix element $\langle P'|\dots|P\rangle$ in Eq. (\ref{TprodEffex}), where $P'-P\equiv l$, and then take the limit $l\to0$ in the final expression. Hence the incoming photon momenta in our computation of $\Gamma^{\mu\nu}_A[k_1, k_3]$ are kept distinct.

Taking the second derivative of the effective action, we obtain
\begin{eqnarray}
{\tilde{\Gamma}}^{\mu\nu}_A[k_1, k_3] &=& \frac{e^2 e_f^2 }{2}\int^\infty_0\frac{dT}{T} ~{\rm Tr_c} \int \mathcal{D}x\int \mathcal{D}\psi 
\Big[ V^\mu_1(k_1)V^\nu_3(k_3) - (\mu\leftrightarrow\nu)\Big] \label{photamp}\\
&\times& \exp\Big\{-\int^T_0 d\tau \Big(\frac{1}{4} \dot{x}^2 + \frac{1}{2}\psi_\mu\dot{\psi}^\mu + ig\dot{x}^\mu A_\mu - ig \psi^\mu \psi^\nu F_{\mu\nu}\Big)\Big\}\,,
\nonumber
\end{eqnarray}
where
\begin{eqnarray}
V^\mu_i(k_i) \equiv \int^T_0 d\tau_i  (\dot{x}^\mu_i + 2i \psi^\mu_i k_j\cdot\psi_j ) e^{ik_i\cdot x_i}\,,
\end{eqnarray}
is the vertex corresponding to the interaction of a  worldline  with the external electromagnetic current, and $x_i\equiv x(\tau_i)$, $\psi_i\equiv \psi(\tau_i)$.

\begin{figure}[htb]
 \begin{center}
 \includegraphics[width=90mm]{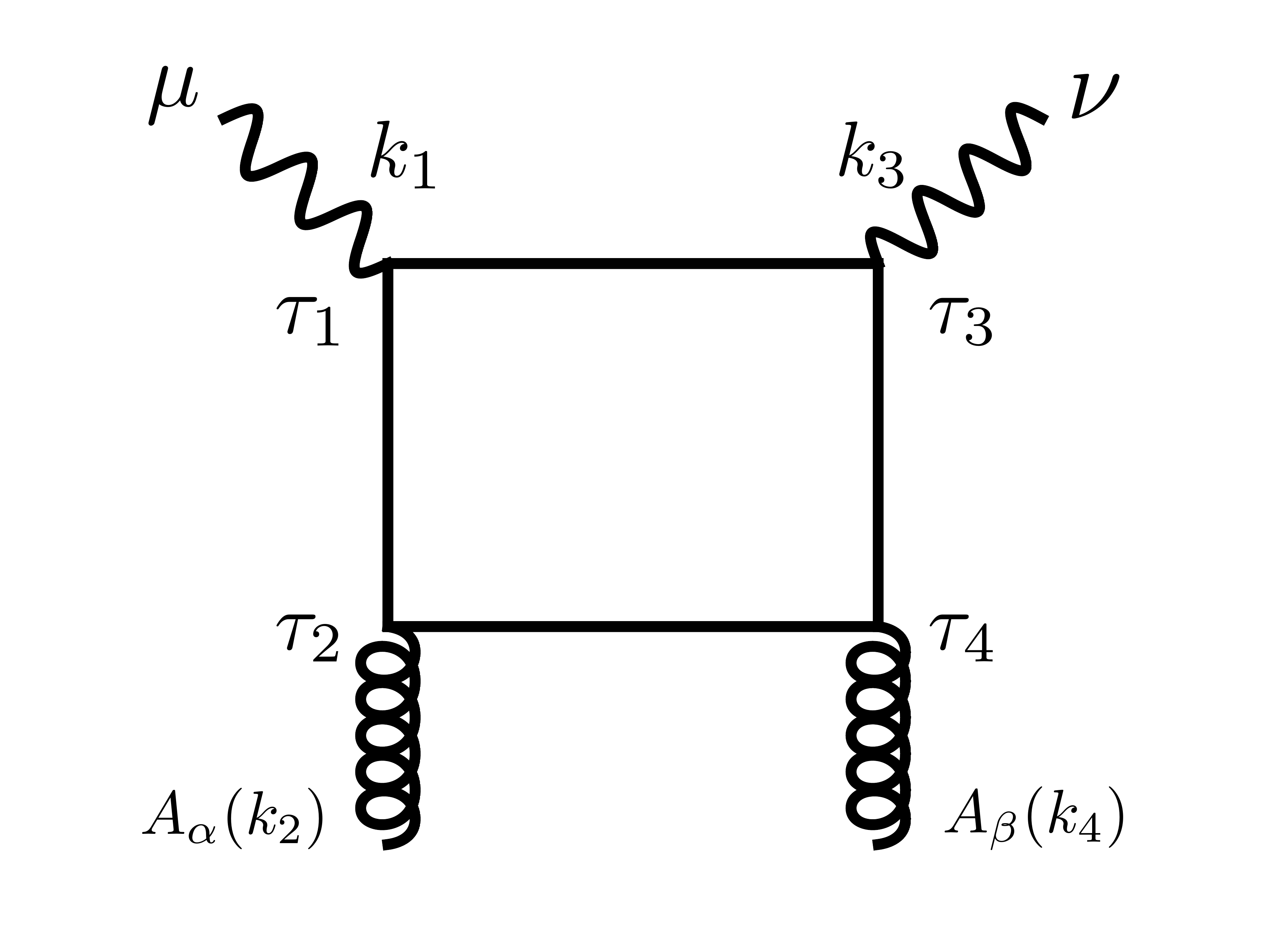}
 \end{center}
 \caption{\label{fig:boxgraph}The box diagram $\Gamma^{\mu\nu\alpha\beta}_A[k_1, k_3, k_2, k_4]$ for polarized DIS.}
 \end{figure}

Expanding the worldline action up to the second order in the background field\footnote{In discussions of the anomaly, it is often convenient to impose Fock-Schwinger gauge $x\cdot A=0$, and expand the result in powers of $F_{\mu\nu}$-- see~\cite{Mueller:2017arw} for instance, for an explicit derivation of the anomaly equation in the worldline formalism. For our discussion of the anomaly, the approach here is sufficient; that it is so is a non-trivial feature of the non-Abelian axial anomaly~\cite{AlvarezGaume:2005qb,Bilal:2008qx}.}, one obtains, 
\begin{eqnarray}
&&{\tilde{\Gamma}}^{\mu\nu}_A[k_1, k_3] = (-ig)^2 \frac{e^2 e_f^2 }{2}\int^\infty_0\frac{dT}{T} ~{\rm Tr_c} \int \mathcal{D}x\int \mathcal{D}\psi~\exp\Big\{-\int^T_0 d\tau \Big(\frac{1}{4} \dot{x}^2 + \frac{1}{2}\psi \cdot\dot{\psi} \Big)\Big\}
\\
&&\times~\Big[ V^\mu_1(k_1)V^\nu_3(k_3) \int^T_0 d\tau_2 \Big( \dot{x}^\alpha_2 A_\alpha(x_2) +  2 \psi^\alpha_2 \psi^\lambda_2 \partial_\lambda A_{\alpha}(x_2)\Big) \int^T_0 d\tau_4 \Big( \dot{x}^\beta_4 A_\beta(x_4) + 2 \psi^\beta_4 \psi^\eta_4 \partial_\eta A_{\beta}(x_4) \Big) - (\mu\leftrightarrow\nu)\Big]\,.
\nonumber
\end{eqnarray}
One can rewrite this further in terms of the Fourier transforms of the background gauge fields
\begin{eqnarray}
A_\alpha(x_2) = \int \frac{d^4k_2}{(2\pi)^4} e^{ik_2\cdot x_2} {\tilde A}_\alpha(k_2);\ \ \ \ \ A_\beta(x_4) = \int \frac{d^4k_4}{(2\pi)^4} e^{ik_4\cdot x_4} {\tilde A}_\beta(k_4)\,,
\label{fourbackground}
\end{eqnarray}
as 
\begin{eqnarray}
&&\Gamma^{\mu\nu}_A[k_1, k_3] = \int \frac{d^4k_2}{(2\pi)^4} \int \frac{d^4k_4}{(2\pi)^4}~
\Gamma^{\mu\nu\alpha\beta}_A[k_1, k_3, k_2, k_4]~ {\rm Tr_c}({\tilde A}_\alpha(k_2) {\tilde A}_\beta(k_4))\,,
\label{Hadgamma}
\end{eqnarray}
where
\begin{eqnarray}
&&\Gamma^{\mu\nu\alpha\beta}_A[k_1, k_3, k_2, k_4] \equiv -\frac{g^2 e^2 e_f^2 }{2}\int^\infty_0\frac{dT}{T} ~ \int \mathcal{D}x\int \mathcal{D}\psi~\exp\Big\{-\int^T_0 d\tau \Big(\frac{1}{4} \dot{x}^2 + \frac{1}{2}\psi\cdot\dot{\psi} \Big)\Big\}
\nonumber\\
&&\times \Big[V^\mu_1(k_1)V^\nu_3(k_3)V^\alpha_2(k_2) V^\beta_4(k_4) - (\mu\leftrightarrow\nu)\Big]\,,
\end{eqnarray}
corresponds to the well-known box diagram of DIS with four incoming momenta $k_i$ shown in Fig. \ref{fig:boxgraph}.

Taking the product of the four worldline interaction vertices $V^\mu_i(k_i)$, and removing the terms proportional to $ \dot{x}^\mu_1\dot{x}^\nu_3$ and $\dot{x}^\alpha_2 \dot{x}^\beta_4$, which do not contribute to the antisymmetric part of the hadron tensor\footnote{Indeed, the interaction of the worldline with external particles through $\dot{x}^\mu_1\dot{x}^\nu_3$ and $\dot{x}^\alpha_2 \dot{x}^\beta_4$ coincides with that of  scalar QED. As a result, it doesn't generate any spin dependent effect.},
we obtain
\begin{eqnarray}
&&\Gamma^{\mu\nu\alpha\beta}_A[k_1, k_3, k_2, k_4] = -\frac{g^2e^2 e_f^2 }{2}\int^\infty_0\frac{dT}{T} ~ \int \mathcal{D}x\int \mathcal{D}\psi~\exp\Big\{-\int^T_0 d\tau \Big(\frac{1}{4} \dot{x}^2 + \frac{1}{2}\psi\cdot \dot{\psi} \Big)\Big\}
\nonumber\\
&&\times \prod^4_{k=1}\int^T_0 d\tau_k~ \Big[\sum^9_{n=1}\mathcal{C}^{\mu\nu\alpha\beta}_{n;(\tau_1,\tau_2,\tau_3,\tau_4)}[k_1, k_3, k_2, k_4] - (\mu\leftrightarrow\nu)\Big]e^{i\sum^4_{i=1}k_i x_i} \,.
\label{GammawithC}
\end{eqnarray}
where the coordinate ($x_i \equiv x(\tau_i)$) and Grassmann variables ($\psi_i\equiv \psi(\tau_i)$) in the coefficients $\mathcal{C}^{\mu\nu\alpha\beta}_{n;(\tau_1,\tau_2,\tau_3,\tau_4)}[k_1, k_3, k_2, k_4]$ depend on the proper time coordinates of the interaction of the worldlines with the external electromagnetic and gauge fields:
\begin{eqnarray}
&&\mathcal{C}^{\mu\nu\alpha\beta}_{1;(\tau_1,\tau_2,\tau_3,\tau_4)}[k_1, k_3, k_2, k_4] = - 4\dot{x}^\nu_3 \psi^\mu_1 \psi_1\cdot k_{1} \dot{x}^\beta_4 \psi^\alpha_2 \psi_2\cdot k_{2};
\nonumber\\
&&\mathcal{C}^{\mu\nu\alpha\beta}_{2;(\tau_1,\tau_2,\tau_3,\tau_4)}[k_1, k_3, k_2, k_4] = -4\dot{x}^\nu_3 \psi^\mu_1 \psi_1\cdot k_{1}  \dot{x}^\alpha_2 \psi^\beta_4 \psi_4\cdot k_{4};
\nonumber\\
&&\mathcal{C}^{\mu\nu\alpha\beta}_{3;(\tau_1,\tau_2,\tau_3,\tau_4)}[k_1, k_3, k_2, k_4] = -4 \dot{x}^\mu_1\psi^\nu_3 \psi_3 \cdot k_{3} \dot{x}^\alpha_2 \psi^\beta_4 \psi_4\cdot k_{4};
\nonumber\\
&&\mathcal{C}^{\mu\nu\alpha\beta}_{4;(\tau_1,\tau_2,\tau_3,\tau_4)}[k_1, k_3, k_2, k_4] = -4 \dot{x}^\mu_1\psi^\nu_3 \psi_3 \cdot k_{3}  \dot{x}^\beta_4 \psi^\alpha_2 \psi_2\cdot k_{2}
\nonumber\\
&&\mathcal{C}^{\mu\nu\alpha\beta}_{5;(\tau_1,\tau_2,\tau_3,\tau_4)}[k_1, k_3, k_2, k_4] = - 8i\dot{x}^\nu_3 \psi^\mu_1 \psi_1\cdot k_{1} \psi^\alpha_2 \psi_2\cdot k_{2} \psi^\beta_4 \psi_4\cdot k_{4};
\nonumber\\
&&\mathcal{C}^{\mu\nu\alpha\beta}_{6;(\tau_1,\tau_2,\tau_3,\tau_4)}[k_1, k_3, k_2, k_4] = - 8i \dot{x}^\mu_1\psi^\nu_3 \psi_3 \cdot k_{3} \psi^\alpha_2 \psi_2\cdot k_{2} \psi^\beta_4 \psi_4\cdot k_{4}
\nonumber\\
&&\mathcal{C}^{\mu\nu\alpha\beta}_{7;(\tau_1,\tau_2,\tau_3,\tau_4)}[k_1, k_3, k_2, k_4] = - 8i \dot{x}^\beta_4 \psi^\alpha_2 \psi_2\cdot k_{2} \psi^\mu_1 \psi_1\cdot k_{1} \psi^\nu_3 \psi_3\cdot k_{3}  ;
\nonumber\\
&&\mathcal{C}^{\mu\nu\alpha\beta}_{8;(\tau_1,\tau_2,\tau_3,\tau_4)}[k_1, k_3, k_2, k_4] = - 8i \dot{x}^\alpha_2 \psi^\beta_4 \psi_4\cdot k_{4} \psi^\mu_1 \psi_1\cdot k_{1} \psi^\nu_3 \psi_3\cdot k_{3} 
\nonumber\\
&&\mathcal{C}^{\mu\nu\alpha\beta}_{9;(\tau_1,\tau_2,\tau_3,\tau_4)}[k_1, k_3, k_2, k_4] = 16 \psi^\mu_1 \psi_1\cdot k_{1} \psi^\nu_3 \psi_3\cdot k_{3} \psi^\alpha_2 \psi_2\cdot k_{2} \psi^\beta_4 \psi_4\cdot k_{4}
\end{eqnarray}

We can further rewrite Eq.~(\ref{GammawithC}) as
\begin{eqnarray}
&&\Gamma^{\mu\nu\alpha\beta}_A[k_1, k_3, k_2, k_4]
\nonumber\\
&&= -\frac{g^2e^2 e_f^2 }{2} (2\pi)^4\delta^{(4)}(\sum^4_{i=1} k_i)\int^\infty_0\frac{dT}{T} 
 \prod^4_{k=1}\int^T_0 d\tau_k~\frac{1}{4\pi^2 T^2} \Big(\sum^9_{n=1}\mathcal{I}^{\mu\nu\alpha\beta}_{n;(\tau_1,\tau_2,\tau_3,\tau_4)}[k_1, k_3, k_2, k_4]\Big)\langle e^{i\sum^4_{i=1}k_i x_i}\rangle\,,
\label{GammawithI}
\end{eqnarray}
where
\begin{eqnarray}
&&\frac{1}{4\pi^2 T^2}~\mathcal{I}^{\mu\nu\alpha\beta}_{n;(\tau_1,\tau_2,\tau_3,\tau_4)}[k_1, k_3, k_2, k_4] \langle e^{i\sum^4_{i=1}k_i x_i}\rangle 
\nonumber\\
&&= \int \mathcal{D}x\int \mathcal{D}\psi~\Big(\mathcal{C}^{\mu\nu\alpha\beta}_{n;(\tau_1,\tau_2,\tau_3,\tau_4)}[k_1, k_3, k_2, k_4] e^{i\sum^4_{i=1}k_i x_i} - (\mu\leftrightarrow\nu)\Big)~\exp\Big\{-\int^T_0 d\tau \Big(\frac{1}{4} \dot{x}^2 + \frac{1}{2}\psi\cdot \dot{\psi} \Big)\Big\} \,.
\label{CtoI}
\end{eqnarray} 
The correlator of the exponential factors in Eq. (\ref{GammawithI}) is obtained as an intermediate step in the computation of the functional integrals in Eq. (\ref{GammawithC}); it can be expressed as 
\begin{eqnarray}
\label{eq:correlator}
&&\langle e^{i\sum^4_{i=1}k_i x_i}\rangle = \exp\Big[
k_{1} \cdot k_{2} G_B(\tau_1, \tau_2) + k_{1} \cdot k_{3} G_B(\tau_1, \tau_3) + k_{1}\cdot k_{4} G_B(\tau_1, \tau_4)
\nonumber\\
&& + k_{2} \cdot k_{3} G_B(\tau_2, \tau_3) + k_{2} \cdot k_{4} G_B(\tau_2, \tau_4) + k_{3} \cdot k_{4} G_B(\tau_3, \tau_4) \Big]\,,
\end{eqnarray}
where  
\begin{eqnarray}
G_B(\tau_i, \tau_j) = |\tau_i - \tau_j| - \frac{(\tau_i - \tau_j)^2}{T}\,,
\end{eqnarray}
is the bosonic worldline propagator~\cite{Strassler:1992zr} on a closed loop of period ${\rm T}$.
This remarkably simple result for the correlator is a generic feature of worldline path integrals and  follows from performing Wick contractions of bosonic and Grassmann worldline propagagators employing techniques pioneered by Bern and Kosower~\cite{Bern:1990ux,Bern:1991aq,Bern:1992ad} and discussed at length in \cite{Schubert:2001he}. 
The explicit expression for one of the coefficients ($\mathcal{I}^{\mu\nu\alpha\beta}_{1;(\tau_1,\tau_2,\tau_3,\tau_4)}[k_1, k_3, k_2, k_4]$) is worked out in Appendix \ref{sec:I1}. 

It is convenient to introduce a reparametrization $\tau = u T$ of the proper time variables, where $u\in [0, 1]$. With this reparametrization, Eq. (\ref{GammawithI}) can be rewritten as 
\begin{eqnarray}
\label{eq:Master}
&&\Gamma^{\mu\nu\alpha\beta}_A[k_1, k_3, k_2, k_4]
\nonumber\\
&&= -\frac{g^2e^2 e_f^2 }{8\pi^2}(2\pi)^4\delta^{(4)}(\sum^4_{i=1} k_i)
 \prod^4_{k=1}\int^1_0 du_k \Big(\sum^9_{n=1}\mathcal{I}^{\mu\nu\alpha\beta}_{n;(u_1,u_2,u_3,u_4)}[k_1, k_3, k_2, k_4]\Big)\int^\infty_0dT~ T \exp\Big[ T\Big( k_{1} \cdot k_{2} G_B(u_1, u_2)
 \nonumber\\
 &&+ k_{1} \cdot k_{3} G_B(u_1, u_3) + k_{1}\cdot k_{4} G_B(u_1, u_4) + k_{2} \cdot k_{3} G_B(u_2, u_3) + k_{2} \cdot k_{4} G_B(u_2, u_4) + k_{3} \cdot k_{4} G_B(u_3, u_4) \Big) \Big]\,.
\end{eqnarray}
The integration over the worldline period $T$ can now be performed easily, and one obtains, 
\begin{eqnarray}
&&\Gamma^{\mu\nu\alpha\beta}_A[k_1, k_3, k_2, k_4]
\nonumber\\
&&= -\frac{g^2e^2 e_f^2 }{8\pi^2}(2\pi)^4\delta^{(4)}(\sum^4_{i=1} k_i)
 \prod^4_{k=1}\int^1_0 du_k \Big(\sum^9_{n=1}\mathcal{I}^{\mu\nu\alpha\beta}_{n;(u_1,u_2,u_3,u_4)}[k_1, k_3, k_2, k_4]\Big)\Big[ - k_{1} \cdot k_{2} G_B(u_1, u_2)
 \nonumber\\
 &&- k_{1} \cdot k_{3} G_B(u_1, u_3) - k_{1}\cdot k_{4} G_B(u_1, u_4) - k_{2} \cdot k_{3} G_B(u_2, u_3) - k_{2} \cdot k_{4} G_B(u_2, u_4) - k_{3} \cdot k_{4} G_B(u_3, u_4) \Big]^{-2}\,,
 \label{GammawithIinu}
\end{eqnarray}

This result for the box diagram is noteworthy because the functional dependence on the external momenta of the gauge fields does not rely on any kinematic assumptions. Further, since the result is insensitive to color or flavor, at this level, it only depends on these external momenta and not on whether the gauge field corresponding to a momentum label is a gluon or a photon. As we will show, this property of the box diagram in the worldline formalism will prove extremely useful. In particular, in the following two sections, we will explore  the structure of the box diagram represented by Eq. (\ref{GammawithIinu}) in the physically interesting Bjorken and Regge asymptotics respectively.  We will show explicitly in both limits that the leading contribution to the box diagram is given by the triangle anomaly.

\begin{figure}[htb]
 \begin{center}
 \includegraphics[width=120mm]{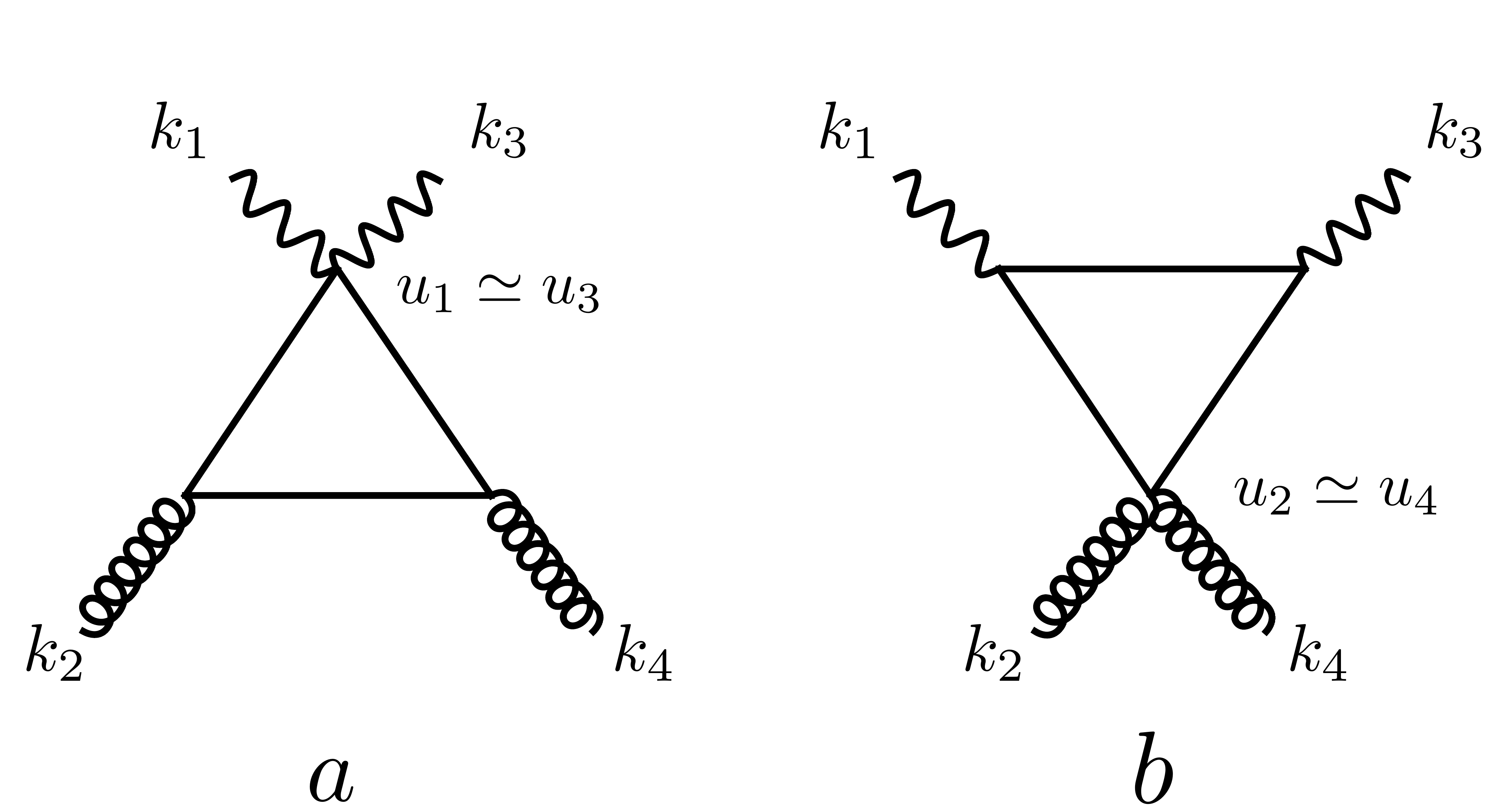}
 \end{center}
 \caption{\label{fig:two_limits}Two ``triangle" limits of the box diagram: a) the Bjorken limit, given by  Eq.~(\ref{GammawithIBj}),  b) the Regge limit given by  Eq.~(\ref{GammawithIRj}).}
 \end{figure}

In the Bjorken limit of QCD, when the virtuality of the incoming photons $Q^2\to\infty$ and $x_{B} = \frac{Q^2}{2P\cdot q}$ is fixed, the distance between the points of interaction of the worldline with the incoming photons ($\tau_1\rightarrow \tau_3$) is defined by a negligibly small number $u_1 - u_3 \sim \Lambda_{\rm QCD}^2/Q^2$, where $\Lambda_{\rm QCD}\approx 200$ MeV is the intrinsic non-perturbative scale of the theory. This limit of the box diagram is illustrated in  Fig.~\ref{fig:two_limits}a. In this limit, 
\begin{eqnarray}
&&\Gamma^{\mu\nu\alpha\beta}_A[k_1, k_3, k_2, k_4]\Big|_{Q^2\to \infty}
\nonumber\\
&&= -\frac{g^2e^2 e_f^2 }{8\pi^2}(2\pi)^4\delta^{(4)}(\sum^4_{i=1} k_i)
 \prod^4_{k=1}\int^1_0 du_k \Big(\sum^9_{n=1}\mathcal{I}^{\mu\nu\alpha\beta}_{n;(u_1,u_2,u_3,u_4)}[k_1, k_3, k_2, k_4]\Big)\Big|_{u_1 = u_3}\Big[ - k_{1} \cdot k_{2} G_B(u_1, u_2)
 \nonumber\\
 &&- k_{1} \cdot k_{3} G_B(u_1, u_3) - k_{1}\cdot k_{4} G_B(u_1, u_4) - k_{2} \cdot k_{3} G_B(u_2, u_3) - k_{2} \cdot k_{4} G_B(u_2, u_4) - k_{3} \cdot k_{4} G_B(u_3, u_4) \Big]^{-2}\,.
\label{GammawithIBj}
\end{eqnarray}
Corrections to this formula are suppressed by a relative power $1/Q^2$.

In a similar fashion, the Regge limit of perturbative QCD (pQCD) is characterized by a fixed virtuality $Q^2\gg \Lambda_{\rm QCD}^2$ and $x_{Bj}\to 0$. In these asymptotics, the interaction of the worldline with the background gluons corresponds to an instantaneous interaction with a shock wave; in the box diagram, this corresponds to $\tau_2\rightarrow \tau_4$, or equivalently, $u_2\simeq u_4$, as shown in Fig.~\ref{fig:two_limits}b. As a result, in this limit,
\begin{eqnarray}
&&\Gamma^{\mu\nu\alpha\beta}_A[k_1, k_3, k_2, k_4]\Big|_{x_{Bj}\to 0}
\nonumber\\
&&= -\frac{g^2e^2 e_f^2 }{8\pi^2}(2\pi)^4\delta^{(4)}(\sum^4_{i=1} k_i)
 \prod^4_{k=1}\int^1_0 du_k \Big(\sum^9_{n=1}\mathcal{I}^{\mu\nu\alpha\beta}_{n;(u_1,u_2,u_3,u_4)}[k_1, k_3, k_2, k_4]\Big)\Big|_{u_2 = u_4}\Big[ - k_{1} \cdot k_{2} G_B(u_1, u_2)
 \nonumber\\
 &&- k_{1} \cdot k_{3} G_B(u_1, u_3) - k_{1}\cdot k_{4} G_B(u_1, u_4) - k_{2} \cdot k_{3} G_B(u_2, u_3) - k_{2} \cdot k_{4} G_B(u_2, u_4) - k_{3} \cdot k_{4} G_B(u_3, u_4) \Big]^{-2}\,.
\label{GammawithIRj}
\end{eqnarray}
In analogy to the Bjorken limit, corrections to this expression are suppressed by a relative power $\sim Q_s^2/M^2$, where $M^2= 2\,xP\cdot q$ is a large scale when $s\rightarrow \infty$. Here $x$ denotes the longitudinal momentum fraction of the proton momentum carried by the background gluon, and $Q_s$ denotes the typical transverse momentum\footnote{We will discuss this emergent scale further in section~\ref{sec:Regge}.} of the background gluons in Regge asymptotics. 

As we noted earlier, the box diagram is only sensitive to the external momentum labels of the external gauge fields. It  will therefore have an identical structure when any two of the proper time values that they correspond to (see Eq.~(\ref{eq:correlator})) are set equal to each other.  Hence the expressions for the Bjorken limit (Eq.~(\ref{GammawithIBj})) and the Regge limit (Eq.~(\ref{GammawithIRj})) can both be understood simply in this formalism as proper time limits of the box diagram;  indeed, as sketched in Fig.~\ref{fig:two_limits}, the triangle structure has a clear visual representation in both asymptotics. This is not the case for the usual OPE language of pQCD where the former is manifest while the latter is not. 

An interesting consequence of our results will be that the structure of the matrix elements for $g_1(x_{\rm Bj},Q^2)$ will be identical in both Bjorken and Regge asymptotics. However the underlying physics of the matrix elements, in particular their QCD evolution, will be quite different in the two kinematics limits.

\section{The triangle anomaly in the Bjorken limit of the box diagram}
\label{sec:Bj}
In this section, we will compute the box diagram of Fig. \ref{fig:boxgraph} in the Bjorken limit. This corresponds to a resolution scale corresponding to a transverse area in the 
proton which vanishes with $1/Q^2\rightarrow \infty$; in the QCD worldline formalism, this corresponds to $u_1 \simeq u_3$ in Eq.~(\ref{eq:Master}), resulting in Eq.~(\ref{GammawithIBj}). 
\subsection{Worldline computation of box diagram in the Bjorken limit}
\label{subsec:Bj-worldline}
 We have explicitly computed the first of the coefficients $\mathcal{I}^{\mu\nu\alpha\beta}_{n;(u_1,u_2,u_3,u_4)}[k_1, k_3, k_2, k_4]$ explicitly in Appendix A; the rest can be computed similarly. The expressions for all of the nine terms are provided in Appendix B. The sum of these, for $u_1=u_3$, is given by  
\begin{eqnarray}
&&\Big(\sum^9_{n=1}\mathcal{I}^{\mu\nu\alpha\beta}_{n;(u_1,u_2,u_3,u_4)}[k_1, k_3, k_2, k_4]\Big)\Big|_{u_1 = u_3} = \epsilon^{\mu\nu\eta}_{\ \ \ \ \kappa} (k_{1\eta} - k_{3\eta})
\nonumber\\
&&\times \Big( k_2 \cdot k_4 \Big[ \dot{G}^2_B(u_1, u_4) - \dot{G}_B(u_1, u_4) \dot{G}_B(u_2, u_4) + \dot{G}_B(u_1, u_2) \big(\dot{G}_B(u_1, u_2) + \dot{G}_B(u_2, u_4) + \dot{G}_B(u_4, u_1) \big)  \Big] \epsilon^{\kappa\alpha\beta\sigma} k_{2\sigma}
\nonumber\\
&&+ k_2\cdot k_4 \Big[ -\dot{G}^2_B(u_1, u_2) - \dot{G}_B(u_1, u_2) \dot{G}_B(u_2, u_4) + \dot{G}_B(u_1, u_4) \big(\dot{G}_B(u_1, u_2) + \dot{G}_B(u_2, u_4) + \dot{G}_B(u_4, u_1)\big) \Big] \epsilon^{\kappa\alpha\beta\sigma} k_{4\sigma}
\nonumber\\
&&+ \Big[ - \dot{G}_B(u_1, u_4) \dot{G}_B(u_2, u_4) +  \dot{G}^2_B(u_1, u_4)
+ \dot{G}_B(u_1, u_2) \big(\dot{G}_B(u_1, u_2) + \dot{G}_B(u_2, u_4) + \dot{G}_B(u_4, u_1)\big) \Big] \epsilon^{\kappa\alpha\sigma\lambda} k^\beta_{2} k_{2\sigma} k_{4\lambda}
\nonumber\\
&&+ \dot{G}^2_B(u_1, u_4) \epsilon^{\kappa\alpha\sigma\lambda} k^\beta_{4} k_{2\sigma} k_{4\lambda}
- \dot{G}^2_B(u_1, u_2) \epsilon^{\kappa\beta\sigma\lambda} k^\alpha_{2} k_{2\sigma} k_{4\lambda} 
\nonumber\\
&&+ \Big[ - \dot{G}_B(u_1, u_2) \dot{G}_B(u_2, u_4) - \dot{G}^2_B(u_1, u_2)
+ \dot{G}_B(u_1, u_4) \big(\dot{G}_B(u_1, u_2) + \dot{G}_B(u_2, u_4) + \dot{G}_B(u_4, u_1)\big) \Big] \epsilon^{\kappa\beta\sigma\lambda} k^\alpha_{4} k_{2\sigma} k_{4\lambda} 
\nonumber\\
&&- \epsilon^{\alpha\lambda\beta\sigma}  k^\kappa_{2} k_{2\lambda} k_{4\sigma} - \epsilon^{\alpha\lambda\beta\sigma} k^\kappa_{4} k_{2\lambda} k_{4\sigma}\Big)\,.
\label{suminBj}
\end{eqnarray}
This expression looks formidable with a large number of Lorentz structures; such structures are also obtained in the well-known perturbative computation of the triangle diagram of the anomaly~\cite{Rosenberg:1962pp,Armillis:2009sm}. We will show explicitly that this expression can be greatly simplified using a few identities and a mass-shell constraint on the background gluons.

We first use the identity\footnote{\label{cyclic} This result can be easily obtained from the standard cyclic identity $v^\mu \epsilon^{\nu\alpha\beta\sigma} + v^\nu \epsilon^{\alpha\beta\sigma\mu} + v^\alpha \epsilon^{\beta\sigma\mu\nu} + v^\beta \epsilon^{\sigma\mu\nu\alpha} + v^\sigma \epsilon^{\mu\nu\alpha\beta} = 0$, where $v^\mu$ is an arbitrary four-vector and a mass-shell condition $k^2_2 = k^2_4= 0$ is imposed on the background gluons.}
\begin{eqnarray}
\label{Levi-Civita}
\epsilon^{\alpha\lambda\beta\sigma}  k^\kappa_{2} k_{2\lambda} k_{4\sigma} = - k^\alpha_{2} \epsilon^{\lambda\beta\sigma\kappa} k_{2\lambda} k_{4\sigma}
- k^\beta_{2} \epsilon^{\sigma\kappa\alpha\lambda}  k_{2\lambda} k_{4\sigma}
- k_{2}\cdot k_{4} \epsilon^{\kappa\alpha\lambda\beta}  k_{2\lambda} \,,
\end{eqnarray}
for the last but one term in Eq. (\ref{suminBj}), and a similar identity for the last term, to reexpress it as 
\begin{eqnarray}
&&\Big(\sum^9_{n=1}\mathcal{I}^{\mu\nu\alpha\beta}_{n;(u_1,u_2,u_3,u_4)}[k_1, k_3, k_2, k_4]\Big)\Big|_{u_1 = u_3} = \epsilon^{\mu\nu\eta}_{\ \ \ \ \kappa} (k_{1\eta} - k_{3\eta})
\nonumber\\
&&\times \Big( \Big[ -1 + \dot{G}^2_B(u_1, u_4) - \dot{G}_B(u_1, u_4) \dot{G}_B(u_2, u_4) + \dot{G}_B(u_1, u_2) \big(\dot{G}_B(u_1, u_2) + \dot{G}_B(u_2, u_4) + \dot{G}_B(u_4, u_1) \big)  \Big]
\nonumber\\
&&\times( k_2 \cdot k_4 \epsilon^{\kappa\alpha\beta\sigma} k_{2\sigma} + \epsilon^{\kappa\alpha\sigma\lambda} k^\beta_{2} k_{2\sigma} k_{4\lambda})
+ \Big[ 1 - \dot{G}^2_B(u_1, u_2) - \dot{G}_B(u_1, u_2) \dot{G}_B(u_2, u_4)
\nonumber\\
&&+ \dot{G}_B(u_1, u_4) \big(\dot{G}_B(u_1, u_2) + \dot{G}_B(u_2, u_4) + \dot{G}_B(u_4, u_1)\big) \Big]
 (k_2\cdot k_4 \epsilon^{\kappa\alpha\beta\sigma} k_{4\sigma} + \epsilon^{\kappa\beta\sigma\lambda} k^\alpha_{4} k_{2\sigma} k_{4\lambda})
\nonumber\\
&&+ \Big[ -1 + \dot{G}^2_B(u_1, u_4) \Big] \epsilon^{\kappa\alpha\sigma\lambda} k^\beta_{4} k_{2\sigma} k_{4\lambda}
+ \Big[ 1 - \dot{G}^2_B(u_1, u_2) \Big] \epsilon^{\kappa\beta\sigma\lambda} k^\alpha_{2} k_{2\sigma} k_{4\lambda} \Big)\, .
\end{eqnarray}

Now recall that the $\alpha,\beta$ open indices on the l.h.s are contracted with the gauge fields, as shown in 
Eq.~(\ref{Hadgamma}).  Since the background gluons are on mass-shell, $k^\beta_{4} A_\beta(k_4) = 0$ and $k^\alpha_{2} A_\alpha(k_2) = 0$. Hence the last two tensorial structures in the above equation don't contribute. Further, the first two structures can be simplified using 
\begin{eqnarray}
k_2 \cdot k_4 \epsilon^{\kappa\alpha\beta\sigma} k_{2\sigma} + \epsilon^{\kappa\alpha\sigma\lambda} k^\beta_{2} k_{2\sigma} k_{4\lambda} \rightarrow  - k^\kappa_{2} \epsilon^{\alpha\beta\sigma\lambda}  k_{2\sigma} k_{4\lambda}\,,
\end{eqnarray}
where we have used $k^\alpha_{2} A_\alpha(k_2) = 0$ to eliminate a term in Eq.~(\ref{Levi-Civita}), and likewise, 
\begin{eqnarray}
k_2\cdot k_4 \epsilon^{\kappa\alpha\beta\sigma} k_{4\sigma} + \epsilon^{\kappa\beta\sigma\lambda} k^\alpha_{4} k_{2\sigma} k_{4\lambda} \rightarrow  k^\kappa_{4} \epsilon^{\alpha\beta\sigma\lambda}  k_{2\sigma} k_{4\lambda}\,.
\end{eqnarray}
As a result, the sum of coefficients simplifies to
\begin{eqnarray}
&&\Big(\sum^9_{n=1}\mathcal{I}^{\mu\nu\alpha\beta}_{n;(u_1,u_2,u_3,u_4)}[k_1, k_3, k_2, k_4]\Big)\Big|_{u_1 = u_3} = - \epsilon^{\mu\nu\eta}_{\ \ \ \ \kappa} (k_{1\eta} - k_{3\eta})
\\
&&\times \Big( \Big[ -1 + \dot{G}^2_B(u_1, u_4) - \dot{G}_B(u_1, u_4) \dot{G}_B(u_2, u_4) + \dot{G}_B(u_1, u_2)\, {\cal X}(u_1,u_2,u_4)   \Big] k^\kappa_{2} \epsilon^{\alpha\beta\sigma\lambda}  k_{2\sigma} k_{4\lambda}
\nonumber\\
&&+ \Big[ -1 + \dot{G}^2_B(u_1, u_2) + \dot{G}_B(u_1, u_2) \dot{G}_B(u_2, u_4) - \dot{G}_B(u_1, u_4)\, {\cal X}(u_1,u_2,u_4) \Big] k^\kappa_{4} \epsilon^{\alpha\beta\sigma\lambda}  k_{2\sigma} k_{4\lambda} \Big)\,,
\nonumber
\end{eqnarray}
where 
\begin{eqnarray}
{\cal X}(u_1,u_2,u_4) =\dot{G}_B(u_1, u_2) + \dot{G}_B(u_2, u_4) + \dot{G}_B(u_4, u_1) \,.
\end{eqnarray}
This expression can be written more compactly as 
\begin{eqnarray}
\Big(\sum^9_{n=1}\mathcal{I}^{\mu\nu\alpha\beta}_{n;(u_1,u_2,u_3,u_4)}[k_1, k_3, k_2, k_4]\Big)\Big|_{u_1 = u_3} &=& - \frac{1}{2}\epsilon^{\mu\nu\eta}_{\ \ \ \ \kappa} (k_{1\eta} - k_{3\eta})
\,\Big[ - 2 + {\cal X}^2(u_1,u_2,u_4)
\nonumber \\
&+&  \dot{G}^2_B(u_1, u_2) - \dot{G}^2_B(u_2, u_4) + \dot{G}^2_B(u_4, u_1)
 \Big] ( k^\kappa_{2} + k^\kappa_{4} ) \epsilon^{\alpha\beta\sigma\lambda}  k_{2\sigma} k_{4\lambda}\,\,.
\end{eqnarray}

One can show that 
\begin{eqnarray}
{\cal X}(u_1,u_2,u_4) \equiv \dot{G}_B(u_1, u_2) + \dot{G}_B(u_2, u_4) + \dot{G}_B(u_4, u_1) = -G_F(u_1, u_2)G_F(u_2, u_4)G_F(u_4, u_1)
\label{gfprod}
\end{eqnarray}
where $G_F(u_i,u_j) = {\rm sign}(u_i - u_j)$ is the fermionic worldline propagator. Using then ${\cal X}^2 =1$ and
\begin{eqnarray}
1 - \dot{G}^2_B(u_i, u_j) = 4 \,G_B(u_i, u_j)\,,
\label{b2dot}
\end{eqnarray}
we obtain 
\begin{eqnarray}
\Big(\sum^9_{n=1}\mathcal{I}^{\mu\nu\alpha\beta}_{n;(u_1,u_2,u_3,u_4)}[k_1, k_3, k_2, k_4]\Big)\Big|_{u_1 = u_3} &=& - 2\epsilon^{\mu\nu\eta}_{\ \ \ \ \kappa} (k_{1\eta} - k_{3\eta})
 \Big[ - G_B(u_1, u_2) + G_B(u_2, u_4) - G_B(u_4, u_1)
 \Big] \nonumber \\
 &\times& ( k^\kappa_{2} + k^\kappa_{4} ) \epsilon^{\alpha\beta\sigma\lambda}  k_{2\sigma} k_{4\lambda}\,\,.
 \label{sumcoefBj}
\end{eqnarray}

Substituting this back to Eq. (\ref{GammawithIBj}),  we obtain the following result for the box diagram in the Bjorken limit of QCD:
\begin{eqnarray}
&&\Gamma^{\mu\nu\alpha\beta}_A[k_1, k_3, k_2, k_4]\Big|_{Q^2\to \infty}
= \frac{g^2e^2 e_f^2 }{4\pi^2}\epsilon^{\mu\nu\eta}_{\ \ \ \ \kappa} (k_{1\eta} - k_{3\eta})(2\pi)^4\delta^{(4)}(\sum^4_{i=1} k_i)
 \prod^4_{k=1}\int^1_0 du_k 
 \nonumber\\
&&\times \Big[ - G_B(u_1, u_2) + G_B(u_2, u_4) - G_B(u_4, u_1)
 \Big] ( k^\kappa_{2} + k^\kappa_{4} ) \epsilon^{\alpha\beta\sigma\lambda}  k_{2\sigma} k_{4\lambda}
\Big[ - k_{1} \cdot k_{2} G_B(u_1, u_2)
 \nonumber\\
 &&- k_{1} \cdot k_{3} G_B(u_1, u_3) - k_{1}\cdot k_{4} G_B(u_1, u_4) - k_{2} \cdot k_{3} G_B(u_2, u_3) - k_{2} \cdot k_{4} G_B(u_2, u_4) - k_{3} \cdot k_{4} G_B(u_3, u_4) \Big]^{-2}\,.
 \label{Gammau}
\end{eqnarray}

\begin{figure}[htb]
 \begin{center}
 \includegraphics[width=140mm]{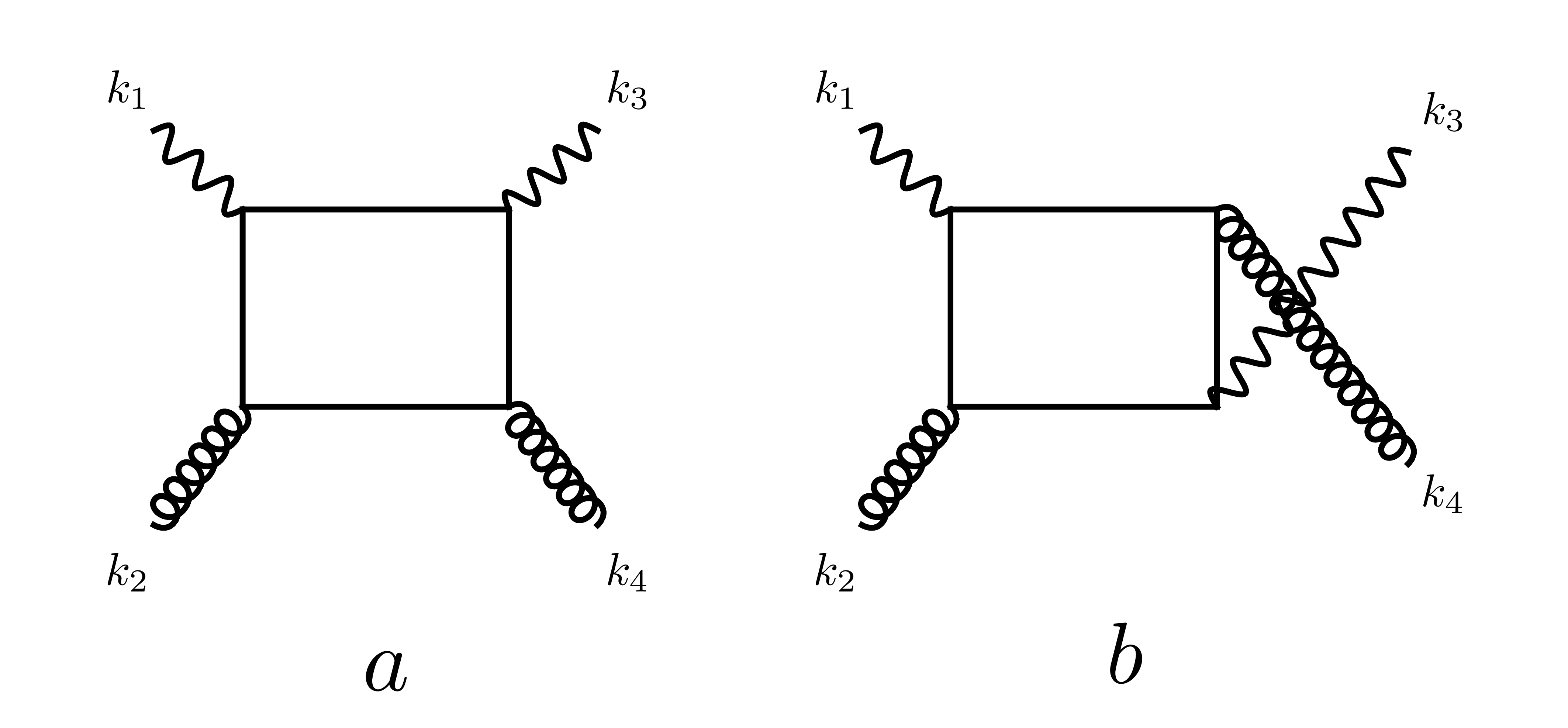}
 \end{center}
 \caption{\label{fig:box_ordering}Two distinct topologies corresponding to the ordering of the proper time coordinates $u_k$.}
 \end{figure}

Finally, we need to integrate over the proper time variables $u_k$ which define the position of the interaction points on the worldline. Using the rotational invariance of the worldline loop~\cite{Schubert:2001he}, it is convenient to fix $u_1 = 0$ (which is equivalent then to $u_1=1$). There are six possible orderings~\cite{Kodaira:1979pa} of the proper time variables $u_k$, which can be split into the two classes  shown in Fig. \ref{fig:box_ordering}a and \ref{fig:box_ordering}b.
 
The first class of diagrams in Fig. \ref{fig:box_ordering}a corresponds to configurations  where there are no gluon insertions between the two electromagnetic currents. There are four such possible orderings: $u_3 > u_2 > u_4$, $u_3 > u_4 > u_2$, $u_2 > u_4 > u_3$, $u_4 > u_2 > u_3$.
By explicit calculation of the integrals over $u_k$ presented below, one finds that all these orderings yield the same contribution with an infrared anomaly pole $\sim \frac{1}{(k_2 + k_4)^2}$. Indeed, as discussed at length in \cite{Usyukina:1992jd}, such graphs have the  generic structure $\frac{1}{(k_2+k_4)^2}\,\frac{1}{(k_1+k_2)^2}\rightarrow \frac{1}{t}\,\frac{1}{Q^2+2xP\cdot q}$, where recall $k_2+k_4=l$,  
$k_1^2= Q^2$ and we define $k_2 = 2\,x P$.

The second class of diagrams in Fig. \ref{fig:box_ordering}b corresponds to a gluon insertion between the two electromagnetic currents (corresponding to a ``cat's eye" topology) and have two possible orderings: $u_2 > u_3 > u_4$ and $u_4 > u_3 > u_2$. These diagrams do not have an infrared pole ! This is because in this case the diagrams have the generic structure $\frac{1}{(k_1+k_4)^2}\,\frac{1}{(k_2+k_3)^2}\rightarrow \frac{1}{Q^2+2xP\cdot q}\,\frac{1}{Q^2+2xP\cdot q}$  and therefore do not have an infrared pole in the forward limit $l_\mu\to 0$. Note that this class of diagrams is suppressed by a factor of $1/Q^2$ relative to those given by Fig. \ref{fig:box_ordering}a and therefore does not contribute  in the Bjorken limit.

Since all four orderings in Fig. \ref{fig:box_ordering}a give the same result, we will fix the ordering of the proper times variables  (multiplying the expression by a factor of $4$) in Eq. (\ref{Gammau}) as
\begin{eqnarray}
&&\Gamma^{\mu\nu\alpha\beta}_A[k_1, k_3, k_2, k_4]\Big|_{Q^2\to \infty}
= 4 \frac{g^2e^2 e_f^2 }{4\pi^2}\epsilon^{\mu\nu\eta}_{\ \ \ \ \kappa} (k_{1\eta} - k_{3\eta})(2\pi)^4\delta^{(4)}(\sum^4_{i=1} k_i)
 \int^1_0 du_3 \int^{u_3}_0 du_4 \int^{u_4}_0 du_2 
 \nonumber\\
&&\times \Big[ - G_B(0, u_2) + G_B(u_2, u_4) - G_B(u_4, 0)
 \Big] ( k^\kappa_{2} + k^\kappa_{4} ) \epsilon^{\alpha\beta\sigma\lambda}  k_{2\sigma} k_{4\lambda}
\Big[ - k_{1} \cdot k_{2} G_B(0, u_2)
 \nonumber\\
 &&- k_{1} \cdot k_{3} G_B(0, u_3) - k_{1}\cdot k_{4} G_B(0, u_4) - k_{2} \cdot k_{3} G_B(u_2, u_3) - k_{2} \cdot k_{4} G_B(u_2, u_4) - k_{3} \cdot k_{4} G_B(u_3, u_4) \Big]^{-2}\,.
 \label{gammaordered}
\end{eqnarray}

We now introduce the change of variables $u_i\rightarrow a_i$, where the latter can be identified as the standard Feynman parameters~\cite{Strassler:1992zr},
\begin{eqnarray}
a_1 = 1 - u_3;\ \ \ a_2 = u_3 - u_4;\ \ \ a_3 = u_4 - u_2;\ \ \ a_4 = u_2\,,
\end{eqnarray}
and rewrite Eq. (\ref{gammaordered}) as
\begin{eqnarray}
&&\Gamma^{\mu\nu\alpha\beta}_A[k_1, k_3, k_2, k_4]\Big|_{Q^2\to \infty}
= 4 \frac{g^2e^2 e_f^2 }{4\pi^2}\epsilon^{\mu\nu\eta}_{\ \ \ \ \kappa} (k_{1\eta} - k_{3\eta})(2\pi)^4\delta^{(4)}(\sum^4_{i=1} k_i)
 \prod^4_{k=1}\int^1_0 da_k ~\delta(1 - \sum^4_{j=1}a_j)
 \nonumber\\
&&\times \Big[ - 2a_2a_4 \Big] ( k^\kappa_{2} + k^\kappa_{4} ) \epsilon^{\alpha\beta\sigma\lambda}  k_{2\sigma} k_{4\lambda}
\Big[ (k_{2} + k_4)^2  a_2a_4 + ( k_{1} + k_2)^2 a_1 a_3 + k^2_{1}  a_1a_4 + k^2_{3} a_1 a_2 \Big]^{-2}\,.
\end{eqnarray}
We have neglected terms $\sim a_1$ in the numerator of this equation; as mentioned previously, these terms, in the Bjorken limit, are  of $O(1/Q^4)$ and are therefore suppressed. To see this more clearly, following \cite{Usyukina:1992jd}, we introduce a further change of variables,
\begin{eqnarray}
a_1 = \alpha;\ \ \ a_i = (1-\alpha)\beta_i,\ \ \ i=2,3,4\,,
\end{eqnarray}
and rewrite the equation as
\begin{eqnarray}
&&\Gamma^{\mu\nu\alpha\beta}_A[k_1, k_3, k_2, k_4]\Big|_{Q^2\to \infty}
= 4 \frac{g^2e^2 e_f^2 }{4\pi^2}\epsilon^{\mu\nu\eta}_{\ \ \ \ \kappa} (k_{1\eta} - k_{3\eta})(2\pi)^4\delta^{(4)}(\sum^4_{i=1} k_i)
 \int^1_0 d\alpha ~\prod^4_{k=2}\int^1_0 d\beta_k ~\delta(1 - \sum_{k=2,3,4}\beta_k ) 
 \nonumber\\
&&\times  \Big[ - 2 \beta_2\beta_4 \Big] ( k^\kappa_{2} + k^\kappa_{4} ) \epsilon^{\alpha\beta\sigma\lambda}  k_{2\sigma} k_{4\lambda}
\Big[ (k_{2} + k_4)^2  (1 - \alpha)\beta_2\beta_4 + ( k_{1} + k_2)^2 \alpha \beta_3 + k^2_{1}  \alpha \beta_4 + k^2_{3} \alpha \beta_2 \Big]^{-2}\,,
\end{eqnarray}

The integral over $\alpha$ can be performed easily, giving
\begin{eqnarray}
\Gamma^{\mu\nu\alpha\beta}_A[k_1, k_3, k_2, k_4]\Big|_{Q^2\to \infty}
&=& - 2\frac{g^2e^2 e_f^2 }{\pi^2}\epsilon^{\mu\nu\eta}_{\ \ \ \ \kappa} (k_{1\eta} - k_{3\eta}) (2\pi)^4\delta^{(4)}(\sum^4_{i=1} k_i)
 ~\frac{( k^\kappa_{2} + k^\kappa_{4} ) \epsilon^{\alpha\beta\sigma\lambda}  k_{2\sigma} k_{4\lambda} }{(k_{2} + k_4)^2  }  
 \nonumber\\
&\times&  \prod^4_{k=2}\int^1_0 d\beta_k ~\delta(1 - \sum_{k=2,3,4}\beta_k ) \frac{1}{ ( k_{1} + k_2)^2 \beta_3 + k^2_{1} \beta_4 + k^2_{3} \beta_2}\,.
\label{Gammawithinfpole}
\end{eqnarray}
Note that the first line on the r.h.s of this equation contains our result (computed explicitly in Appendix C) for the triangle diagram in Eq.~(\ref{eq:VVA1}). 
Specifically, we see that Eq. (\ref{Gammawithinfpole}) has the infrared anomaly pole $\frac{1}{(k_{2} + k_4)^2}\equiv\frac{1}{t}$, which doesn't depend on the integration over $\beta_k$. Hence it is safe to take the forward limit in the integrals over $\beta_k$, and in particular, the forward limit relation between the  momenta of the incoming virtual photons: $k^2_3 = k^2_1$. One can then simplify the equation to read,
\begin{eqnarray}
\Gamma^{\mu\nu\alpha\beta}_A[k_1, k_3, k_2, k_4]\Big|_{Q^2\to \infty}
&=& - 2\frac{g^2e^2 e_f^2 }{\pi^2}\epsilon^{\mu\nu\eta}_{\ \ \ \ \kappa} (k_{1\eta} - k_{3\eta}) (2\pi)^4\delta^{(4)}(\sum^4_{i=1} k_i)
 ~\frac{( k^\kappa_{2} + k^\kappa_{4} ) \epsilon^{\alpha\beta\sigma\lambda}  k_{2\sigma} k_{4\lambda} }{(k_{2} + k_4)^2  }  
 \nonumber\\
&\times&  \prod^4_{k=2}\int^1_0 d\beta_k ~\delta(1 - \sum_{k=2,3,4}\beta_k ) \frac{1}{ 2 k_1 \cdot k_2 \beta_3   + k^2_{1} }\,.
\end{eqnarray}
Next, one can straightforwardly perform the integration over the $\beta_k$ variables, which gives
\begin{eqnarray}
\Gamma^{\mu\nu\alpha\beta}_A[k_1, k_3, k_2, k_4]\Big|_{Q^2\to \infty}
&=& - 2\frac{g^2e^2 e_f^2 }{\pi^2}\epsilon^{\mu\nu\eta}_{\ \ \ \ \kappa} (k_{1\eta} - k_{3\eta}) (2\pi)^4\delta^{(4)}(\sum^4_{i=1} k_i)
 ~\frac{( k^\kappa_{2} + k^\kappa_{4} ) \epsilon^{\alpha\beta\sigma\lambda}  k_{2\sigma} k_{4\lambda} }{(k_{2} + k_4)^2  }  
 \nonumber\\
&\times&  \frac{1}{2 k_1 \cdot k_2}\Big[ \big( 1  + \frac{k^2_{1}}{2 k_1 \cdot k_2} \big)\ln\Big[ \frac{2 k_1 \cdot k_2 + k^2_{1}}{k^2_{1}}\Big]  - 1 \Big]\,,
\label{Gammaaftb}
\end{eqnarray}
where we used the identity
\begin{eqnarray}
\int^1_0 \frac{dx}{ax + b} = \frac{1}{a}\ln\Big[\frac{a+b}{b}\Big]\,,
\end{eqnarray}
employed in standard computations of the box diagram~\cite{tHooft:1978jhc}.

Now substituting the above expression into the box diagram (Eq. (\ref{Hadgamma})), we can write the antisymmetric piece of the polarization tensor in the Bjorken limit as 
\begin{eqnarray}
\label{Gafin}
&&\Gamma^{\mu\nu}_A[k_1, k_3]\Big|_{Q^2\to \infty} = -\frac{g^2e^2 e_f^2 }{\pi^2}\epsilon^{\mu\nu\eta}_{\ \ \ \ \kappa} (k_{1\eta} - k_{3\eta}) ~ \frac{ k^\kappa_{2} + k^\kappa_{4} }{(k_{2} + k_4)^2  }
\nonumber \\
&&\times \int \frac{d^4k_2}{(2\pi)^4} \int \frac{d^4k_4}{(2\pi)^4}~
(2\pi)^4 \delta^{(4)}(\sum^4_{i=1} k_i)
~  \frac{1}{2 k_1 \cdot k_2}\Big[ \big( 1  + \frac{k^2_{1}}{2 k_1 \cdot k_2} \big)\ln\Big[ \frac{2 k_1 \cdot k_2 + k^2_{1}}{k^2_{1}}\Big]  - 1 \Big]~ {\rm Tr_c}(F_{\alpha\beta}(k_2) \tilde{F}^{\alpha\beta}(k_4))\,.
\end{eqnarray}
In arriving at this expression, we have generalized the expression in terms of derivatives of the background fields to express it in terms of the field strength tensors $F^{\mu\nu}$ and its dual ${\tilde F}^{\mu\nu}$, where recall $\tilde{F}_{\mu\nu} = \frac{1}{2}\,\epsilon_{\mu\nu\rho\sigma}F^{\rho\sigma}$.

Finally, substituting Eq. (\ref{Gafin}) into the antisymmetric piece of the hadron tensor (Eq.~(\ref{WModFu})), we obtain
\begin{eqnarray}
i\tilde{W}_f^{\mu\nu}(q, P, S) \Big|_{Q^2\to \infty}
&=& - \frac{g^2 e_f^2 }{\pi^3}\,{\rm Im}~ \epsilon^{\mu\nu\eta}_{\ \ \ \ \kappa} q_{\eta} ~ 
~ \int \frac{d^4k}{(2\pi)^4}~
~  \frac{1}{ 2 q \cdot k}\Big[ \Big( 1  + \frac{q^2}{2 q \cdot k} \Big)\ln\Big[ \frac{2 q \cdot k + q^2}{q^2}\Big]  - 1 \Big]
\nonumber\\
&\times& \int d^4 z\, e^{-ik z } \lim_{l_\kappa \to 0}\frac{ l^\kappa }{l^2  }\langle P^\prime,S| {\rm Tr_c} F_{\alpha\beta}(z) \tilde{F}^{\alpha\beta}(0) |P,S\rangle\,,
\label{WantisimA}
\end{eqnarray}
where ${\tilde W}^{\mu\nu}(q, P, S)=\sum_f\tilde{W}_f^{\mu\nu}(q, P, S)$ and  ${P^\prime}^\kappa-P^\kappa = l^\kappa$. 
In writing this result, we used Eq. (\ref{fourbackground}), integrated over intermediate momenta and coordinates, and performed an analytical continuation of the expression to Minkowski space-time.

Following  \cite{Leader:2001gr}, we will make a high energy approximation and write $k^\mu \approx (k\cdot n) P^\mu$, where $n$ is a dimensionful vector such that $n^2 = 0$ and $n\cdot P = 1$. As a result, we can rewrite the equation as
\begin{eqnarray}
i\tilde{W}_f^{\mu\nu}(q, P, S) \Big|_{Q^2\to \infty}
&=& - \frac{g^2 e_f^2 }{\pi^3}\,{\rm Im} \frac{1}{ 2 P \cdot q} \epsilon^{\mu\nu\eta}_{\ \ \ \ \kappa} q_{\eta} ~
 \int^1_0 dx 
~  \frac{1}{x}\Big[ \Big( 1  - \frac{x_B}{x} \Big)\ln\Big[ \frac{x - x_B}{ - x_B} \Big]  - 1 \Big]
\nonumber\\
&\times& \int \frac{d^4k}{(2\pi)^4}~\delta(x - k\cdot n) \lim_{l_\kappa \to 0}\frac{ l^\kappa }{l^2  } \int d^4 z\, e^{-ik z } \langle P^\prime,S| {\rm Tr_c} F_{\alpha\beta}(z) \tilde{F}^{\alpha\beta}(0) |P,S\rangle\,. 
\end{eqnarray}
 The imaginary part of the expression on the r.h.s is obtained from the identity $\ln(x) = \ln(|x|) + i\pi$, which also requires that we impose $x\geq x_B$. Hence,
\begin{eqnarray}
i\tilde{W}_f^{\mu\nu}(q, P, S) \Big|_{Q^2\to \infty}
&=& \frac{g^2 e_f^2 }{\pi^2} \frac{1}{ 2 P \cdot q} \epsilon^{\mu\nu\eta}_{\ \ \ \ \kappa} q_{\eta} ~ 
  \int^1_{x_B} \frac{dx}{x} 
~ \Big( 1  - \frac{x_B}{x} \Big)
\nonumber\\
&\times& \int \frac{d^4k}{(2\pi)^4}~\delta(x - k\cdot n) \lim_{l_\kappa \to 0}\frac{ l^\kappa }{l^2  }\,\int d^4 z e^{-ik z } \langle P^\prime,S| {\rm Tr_c} F_{\alpha\beta}(z) \tilde{F}^{\alpha\beta}(0) |P,S\rangle\,.
\end{eqnarray}
Integrating over $k^\mu$ further yields,
\begin{eqnarray}
i\tilde{W}_f^{\mu\nu}(q, P, S) \Big|_{Q^2\to \infty}
= \frac{4}{ P \cdot q} \epsilon^{\mu\nu\eta}_{\ \ \ \ \kappa} q_{\eta} ~
 \int^1_{x_B} \frac{dx}{x} 
~ \Big( 1  - \frac{x_B}{x} \Big) \int \frac{d\xi}{2\pi} e^{-i\xi x } \lim_{l_\kappa \to 0}\frac{ l^\kappa }{l^2  } \langle P^\prime,S| \frac{\alpha_s e_f^2}{2\pi} {\rm Tr_c} F_{\alpha\beta}(\xi n) 
\tilde{F}^{\alpha\beta}(0) |P,S\rangle\,,
\nonumber\\
\label{WAresBj}
\end{eqnarray}
where the r.h.s is rewritten in terms of the matrix element of the nonlocal operator 
\begin{eqnarray}
\int \frac{d\xi}{2\pi} e^{-i\xi x }\lim_{l_\kappa \to 0}\frac{ l^\kappa }{l^2  }  \langle P^\prime,S| \frac{\alpha_s e_f^2}{2\pi} {\rm Tr_c} F_{\alpha\beta}(\xi n) \tilde{F}^{\alpha\beta}(0) |P,S\rangle\,.
\end{eqnarray}
Comparing Eq. (\ref{WAresBj}) with the general tensorial decomposition of the antisymmetric part of the hadron tensor in Eq.~(\ref{WA}), allows us to extract our final result for $g_1(x_B, Q^2)$ in the Bjorken limit:
\begin{eqnarray}
\label{eq:g1-Bj-final}
 S^\mu g_1(x_B, Q^2)\Big|_{Q^2\to \infty} = \sum_f e_f^2 \frac{\alpha_s}{i\pi M_N}
 \int^1_{x_B} \frac{dx}{x} 
~ \Big( 1  - \frac{x_B}{x} \Big) \int \frac{d\xi}{2\pi} e^{-i\xi x }\lim_{l_\mu \to 0} \frac{ l^\mu }{l^2  }\langle P^\prime,S| {\rm Tr_c} F_{\alpha\beta}(\xi n) \tilde{F}^{\alpha\beta}(0) |P,S\rangle\,.
\label{resultBj}
\end{eqnarray}

Finally, as noted in the introduction, the first moment of $g_1(x_B,Q^2)$ is simply related to $\Sigma(Q^2)$ and is given by 
\begin{eqnarray}
 S^\mu\int^1_0 dx_B \,g_1(x_B, Q^2)\Big|_{Q^2\to \infty} = \sum_f e_f^2 \frac{\alpha_S}{2 i\pi M_N}\,
 \lim_{l_\mu \to 0} \frac{ l^\mu }{l^2  } \langle P^\prime ,S| {\rm Tr_c} F_{\alpha\beta}(0) \tilde{F}^{\alpha\beta}(0) |P,S\rangle\,.
\end{eqnarray}
Note that this expression is the first term on the r.h.s of Eq.~(5.19) in Ref.~\cite{Jaffe:1989jz}.  Clearly, this differs from Eq.~(\ref{eq:sigma-anomaly}) because, as discussed in the introduction, the contribution from the pseudoscalar sector is not included here.

\subsection{Discussion of Bjorken limit result in Eq.~(\ref{eq:g1-Bj-final})}
\label{subsec:Bj-discussion}

The result for $g_1(x_B, Q^2)$ is one of the principal results of this paper. It shows that the box diagram for polarized DIS in the Bjorken limit is dominated by the triangle anomaly and if not regulated appropriately will diverge in the forward limit $l^\kappa\rightarrow 0$. Though nonlocal, the nonlocality must be interpreted as a smearing of the operator corresponding to the bare topological charge density $Q_B = \frac{\alpha_S}{8\pi} {\rm Tr}\left(F {\tilde F}\right)$. This is because our derivation parallels exactly the derivation (worked out in Appendix~\ref{sec:triangle}) of the triangle anomaly in the worldline formalism. Indeed, as noted earlier, the first line of Eq.~(\ref{Gammawithinfpole}) already contains the infrared pole of the anomaly, with the terms on the second line only giving finite contributions. 

As we will discuss further in Papers II\,\& III, this operator $Q_B$ undergoes renormalization with evolution  and mixes, via the anomaly, with the $t$-channel exchange of a massless isosinglet pseudoscalar $\eta_0$ to generate the massive $\eta^\prime$ meson. Thus as emphasized by Veneziano~\cite{Veneziano:1989ei}, how one recovers a finite result for $g_1(x_B,Q^2)$ in the forward limit $l^\kappa\rightarrow 0$ is deeply tied to the resolution of the $U_A(1)$ problem in QCD. This is independent of whether the underlying dynamical mechanism is  due to instantons or other nonperturbative phenomena and a consequence of anomalous chiral Ward identities~\cite{Shore:2007yn}.

With regard to the underlying dynamical mechanism, we should emphasize that the computation of the matrix element on the r.h.s of the above expression is nontrivial. Firstly, if we  naively take the forward limit, the matrix element vanishes. One way to properly interpret this matrix element (along with the factor $l^\mu/l^2$) is to consider its path integral realization as a convolution of the operator and the density matrix of states. As is well known, the  anomaly arises from the measure of the path integral~\cite{Fujikawa:1979ay}. In the worldline formalism, this contribution of the measure can be reexpressed as the imaginary part of the one loop effective action (or equivalently, the phase of the fermion determinant). Thus the  effect of the anomaly is distinct from that of the real part of the effective action, which is, for instance, responsible for the equations of motion ~\cite{Mueller:2017arw,Mueller:2019gjj}. 

In addition to the anomaly contribution,  the imaginary part of the effective action contains a Wess-Zumino-Witten term for the $\eta_0$, as noted in the introduction, that will cancel the pole of the anomaly in the computation of the path integral~\cite{DHoker:1995aat}. This key point will be discussed further in Paper II. Our perspective is  similar to that of Shore, Narison and  Veneziano~\cite{Narison:1994hv,Shore:2007yn, Narison:1998aq}, who employed the Wess-Zumino effective action~\cite{Wess:1971yu} in their studies to obtain the same results for the first moment of $g_1(x_B,Q^2)$.

To discuss further the implications of our result, observe  that the matrix element in Eq.~(\ref{eq:g1-Bj-final}) has a distinct tensorial structure from that in the polarized gluon distribution\footnote{We omit writing the gauge link between $F^{+\mu}$ and ${\tilde F}^+_{\ \mu}$ required to ensure gauge invariance.}, which is defined as~\cite{Kodaira:1998jn}
\begin{eqnarray}
\label{eq:DeltaGx_B}
\Delta G(x_B,Q^2) &=& \frac{2i}{x_B}\int \frac{d\xi}{2\pi}\,e^{-i\xi x_B} \langle P,S| {\rm Tr_c} n_\alpha F^{\alpha\mu}(\xi n) n^\beta \tilde{F}_{\beta\mu}(0) |P,S\rangle\nonumber \\
&\equiv& \frac{2i}{x_B (P^+)^2}\int \frac{d\xi}{2\pi}\,e^{-i\xi x_B} \langle P,S| {\rm Tr_c} F^{+\mu}(\xi n) \tilde{F}^+_{\ \mu}(0) |P,S\rangle\,.
\end{eqnarray}
More importantly, as we have argued, $g_1(x_B,Q^2)$ in Eq.~(\ref{eq:g1-Bj-final}) is dominated entirely by the anomaly contribution from the imaginary part of the worldline effective action in the path integral for the matrix element. It is therefore unclear how relate it to the r.h.s of the above expression\footnote{We thank Bob Jaffe for bringing his Varenna lectures (unpublished) to our attention, where this point is discussed at some length.} as is often done in perturbative computations in the literature~\cite{Zijlstra:1993sh,Ball:1995ye,Blumlein:1996hb}. Further clarity\footnote{We thank Y. Hatta for an interesting discussion on this point.} on this important issue can be obtained from the explicit computation of the matrix elements for $g_1(x_B,Q^2)$ and $\Delta G(x_B,Q^2)$.

Our results are consistent with perturbative computations of the renormalization group evolution of $\Sigma(Q^2)$ with $Q^2$. 
In pQCD computations, the RG evolution of $\Sigma(Q^2)$ can mix that of the first moment $\Delta G(Q^2)$ of Eq.~(\ref{eq:DeltaGx_B}), the matrix element of the other isosinglet twist two operator in polarized DIS. Their combined evolution is described by a two-by-two matrix of splitting functions of polarized quarks splitting into softer polarized quarks or gluons and likewise, polarized gluons splitting into softer polarized gluons or quarks. These splitting functions were  computed to leading order (LO) in \cite{Ahmed:1976ee, Altarelli:1977zs,Kodaira:1979pa}, to next-to-leading order (NLO) in \cite{Mertig:1995ny, Vogelsang:1995vh,Vogelsang:1996im} and to next-to-next-to-leading order (NNLO) in 
\cite{Moch:2014sna,Moch:2015usa}. As shown already for the leading order splitting functions in \cite{Altarelli:1990jp}, but remarkably also for the values of the splitting functions computed to NNLO accuracy~\cite{deFlorian:2019egz},  $\Sigma(Q^2)$ does not mix with $\Delta G(Q^2)$. This is precisely what we would expect since $\Sigma(Q^2)$ has a distinct topological structure determined by the chiral anomaly alone. On the other hand, while $\Delta G$ does not influence the evolution of $\Sigma(Q^2)$, the converse needn't be true, as also observed in the pQCD computations. We would argue that this is because $\Delta G$ is sensitive to fermion  zero modes that are influenced by the anomaly. 

\section{The triangle anomaly in the Regge limit of the box diagram}
\label{sec:Regge}
We  now turn our attention to the computation of the box diagram in the other interesting asymptotics of QCD, the Regge limit of $x_B\ll 1$ for fixed $Q^2\gg \Lambda_{\rm QCD}^2$.  In this regime, the quarks in the box diagram suffer a nearly instantaneous shock wave interaction with  the background gluons, which corresponds to taking the limit $u_2\rightarrow u_4$ in Eq.~(\ref{GammawithIinu}), given by Eq.~(\ref{GammawithIRj}). As we shall see, the derivation closely parallels that of our discussion of the Bjorken limit, with some small but important differences that we will draw the reader's attention to. 

\subsection{Worldline computation of the box diagram in the Regge limit}
\label{subsec:Rj-worldline}

A cursory examination reveals that taking the limit $u_2\rightarrow u_4$ in the box diagram is very similar to taking  $u_1\rightarrow u_3$ in the Bjorken limit. Thus one can obtain the sum 
of the coefficients in Eq. (\ref{GammawithIRj}) by borrowing Eq. (\ref{suminBj}) from the previous section and making the following substitutions:  $\mu\leftrightarrow\alpha$, $\nu\leftrightarrow\beta$, $k_1\leftrightarrow k_2$, $k_3\leftrightarrow k_4$, $u_1 \leftrightarrow u_2$, $u_3\leftrightarrow u_4$:
\begin{eqnarray}
\label{eq:Regge-tensor-sum}
&&\Big(\sum^9_{n=1}\mathcal{I}^{\mu\nu\alpha\beta}_{n;(u_1,u_2,u_3,u_4)}[k_1, k_3, k_2, k_4]\Big)\Big|_{u_2 = u_4} = \epsilon^{\alpha\beta\eta}_{\ \ \ \ \kappa} (k_{2\eta} - k_{4\eta})
\nonumber\\
&&\times \Big( k_1 \cdot k_3 \Big[ \dot{G}^2_B(u_2, u_3) - \dot{G}_B(u_2, u_3) \dot{G}_B(u_1, u_3) + \dot{G}_B(u_2, u_1) \big(\dot{G}_B(u_2, u_1) + \dot{G}_B(u_1, u_3) + \dot{G}_B(u_3, u_2) \big)  \Big] \epsilon^{\kappa\mu\nu\sigma} k_{1\sigma}
\nonumber\\
&&+ k_1\cdot k_3 \Big[ -\dot{G}^2_B(u_2, u_1) - \dot{G}_B(u_2, u_1) \dot{G}_B(u_1, u_3) + \dot{G}_B(u_2, u_3) \big(\dot{G}_B(u_2, u_1) + \dot{G}_B(u_1, u_3) + \dot{G}_B(u_3, u_2)\big) \Big] \epsilon^{\kappa\mu\nu\sigma} k_{3\sigma}
\nonumber\\
&&+ \Big[ - \dot{G}_B(u_2, u_3) \dot{G}_B(u_1, u_3) +  \dot{G}^2_B(u_2, u_3)
+ \dot{G}_B(u_2, u_1) \big(\dot{G}_B(u_2, u_1) + \dot{G}_B(u_1, u_3) + \dot{G}_B(u_3, u_2)\big) \Big] \epsilon^{\kappa\mu\sigma\lambda} k^\nu_{1} k_{1\sigma} k_{3\lambda}
\nonumber\\
&&+ \dot{G}^2_B(u_2, u_3) \epsilon^{\kappa\mu\sigma\lambda} k^\nu_{3} k_{1\sigma} k_{3\lambda}
- \dot{G}^2_B(u_2, u_1) \epsilon^{\kappa\nu\sigma\lambda} k^\mu_{1} k_{1\sigma} k_{3\lambda}
\nonumber\\
&&+ \Big[ - \dot{G}_B(u_2, u_1) \dot{G}_B(u_1, u_3) - \dot{G}^2_B(u_2, u_1)
+ \dot{G}_B(u_2, u_3) \big(\dot{G}_B(u_2, u_1) + \dot{G}_B(u_1, u_3) + \dot{G}_B(u_3, u_2)\big) \Big] \epsilon^{\kappa\nu\sigma\lambda} k^\mu_{3} k_{1\sigma} k_{3\lambda}
\nonumber\\
&&- \epsilon^{\mu\lambda\nu\sigma}  k^\kappa_{1} k_{1\lambda} k_{3\sigma} - \epsilon^{\mu\lambda\nu\sigma} k^\kappa_{3} k_{1\lambda} k_{3\sigma}\Big)\,.
\end{eqnarray}
In analogy to the previous derivation, this expression can be simplified by  employing the cyclic identity for the Levi-Civita tensor (in footnote~\ref{cyclic}), which gives,
\begin{eqnarray}
\label{eq:LC_R1}
- \epsilon^{\mu\lambda\nu\sigma}  k^\kappa_{1} k_{1\lambda} k_{3\sigma} = - k_{1}\cdot k_{3} \epsilon^{\kappa\mu\nu\sigma} k_{1\sigma} - \epsilon^{\kappa\mu\sigma\lambda} k^\nu_{1} k_{1\sigma} k_{3\lambda} + \epsilon^{\kappa\nu\sigma\lambda} k^\mu_{1} k_{1\sigma} k_{3\lambda} + k^2_{1} \epsilon^{\nu\sigma\kappa\mu} k_{3\sigma}\,,
\end{eqnarray}
and
\begin{eqnarray}
\label{eq:LC_R2}
&&- \epsilon^{\mu\lambda\nu\sigma} k^\kappa_{3} k_{1\lambda} k_{3\sigma} = k_{1}\cdot k_{3} \epsilon^{\kappa\mu\nu\sigma} k_{3\sigma} + \epsilon^{\kappa\nu\sigma\lambda} k^\mu_{3} k_{1\sigma} k_{3\lambda} - \epsilon^{\kappa\mu\sigma\lambda} k^\nu_{3} k_{1\sigma} k_{3\lambda} - k^2_{3} \epsilon^{\kappa\mu\nu\sigma} k_{1\sigma}\,.
\end{eqnarray}
Note that unlike  Eq.~(\ref{Levi-Civita}), where the term corresponding to the virtuality of the background gluons was set to be $k_2^2=k_4^2=0$ in the Bjorken limit, here the analogous terms give $k_1^3= k_3^2=Q^2\neq 0$.
With Eqs.~(\ref{eq:LC_R1}) and (\ref{eq:LC_R2}), we can rewrite Eq.~(\ref{eq:Regge-tensor-sum}) as
\begin{eqnarray}
&&\Big(\sum^9_{n=1}\mathcal{I}^{\mu\nu\alpha\beta}_{n;(u_1,u_2,u_3,u_4)}[k_1, k_3, k_2, k_4]\Big)\Big|_{u_2 = u_4} = \epsilon^{\alpha\beta\eta}_{\ \ \ \ \kappa} (k_{2\eta} - k_{4\eta})
\nonumber\\
&&\times \Big( \Big[ 1 - \dot{G}^2_B(u_2, u_3) + \dot{G}_B(u_2, u_3) \dot{G}_B(u_1, u_3) - \dot{G}_B(u_2, u_1) \big(\dot{G}_B(u_2, u_1) + \dot{G}_B(u_1, u_3) + \dot{G}_B(u_3, u_2) \big)  \Big] k^\kappa_{1} \epsilon^{\mu\sigma\lambda\nu} k_{1\sigma} k_{3\lambda}
\nonumber\\
&&+ \Big[ -1 + \dot{G}^2_B(u_2, u_1) + \dot{G}_B(u_2, u_1) \dot{G}_B(u_1, u_3) - \dot{G}_B(u_2, u_3) \big(\dot{G}_B(u_2, u_1) + \dot{G}_B(u_1, u_3) + \dot{G}_B(u_3, u_2)\big) \Big]k^\kappa_{3} \epsilon^{\nu\sigma\lambda\mu} k_{1\sigma} k_{3\lambda}
\nonumber\\
&&+ \Big[ - \dot{G}^2_B(u_2, u_1) - \dot{G}_B(u_2, u_1) \dot{G}_B(u_1, u_3) +  \dot{G}_B(u_2, u_3) \dot{G}_B(u_2, u_1) + \dot{G}_B(u_2, u_3) \dot{G}_B(u_1, u_3) \Big] \epsilon^{\kappa\mu\sigma\lambda} k^\nu_{3} k_{1\sigma} k_{3\lambda}
 \nonumber\\
 &&+ \Big[ \dot{G}^2_B(u_2, u_3) - \dot{G}_B(u_2, u_3) \dot{G}_B(u_1, u_3) 
 + \dot{G}_B(u_2, u_1) \dot{G}_B(u_1, u_3) + \dot{G}_B(u_2, u_1) \dot{G}_B(u_3, u_2) \Big] \epsilon^{\kappa\nu\sigma\lambda} k^\mu_{1} k_{1\sigma} k_{3\lambda} 
 \nonumber\\
 &&+\Big[ \dot{G}^2_B(u_2, u_3) - \dot{G}_B(u_2, u_3) \dot{G}_B(u_1, u_3) + \dot{G}_B(u_2, u_1) \big(\dot{G}_B(u_2, u_1) + \dot{G}_B(u_1, u_3) + \dot{G}_B(u_3, u_2) \big)  \Big] k^2_{1} \epsilon^{\nu\sigma\kappa\mu} k_{3\sigma}
 \nonumber\\
&&+ \Big[ \dot{G}^2_B(u_2, u_1) + \dot{G}_B(u_2, u_1) \dot{G}_B(u_1, u_3) - \dot{G}_B(u_2, u_3) \big(\dot{G}_B(u_2, u_1) + \dot{G}_B(u_1, u_3) + \dot{G}_B(u_3, u_2)\big) \Big] k^2_{3} \epsilon^{\mu\kappa\nu\sigma} k_{1\sigma}
\end{eqnarray}

From gauge invariance, the terms proportional to $\sim k^\mu_1$ and $\sim k^\nu_3$ do not  contribute to the final answer. Further, the terms proportional to  $\sim k^2_1$ and $\sim k^2_3$ will be proportional to $ \frac{Q^2}{2P\cdot q} \equiv x_B$ in the forward limit and are  suppressed, relative to the other terms, in Regge asymptotics.  The leading contributions are therefore given by the first two lines in the previous equation, which can be written in a greatly simplified form as 
\begin{eqnarray}
&&\Big(\sum^9_{n=1}\mathcal{I}^{\mu\nu\alpha\beta}_{n;(u_1,u_2,u_3,u_4)}[k_1, k_3, k_2, k_4]\Big)\Big|_{u_2 = u_4} = - \frac{1}{2} \epsilon^{\alpha\beta\eta}_{\ \ \ \ \kappa} (k_{2\eta} - k_{4\eta})
\nonumber \\
&&\times \Big[ -2 + \dot{G}^2_B(u_2, u_1) - \dot{G}^2_B(u_1, u_3) + \dot{G}^2_B(u_3, u_2)
+ \big(\dot{G}_B(u_2, u_1) + \dot{G}_B(u_1, u_3) + \dot{G}_B(u_3, u_2) \big)^2 \Big] 
(k^\kappa_{1} + k^\kappa_{3}) \epsilon^{\mu\nu\sigma\lambda} k_{1\sigma} k_{3\lambda}\,.
\nonumber\\
\end{eqnarray}

Using Eqs. (\ref{gfprod}) and (\ref{b2dot}) to further simplify the expression in the brackets, we obtain 
\begin{eqnarray}
\Big(\sum^9_{n=1}\mathcal{I}^{\mu\nu\alpha\beta}_{n;(u_1,u_2,u_3,u_4)}[k_1, k_3, k_2, k_4]\Big)\Big|_{u_2 = u_4} 
&=& - 2 \epsilon^{\alpha\beta\eta}_{\ \ \ \ \kappa} (k_{2\eta} - k_{4\eta})
 \Big[ - G_B(u_2, u_1) + G_B(u_1, u_3) - G_B(u_3, u_2) \Big] \nonumber \\
 &\times& (k^\kappa_{1} + k^\kappa_{3}) \epsilon^{\mu\nu\sigma\lambda} k_{1\sigma} k_{3\lambda}\,.
\end{eqnarray}
Finally,  we can reexpress this result as\footnote{We employ here energy-momentum conservation ($k_1^\kappa+k_3^\kappa = - k_2^\kappa - k_4^\kappa$), and the identity $\epsilon^{\mu\nu\sigma\lambda} k_{1\sigma}k_{3\sigma} = \frac{1}{2}\epsilon^{\mu\nu\sigma\lambda} (k_1-k_3)_\sigma (k_1+k_3)_\lambda$  (likewise, $\epsilon^{\alpha\beta\sigma\lambda} (k_2-k_4)_\sigma (k_2+k_4)_\lambda = 2 \epsilon^{\alpha\beta\sigma\lambda} k_{2\sigma} k_{4\lambda}$).} 
\begin{eqnarray}
\Big(\sum^9_{n=1}\mathcal{I}^{\mu\nu\alpha\beta}_{n;(u_1,u_2,u_3,u_4)}[k_1, k_3, k_2, k_4]\Big)\Big|_{u_2 = u_4} 
\label{sumofcoefRj}
&=& - 2 \epsilon^{\mu\nu\eta}_{\ \ \ \ \kappa} (k_{1\eta}  -  k_{3\eta})
 \Big[ - G_B(u_2, u_1) + G_B(u_1, u_3) - G_B(u_3, u_2) \Big] \nonumber \\
 &\times& (k^\kappa_{2}  + k^\kappa_{4} ) \epsilon^{\alpha\beta\sigma\lambda} k_{2\sigma} k_{4\lambda}\,.
\end{eqnarray}
This expression has a very similar structure when compared to Eq. (\ref{sumcoefBj}) that we derived in the Bjorken limit.

Substituting this sum of coefficients into the expression in Eq.~(\ref{GammawithIRj}) for the box diagram in the Regge limit yields
\begin{eqnarray}
&&\Gamma^{\mu\nu\alpha\beta}_A[k_1, k_3, k_2, k_4]\Big|_{x_{Bj}\to 0} = \frac{g^2e^2 e_f^2 }{4\pi^2} \epsilon^{\mu\nu\eta}_{\ \ \ \ \kappa} (k_{1\eta}  -  k_{3\eta}) (2\pi)^4\delta^{(4)}(\sum^4_{i=1} k_i) \prod^4_{k=1}\int^1_0 du_k 
 \nonumber\\
 &&\times \Big[ - G_B(u_2, u_1) + G_B(u_1, u_3) - G_B(u_3, u_2) \Big] (k^\kappa_{2}  + k^\kappa_{4} ) \epsilon^{\alpha\beta\sigma\lambda} k_{2\sigma} k_{4\lambda} \Big[ - k_{1} \cdot k_{2} G_B(u_1, u_2)
\nonumber\\
 &&- k_{1} \cdot k_{3} G_B(u_1, u_3) - k_{1}\cdot k_{4} G_B(u_1, u_4) - k_{2} \cdot k_{3} G_B(u_2, u_3) - k_{2} \cdot k_{4} G_B(u_2, u_4) - k_{3} \cdot k_{4} G_B(u_3, u_4) \Big]^{-2}\,.
\end{eqnarray}
Using rotational invariance of the worldline, we fix $u_2 = 0$. In exact analogy to our calculation in the Bjorken limit, there are six possible orderings of the proper time variables $u_k$ and these can be split, as previously,
into the two topologies corresponding to Figs. \ref{fig:box_ordering}a and \ref{fig:box_ordering}b. As noted, the two contributions corresponding to Fig. \ref{fig:box_ordering}b have the structure $\frac{1}{(k_1+k_4)^2} \frac{1}{(k_2+k_3)^2}\rightarrow \frac{1}{(Q^2+2\,xP\cdot q)^2}$  while the four contributions corresponding to  Fig. \ref{fig:box_ordering}a are identical, having the structure $\frac{1}{l^2(Q^2+2\,xP\cdot q)}$ and contain the infrared pole in $l^2$. Since the ``cat's eye" diagrams are finite and in addition have a relative suppression $l^2/M^2$, where $M^2\equiv 2x P\cdot q\gg Q^2$ in Regge asymptotics\footnote{The longitudinal momentum fraction $x$ of the background gluons can be taken to be small but finite as $x_B\rightarrow 0$.}, we will consider only the four diagrams corresponding to Fig. \ref{fig:box_ordering}a henceforth.

Since the contributions of all for diagrams are identical, it is sufficient to examine one particular ordering of the proper time variables, which we will choose to be $u_4 > u_3 > u_1$ and multiply the expression by a factor of 4. We then get 
\begin{eqnarray}
&&\Gamma^{\mu\nu\alpha\beta}_A[k_1, k_3, k_2, k_4]\Big|_{x_{Bj}\to 0} = 4\frac{g^2e^2 e_f^2 }{4\pi^2} \epsilon^{\mu\nu\eta}_{\ \ \ \ \kappa} (k_{1\eta}  -  k_{3\eta}) (2\pi)^4\delta^{(4)}(\sum^4_{i=1} k_i) \int^1_0 du_4 \int^{u_4}_0 du_3  \int^{u_3}_0 du_1 
 \nonumber\\
 &&\times \Big[ - G_B(0, u_1) + G_B(u_1, u_3) - G_B(u_3, 0) \Big] (k^\kappa_{2}  + k^\kappa_{4} ) \epsilon^{\alpha\beta\sigma\lambda} k_{2\sigma} k_{4\lambda} \Big[ - k_{1} \cdot k_{2} G_B(u_1, 0)
\nonumber\\
 &&- k_{1} \cdot k_{3} G_B(u_1, u_3) - k_{1}\cdot k_{4} G_B(u_1, u_4) - k_{2} \cdot k_{3} G_B(0, u_3) - k_{2} \cdot k_{4} G_B(0, u_4) - k_{3} \cdot k_{4} G_B(u_3, u_4) \Big]^{-2}\,.
\end{eqnarray}

Introducing the Feynman parameters, 
\begin{eqnarray}
a_1 = 1 - u_4;\ \ \ a_2 = u_4 - u_3;\ \ \ a_3 = u_3 - u_1;\ \ \ a_4 = u_1\,,
\end{eqnarray}
and using the explicit form of the boson worldline propagator, 
\begin{eqnarray}
G_B(u_i, u_j) = |u_i - u_j| - (u_i - u_j)^2\,,
\end{eqnarray}
we can rewrite the equation as
\begin{eqnarray}
&&\Gamma^{\mu\nu\alpha\beta}_A[k_1, k_3, k_2, k_4]\Big|_{x_{Bj}\to 0} = 4\frac{g^2e^2 e_f^2 }{4\pi^2} \epsilon^{\mu\nu\eta}_{\ \ \ \ \kappa} (k_{1\eta}  -  k_{3\eta}) (2\pi)^4\delta^{(4)}(\sum^4_{i=1} k_i) \prod^4_{k=1}\int^1_0 da_k ~\delta(1 - \sum^4_{j=1}a_j)
 \nonumber\\
 &&\times \Big[ - 2a_2a_4  \Big] (k^\kappa_{2}  + k^\kappa_{4} ) \epsilon^{\alpha\beta\sigma\lambda} k_{2\sigma} k_{4\lambda} \Big[ (k_{2} + k_4)^2 a_2 a_4 + (k_{1} + k_2)^2 a_1a_3 + k^2_{1} a_3 a_4 + k^2_{2} a_1a_4 + k^2_3 a_2a_3 + k^2_4 a_1a_2 \Big]^{-2}\,.
\end{eqnarray}
where in the numerator we neglect terms\footnote{We remind the reader that due to rotational invariance of the worldline, the proper time coordinates $u_2 = 0$ and $u_2 = 1$ correspond to the same point.} proportional to $a_1 \equiv u_2 - u_4$. In  the denominator, in contrast to our discussion in the Bjorken limit, we will keep the dependence on the virtuality of the background gluons $k^2_{2}$ and $k^2_4$; these  regulate the integrals over the $\beta$ variables.

We next make the change of variables,
\begin{eqnarray}
a_1 = \alpha;\ \ \ a_i = (1-\alpha)\beta_i,\ \ \ i=2,3,4\,,
\end{eqnarray}
which yields
\begin{eqnarray}
&&\Gamma^{\mu\nu\alpha\beta}_A[k_1, k_3, k_2, k_4]\Big|_{x_{Bj}\to 0}
\\
&&= 4\frac{g^2e^2 e_f^2 }{4\pi^2} \epsilon^{\mu\nu\eta}_{\ \ \ \ \kappa} (k_{1\eta}  -  k_{3\eta}) (2\pi)^4\delta^{(4)}(\sum^4_{i=1} k_i) \int^1_0 d\alpha \prod^4_{k=2}\int^1_0 d\beta_k ~ \delta(1 - \sum^4_{j=2}\beta_j) \Big[ - 2 \beta_2 \beta_4  \Big] (k^\kappa_{2}  + k^\kappa_{4} ) \epsilon^{\alpha\beta\sigma\lambda} k_{2\sigma} k_{4\lambda}
\nonumber\\
 &&\times  \Big[ (k_{2} + k_4)^2 (1-\alpha)\beta_2 \beta_4 + (k_{1} + k_2)^2 \alpha\beta_3 + k^2_{1} (1-\alpha)\beta_3 \beta_4 + k^2_{2} \alpha \beta_4 + k^2_3 (1-\alpha)\beta_2\beta_3 + k^2_4 \alpha \beta_2 \Big]^{-2}\,.
 \nonumber
\end{eqnarray}
where in the numerator we neglect terms\footnote{These terms are of order 
$Q^2/M^2\ll 1$ in Regge asymptotics.} proportional to $\alpha = u_2 - u_4$.

The integration over $\alpha$ yields
\begin{eqnarray}
&&\Gamma^{\mu\nu\alpha\beta}_A[k_1, k_3, k_2, k_4]\Big|_{x_{Bj}\to 0} = - 2 \frac{g^2e^2 e_f^2 }{\pi^2} \epsilon^{\mu\nu\eta}_{\ \ \ \ \kappa} (k_{1\eta}  -  k_{3\eta}) (2\pi)^4\delta^{(4)}(\sum^4_{i=1} k_i) ~(k^\kappa_{2}  + k^\kappa_{4} ) \epsilon^{\alpha\beta\sigma\lambda} k_{2\sigma} k_{4\lambda}
\label{Rjfrac}\\
 &&\times  \prod^4_{k=2}\int^1_0 d\beta_k ~ \delta(1 - \sum^4_{j=2}\beta_j) \frac{\beta_2 \beta_4}{(k_{2} + k_4)^2 \beta_2 \beta_4 + k^2_{1} \beta_3 \beta_4 + k^2_3 \beta_2 \beta_3} \times \frac{1}{ (k_{1} + k_2)^2 \beta_3  + k^2_{2} \beta_4 + k^2_4 \beta_2}\,.
 \nonumber
\end{eqnarray}
We will now equate $k^2_3 = k^2_1$ and $k^2_4 = k^2_2$, which are both valid in the forward limit. As we will see, these substitutions don't affect the infrared $1/l^2$ pole in  the final answer. Hence, 
\begin{eqnarray}
&&\Gamma^{\mu\nu\alpha\beta}_A[k_1, k_3, k_2, k_4]\Big|_{x_{Bj}\to 0} = - 2 \frac{g^2e^2 e_f^2 }{\pi^2} \epsilon^{\mu\nu\eta}_{\ \ \ \ \kappa} (k_{1\eta}  -  k_{3\eta}) (2\pi)^4\delta^{(4)}(\sum^4_{i=1} k_i) ~(k^\kappa_{2}  + k^\kappa_{4} ) \epsilon^{\alpha\beta\sigma\lambda} k_{2\sigma} k_{4\lambda}
\\
 &&\times  \prod^4_{k=2}\int^1_0 d\beta_k ~ \delta(1 - \sum^4_{j=2}\beta_j) \frac{\beta_2 \beta_4}{(k_{2} + k_4)^2 \beta_2 \beta_4 + k^2_{1} \beta_3 (1 - \beta_3 )} \times \frac{1}{ (2k_1\cdot k_2 + k^2_{1})\beta_3 + k^2_2 }\,.
 \nonumber
\end{eqnarray}

The integration over the variable $\beta_3$ is dominated by small values $\beta_3\sim \frac{k^2_2}{2k_1\cdot k_2}$, where we take into account the fact that in the Regge limit $2k_1\cdot k_2 + k^2_1 \simeq 2k_1\cdot k_2$. One can therefore write,
\begin{eqnarray}
\Gamma^{\mu\nu\alpha\beta}_A[k_1, k_3, k_2, k_4]\Big|_{x_{Bj}\to 0}
 &=& - 2\, \frac{g^2e^2 e_f^2 }{\pi^2} \epsilon^{\mu\nu\eta}_{\ \ \ \ \kappa} (k_{1\eta}  -  k_{3\eta}) (2\pi)^4\delta^{(4)}(\sum^4_{i=1} k_i) ~ \frac{k^\kappa_{2}+ k^\kappa_{4}}{(k_{2} + k_4)^2 } \epsilon^{\alpha\beta\sigma\lambda} k_{2\sigma} k_{4\lambda}\nonumber \\
 &\times& \int^1_0 d\beta_3 ~ \frac{1}{ 2k_1\cdot k_2 \beta_3 + k^2_2 }\,.
\end{eqnarray}
As promised, the infrared pole $l^2 = (k_2+k_4)^2$ is manifest in this expression. 
The integration over $\beta_3$ is easily performed, giving
\begin{eqnarray}
\Gamma^{\mu\nu\alpha\beta}_A[k_1, k_3, k_2, k_4]\Big|_{x_{Bj}\to 0} 
= - 2 \frac{g^2e^2 e_f^2 }{\pi^2} \epsilon^{\mu\nu\eta}_{\ \ \ \ \kappa} (k_{1\eta}  -  k_{3\eta}) (2\pi)^4\delta^{(4)}(\sum^4_{i=1} k_i) ~ \frac{k^\kappa_{2}+ k^\kappa_{4}}{(k_{2} + k_4)^2 } \epsilon^{\alpha\beta\sigma\lambda} k_{2\sigma} k_{4\lambda}~ \frac{1}{2k_1\cdot k_2}\ln\Big[\frac{2k_1\cdot k_2 + k^2_2}{k^2_2}\Big]\,.\nonumber\\
\end{eqnarray}

Substituting this result into Eqs. (\ref{Hadgamma}) and (\ref{WModFu}), performing the integrals over intermediate coordinates and momenta, after analytical continuation to Minkowski space-time, we obtain
\begin{eqnarray}
i\tilde{W}_f^{\mu\nu}(q, P, S) = - \frac{g^2 e_f^2 }{\pi^3} \epsilon^{\mu\nu\eta}_{\ \ \ \ \kappa} q_{\eta}\,
 {\rm Im} \int \frac{d^4k}{(2\pi)^4} ~ \frac{1}{2\,k\cdot q}\ln\Big[\frac{2q\cdot k + k^2}{k^2}\Big]~  \int d^4z\, e^{-ik z} \lim_{l^\kappa\to 0} \frac{l^\kappa}{l^2 } \langle P',S| {\rm Tr_c}( F_{\alpha\beta}(z) \tilde{F}^{\alpha\beta}(0)) |P,S\rangle\,,\nonumber \\
\end{eqnarray}
where $k\equiv k_2$ is the virtuality of the background gluons, and $l = P' - P$.

We will now take into account the fact that the virtuality of the background gluons in the Regge limit is transverse, $k^2\simeq - k^2_\perp$. This is analogous to the Bjorken limit expression in Eq.~(\ref{WantisimA}), where we had $q^2 = -Q^2$. We can similarly take the imaginary part of the logarithm, which gives   
\begin{eqnarray}
&&i\tilde{W}_f^{\mu\nu}(q, P, S) =  \frac{g^2 e_f^2 }{\pi^2} \epsilon^{\mu\nu\eta}_{\ \ \ \ \kappa} q_{\eta}
 \int \frac{d^4k}{(2\pi)^4} ~ \frac{1}{2\,k\cdot q}  \int d^4z \,e^{-ik z} \lim_{l^\kappa\to 0} \frac{l^\kappa}{l^2 } \langle P',S| {\rm Tr_c}( F_{\alpha\beta}(z) \tilde{F}^{\alpha\beta}(0)) |P,S\rangle\,.
 \label{WRjbefintk}
\end{eqnarray}

The rest of the derivation follows exactly along the lines of Section~\ref{sec:Bj}.  Assuming that $k\approx (k\cdot n)P^\mu$, introducing the variable $x$ through the identity $\int dx \int d\xi e^{i(x - (k\cdot n))\xi} = 2\pi$, and performing the integration over $k$ in Eq. (\ref{WRjbefintk}), we obtain 
\begin{eqnarray}
&&i\tilde{W}_f^{\mu\nu}(q, P, S) =  \frac{4}{P\cdot q} \epsilon^{\mu\nu\eta}_{\ \ \ \ \kappa} q_{\eta} \int^1_{x_B} \frac{dx}{x} ~  \int \frac{d\xi}{2\pi} e^{-i \xi x} \lim_{l^\kappa\to 0} \frac{l^\kappa}{l^2 } \langle P',S| \frac{\alpha_s e_f^2 }{2\pi} {\rm Tr_c} F_{\alpha\beta}(\xi n ) \tilde{F}^{\alpha\beta}(0) |P,S\rangle.
\end{eqnarray}
Comparing our result with the antisymmetric part of the hadron tensor in Eq.~(\ref{WA}), we obtain finally, 
\begin{eqnarray}
S^\mu g_1(x_B, Q^2)\Big|_{x_{\rm Bj}\rightarrow 0} = \sum_f e_f^2 \frac{\alpha_s}{i\pi M_N}
 \int^1_{x_B} \frac{dx}{x} 
 \int \frac{d\xi}{2\pi} e^{-i\xi x }\lim_{l_\mu \to 0} \frac{ l^\mu }{l^2  }\langle P^\prime,S| {\rm Tr_c} F_{\alpha\beta}(\xi n) \tilde{F}^{\alpha\beta}(0) |P,S\rangle\,.
\label{resultRj}
\end{eqnarray}

\subsection{Discussion of Eq.~(\ref{resultRj})}
\label{subsec:Rj-discussion}
Eq.~(\ref{resultRj}) is the other key result of this paper. It is  valid in Regge asymptotics, where $2k\cdot q \gg Q^2 \to x > x_B$ and $x_B \to 0$. With this in mind, if we compare Eqs. (\ref{resultRj}) and (\ref{resultBj}), the result for $g_1(x_B,Q^2)$ is formally identical in the two limits. 

However both qualitatively and quantitatively, the results for $g_1(x_B,Q^2)$ can be quite different in the two limits because the scales controlling the off-forward matrix element of the topological charge density represent very different physics. In the Bjorken limit, this scale is set by $Q^2$ while in the Regge limit, it is likely set by an emergent semi-hard saturation scale $Q_s(x)$ that controls the virtuality of the background gluons $k^2 = -k_\perp^2 = -Q_s^2$. The saturation scale is a dynamical close packing scale corresponding to the maximal occupancy ($\sim 1/\alpha_s$) of gluon modes with transverse momenta $k_\perp \leq Q_s(x)$ at a given $x$~\cite{Gelis:2010nm,Kovchegov:2012mbw}. 

The scale evolution of the matrix element in Regge asymptotics will depend on the small $x$ evolution within the shock wave that describes spin diffusion from large $x$ in the polarized proton target to the DIS probe. As we noted in the introduction, there is an extensive literature studying spin diffusion in the Regge asymptotics of perturbative QCD~\cite{Kirschner:1983di,Bartels:1995iu,Bartels:1996wc,Kovchegov:2015pbl,Kovchegov:2016weo,Kovchegov:2017jxc,Chirilli:2018kkw,Boussarie:2019icw,Cougoulic:2019aja,Kovchegov:2020hgb,Cougoulic:2020tbc}.
How the physics of $U_A(1)$ breaking  discussed briefly in Sec.~\ref{subsec:Bj-discussion} is realized by small $x$ partons is an open question of great interest. Interestingly some of this nonperturbative dynamics can be explored in weak coupling since $\alpha_s(Q_s)\ll 1$ in Regge asymptotics. 
In our forthcoming work, we will write down the small $x$ effective action that is consistent with the anomalous chiral Ward identities (reflecting  the physics of the anomaly) and describe its QCD evolution.

\section{Summary and outlook}
In this paper, we employed the worldline formalism, we developed previously for unpolarized DIS in \cite{Tarasov:2019rfp}, to compute the box diagram of polarized DIS in the Bjorken and Regge asymptotics of QCD. In particular, we find that the computation of the box diagram is nearly identical in both asymptotic limits. This is remarkable from the point of view of standard computational techniques that employ the operator product expansion, since it is well known that the OPE cannot be applied straightforwardly in Regge asymptotics~\cite{Mueller:1996hm}.  

We find that in both asymptotics, the matrix element for the  $g_1(x_B,Q^2)$ structure function is identically controlled by the triangle anomaly, which has an infrared pole in the forward scattering limit. As we discussed at length in the introduction, the cancellation of this pole involves a subtle interplay of perturbative and nonperturbative physics that is deeply related to the $U_A(1)$ problem in QCD. As a further consequence, our results bring up important questions regarding the applicability of QCD factorization to observables that are sensitive to the anomaly.  

Though the matrix element for $g_1$ is formally identical in both Bjorken and Regge limits, the underlying physics is quite different. Some of these qualitative differences are already evident from our derivations in Secs.~\ref{sec:Bj} and \ref{sec:Regge}. In paper II in this series,  we will derive the small $x$ effective action that follows from the cancellation of the infrared pole in the matrix element of the anomaly. This effective action,  consistent with anomalous chiral Ward identities, is controlled by two dimensionful scales in Regge asymptotics. The first is the color charge squared per unit area, which is proportional to the saturation scale $Q_s^2$~\cite{McLerran:1993ni,McLerran:1993ka,McLerran:1994vd}, while the second is the pure Yang-Mills topological susceptibility~\cite{Veneziano:1989ei,Shore:1990zu,Shore:1991dv}. The physics of the former has been discussed extensively in the context of small $x$ physics with the framework of the Color Glass Condensate Effective Field Theory~\cite{Gelis:2010nm}.

The latter is often tied to the physics of instantons in QCD, though this is not necessarily the case. An excellent review of the many subtleties involved can be found in~\cite{Schafer:1996wv}. More generally, one can argue that the fundamental origin of this scale has to do with the description of the  QCD vacuum as energy degenerate  $\theta$-vacua, each corresponding to distinct integer valued Chern-Simons number. We will argue that the dynamics governing helicity evolution is governed by over the barrier sphaleron transitions that are enhanced by the large dynamical saturation scale. This is analogous to the important role played by  sphaleron transitions in electroweak baryogenesis~\cite{Kuzmin:1985mm} and in QCD at  finite 
temperature~\cite{McLerran:1990de,Moore:2010jd}. Such an interplay between the saturation scale and rate of topological transitions has been studied recently for highly occupied Yang-Mills fields off-equilibrium~\cite{Mace:2016svc}.

Paper III will discuss the renormalization group evolution of the helicity dependent effective action.  With decreasing $x_B$ (or increasing energy), studies in the CGC EFT suggest that the background classical (high occupancy) fields have a lumpier 
structure with radii $\sim 1/Q_s$; as shown in \cite{Mace:2016svc} for off-equilibrium Yang-Mills fields, these lumps are also associated with a faster rate of topological transitions that flip the helicity of left handed quarks to right handed quarks or vice versa. Our conjecture, which will be explored at greater length in Paper III, is that spin diffusion in this ``disordered medium" is rapid, leading to a strong damping of $g_1(x_B,Q^2)$ and other spin-dependent observables that are sensitive to the anomaly~\cite{Shore:2007yn}. 
Because $\alpha_s(Q_s)<<1$ in Regge asymptotics, the rate of this damping can be computed in  a weak coupling framework and compared to the results of polarized DIS experiments at the EIC. Such experiments therefore have the potential to uncover novel dynamical features associated with the topology of the QCD  vacuum and its interplay with the physics of gluon saturation. 

\section*{Acknowledgements}
We are grateful to Werner Vogelsang for several very informative discussions on many aspects of the spin puzzle and for a close reading of the manuscript. We would also like to thank Gia Dvali for inspiring discussions, in particular with regard to the wider implications of our work. 
We  thank Jochen Bartels, Yoshitaka Hatta, Bob Jaffe, Jamal Jalilian-Marian, Xiangdong Ji, Yuri Kovchegov, Daniel Pitonyak, Matt Sievert, George Sterman and Feng Yuan for useful discussions on  polarized DIS relevant to this work. Fabian Rennecke and Rob Pisarski are owed thanks for their insights on instantons and the chiral anomaly. 
R.V.'s work is supported by the U.S. Department of Energy under Contract No. DE-SC0012704 and by the US DOE within the framework of the TMD Theory Topical Collaboration. His work was also supported in part by an LDRD grant from Brookhaven Science Associates. A.T.'s work is supported by the U.S. Department of Energy, Office of Science, Office of Nuclear Physics under Award No. DE-SC0004286.
\appendix

\section{Computation  of the coefficient $\mathcal{I}^{\mu\nu\alpha\beta}_{1;(\tau_1,\tau_2,\tau_3,\tau_4)}[k_1, k_3, k_2, k_4]$ in Eq. (\ref{CtoI})\label{sec:I1}}

In this appendix, we will provide a detailed computation of the first of nine terms in the coefficient that appears in the worldline expression (Eq. (\ref{CtoI})) for the box diagram for polarized DIS:
\begin{eqnarray}
\frac{1}{4\pi^2 T^2}~\mathcal{I}^{\mu\nu\alpha\beta}_{1;(\tau_1,\tau_2,\tau_3,\tau_4)}[k_1, k_3, k_2, k_4] \langle e^{i\sum^4_{i=1}k_i x_i}\rangle &=&  \int \mathcal{D}x\int \mathcal{D}\psi 
 \Big(\mathcal{C}^{\mu\nu\alpha\beta}_{1;(\tau_1,\tau_2,\tau_3,\tau_4)}[k_1, k_3, k_2, k_4] e^{i\sum^4_{i=1}k_i x_i} - (\mu\leftrightarrow\nu)\Big)\nonumber\\
&\times& \exp\Big\{-\int^T_0 d\tau \Big(\frac{1}{4} \dot{x}^2 + \frac{1}{2}\psi_\mu\dot{\psi}^\mu \Big)\Big\}\,,
\end{eqnarray}
Here 
$\langle\dots\rangle$ on the l.h.s denotes the Wick contraction of the worldline trajectories that we discussed in the main text and $\mathcal{C}^{\mu\nu\alpha\beta}_{1;(\tau_1,\tau_2,\tau_3,\tau_4)}[k_1, k_3, k_2, k_4] = - 4\dot{x}^\nu_3 \psi^\mu_1 \psi_1\cdot k_{1}\dot{x}^\beta_4 \psi^\alpha_2 \psi_2\cdot k_{2}$. The coefficients $C_2\rightarrow C_4$ can be  obtained by a simple permutation of the momentum and proper time labels of our result below and the results for $C_5\rightarrow C_9$ are obtained from a straightforward generalization of our discussion. The results for $\mathcal{I}^{\mu\nu\alpha\beta}_{2,\cdots,9;(\tau_1,\tau_2,\tau_3,\tau_4)}$ are given in Appendix B.

We begin by writing the expression above as\footnote{The details of calculation of worldline functional integrals can be found in Ref. \cite{Schubert:2001he}.}
\begin{eqnarray}
\frac{1}{4\pi^2 T^2}~\mathcal{I}^{\mu\nu\alpha\beta}_{1;(\tau_1,\tau_2,\tau_3,\tau_4)}[k_1, k_3, k_2, k_4] \langle e^{i\sum^4_{i=1}k_i x_i}\rangle
= (- 4) (4\pi T)^{-2} 4 \Big( \langle \dot{x}^\nu_3 \dot{x}^\beta_4 e^{i\sum^4_{i=1}k_i x_i} \rangle \langle \psi^\mu_1 \psi^\rho_1  \psi^\alpha_2 \psi^\lambda_2 \rangle k_{1\rho} k_{2\lambda}   - (\mu\leftrightarrow\nu)\Big) \, .\nonumber \\
\end{eqnarray}

We will now compute the Wick contractions of the worldline trajectories. Starting with 
$\langle \dot{x}^\nu_3 \dot{x}^\beta_4 e^{ik_1 x_1} e^{ik_3 x_3} e^{ik_2 x_2} e^{ik_4 x_4} \rangle$ and performing the 
 Wick contractions, we can write it  as 
\begin{eqnarray}
\label{eq:Wick1}
\langle \dot{x}^\nu_3 \dot{x}^\beta_4 e^{ik_1 x_1} e^{ik_3 x_3} e^{ik_2 x_2} e^{ik_4 x_4} \rangle\,
&&= \langle \dot{x}^\nu_3 \dot{x}^\beta_4 \rangle \, e^{ik_1 x_1} e^{ik_3 x_3} e^{ik_2 x_2} e^{ik_4 x_4} \nonumber\\
&&+\,\langle \dot{x}^\nu_3 e^{ik_1 x_1} \rangle \langle \dot{x}^\beta_4  e^{ik_3 x_3}\rangle \, e^{ik_2 x_2} e^{ik_4 x_4} +\langle \dot{x}^\nu_3 e^{ik_1 x_1} \rangle \langle \dot{x}^\beta_4 e^{ik_2 x_2} \rangle \,e^{ik_3 x_3} e^{ik_4 x_4} 
\nonumber\\
&&+\,\langle \dot{x}^\nu_3 e^{ik_2 x_2} \rangle \langle \dot{x}^\beta_4 e^{ik_1 x_1}\rangle \,e^{ik_3 x_3} e^{ik_4 x_4} + \langle \dot{x}^\nu_3 e^{ik_2 x_2} \rangle \langle \dot{x}^\beta_4 e^{ik_3 x_3} \rangle \,e^{ik_1 x_1} e^{ik_4 x_4} 
\nonumber\\
&&+\,\langle \dot{x}^\nu_3 e^{ik_4 x_4} \rangle \langle \dot{x}^\beta_4 e^{ik_1 x_1}\rangle \, e^{ik_3 x_3} e^{ik_2 x_2} 
+\langle \dot{x}^\nu_3 e^{ik_4 x_4} \rangle \langle \dot{x}^\beta_4 e^{ik_3 x_3} \rangle\, e^{ik_1 x_1} e^{ik_2 x_2}\nonumber \\
&&+\,\langle \dot{x}^\nu_3 e^{ik_4 x_4} \rangle \langle \dot{x}^\beta_4 e^{ik_2 x_2} \rangle \,e^{ik_1 x_1} e^{ik_3 x_3} \,.
\end{eqnarray}

Using the worldline identities~\cite{Schubert:2001he}
\begin{eqnarray}
\langle y^\mu(\tau_1) e^{ik\cdot y(\tau_2)}\rangle &=& i \langle y^\mu(\tau_1) y^\nu(\tau_2)\rangle k_\nu e^{ik\cdot y(\tau_2)}\,,\nonumber\\
\langle y^\mu(\tau_1)y^\nu(\tau_2)\rangle &=& - g^{\mu\nu}G_B(\tau_1, \tau_2)\,,
\end{eqnarray}
in the expression above, yields
\begin{eqnarray}
&&\langle \dot{x}^\nu_3 \dot{x}^\beta_4 e^{ik_1 x_1} e^{ik_3 x_3} e^{ik_2 x_2} e^{ik_4 x_4} \rangle
\nonumber\\
&&= \Big[ - g^{\nu\beta}\frac{\partial^2}{\partial \tau_3 \partial \tau_4}G_B(\tau_3, \tau_4) 
\nonumber\\
&&- k_{1\zeta}k_{3\xi} g^{\nu\zeta}g^{\beta\xi} \dot{G}_B(\tau_3, \tau_1) \dot{G}_B(\tau_4, \tau_3) - k_{1\zeta}k_{2\xi} g^{\nu\zeta}g^{\beta\xi}\dot{G}_B(\tau_3, \tau_1) \dot{G}_B(\tau_4, \tau_2)
\nonumber\\
&&- k_{2\zeta} k_{1\xi}g^{\nu\zeta} g^{\beta\xi}\dot{G}_B(\tau_3, \tau_2) \dot{G}_B(\tau_4, \tau_1) - k_{2\zeta} k_{3\xi}g^{\nu\zeta}g^{\beta\xi} \dot{G}_B(\tau_3, \tau_2) \dot{G}_B(\tau_4, \tau_3) 
\nonumber\\
&&- k_{4\zeta} k_{1\xi} g^{\nu\zeta}g^{\beta\xi}\dot{G}_B(\tau_3, \tau_4) \dot{G}_B(\tau_4, \tau_1) - k_{4\zeta} k_{3\xi} g^{\nu\zeta}g^{\beta\xi}\dot{G}_B(\tau_3, \tau_4) \dot{G}_B(\tau_4, \tau_3) - k_{4\zeta} k_{2\xi}g^{\nu\zeta}g^{\beta\xi} \dot{G}_B(\tau_3, \tau_4) \dot{G}_B(\tau_4, \tau_2) \Big]
\nonumber\\
&&\times \langle e^{ik_1 x_1} e^{ik_3 x_3} e^{ik_2 x_2} e^{ik_4 x_4} \rangle \, .
\end{eqnarray}
This can be written more simply as 
\begin{eqnarray}
\label{eq:boson-PW}
&&\langle \dot{x}^\nu_3 \dot{x}^\beta_4 e^{ik_1 x_1} e^{ik_3 x_3} e^{ik_2 x_2} e^{ik_4 x_4} \rangle
\nonumber\\
&&= \Big[ - g^{\nu\beta}\frac{\partial^2}{\partial \tau_3 \partial \tau_4}G_B(\tau_3, \tau_4) 
- k^\nu_{1}k^\beta_{3} \dot{G}_B(\tau_3, \tau_1) \dot{G}_B(\tau_4, \tau_3) - k^\nu_{1}k^\beta_{2} \dot{G}_B(\tau_3, \tau_1) \dot{G}_B(\tau_4, \tau_2)
\nonumber\\
&&- k^\nu_{2} k^\beta_{1} \dot{G}_B(\tau_3, \tau_2) \dot{G}_B(\tau_4, \tau_1) - k^\nu_{2} k^\beta_{3} \dot{G}_B(\tau_3, \tau_2) \dot{G}_B(\tau_4, \tau_3) 
\nonumber\\
&&- k^\nu_{4} k^\beta_{1} \dot{G}_B(\tau_3, \tau_4) \dot{G}_B(\tau_4, \tau_1) - k^\nu_{4} k^\beta_{3} \dot{G}_B(\tau_3, \tau_4) \dot{G}_B(\tau_4, \tau_3) - k^\nu_{4} k^\beta_{2} \dot{G}_B(\tau_3, \tau_4) \dot{G}_B(\tau_4, \tau_2) \Big]
\nonumber\\
&&\times \langle e^{ik_1 x_1} e^{ik_3 x_3} e^{ik_2 x_2} e^{ik_4 x_4} \rangle\, .
\end{eqnarray}

We next consider the Grassmann Wick contractions,
\begin{eqnarray}
&&\langle \psi^\mu_1 \psi^\rho_1  \psi^\alpha_2 \psi^\lambda_2 \rangle
= -\langle \psi^\mu_1 \psi^\alpha_2\rangle \langle \psi^\rho_1 \psi^\lambda_2 \rangle + \langle \psi^\mu_1 \psi^\lambda_2 \rangle \langle \psi^\rho_1  \psi^\alpha_2 \rangle\,.
\label{eq:Grass-contract}
\end{eqnarray}
Using the identity,
\begin{eqnarray}
\langle \psi^\mu(\tau_1) \psi^\nu(\tau_2)\rangle = \frac{1}{2}g^{\mu\nu}G_F(\tau_1,\tau_2) \equiv \frac{1}{2}g^{\mu\nu}{\rm sign}(\tau_1 - \tau_2)\,,
\label{wickgrassm}
\end{eqnarray}
we obtain
\begin{eqnarray}
&&\langle \psi^\mu_1 \psi^\rho_1  \psi^\alpha_2 \psi^\lambda_2 \rangle
= \frac{1}{4}(-g^{\mu\alpha} g^{\rho\lambda} + g^{\mu\lambda} g^{\rho\alpha})\,.
\end{eqnarray}
 Contracting the product of two Levi-Civita tensors, which gives, 
\begin{eqnarray}
\epsilon^{\alpha\beta\mu\nu}\epsilon_{\alpha\beta\rho\sigma} = -2(\delta^\mu_\rho\delta^\nu_\sigma - \delta^\mu_\sigma \delta^\nu_\rho)
\end{eqnarray}
or equivalently,
\begin{eqnarray}
\label{eq:LC1}
\frac{1}{8}\epsilon^{\mu\rho\kappa\eta}\epsilon^{\alpha\lambda}_{\ \ \ \kappa\eta} = \frac{1}{4}(- g^{\mu\alpha}g^{\rho\lambda} + g^{\mu\lambda} g^{\rho\alpha})
\end{eqnarray}
allows us to express Eq.~(\ref{eq:Grass-contract}) in terms of this product as,
\begin{eqnarray}
\label{eq:Grass-LC}
\langle \psi^\mu_1 \psi^\rho_1  \psi^\alpha_2 \psi^\lambda_2 \rangle = \frac{1}{8}\epsilon^{\mu\rho\kappa\eta}\epsilon^{\alpha\lambda}_{\ \ \ \kappa\eta}\,\,.
\end{eqnarray}

Substituting Eqs.~(\ref{eq:boson-PW}) and (\ref{eq:Grass-LC}) back into expression for $\mathcal{I}^{\mu\nu\alpha\beta}_{1;(\tau_1,\tau_2,\tau_3,\tau_4)}[k_1, k_3, k_2, k_4]$, we obtain:
\begin{eqnarray}
&&\frac{1}{4\pi^2 T^2}~\mathcal{I}^{\mu\nu\alpha\beta}_{1;(\tau_1,\tau_2,\tau_3,\tau_4)}[k_1, k_3, k_2, k_4] \langle e^{i\sum^4_{i=1}k_i x_i}\rangle
\\
&&= (- 4) (4\pi T)^{-2} 4 \Big(\Big[ - g^{\nu\beta}\frac{\partial^2}{\partial \tau_3 \partial \tau_4}G_B(\tau_3, \tau_4) 
- k^\nu_{1}k^\beta_{3} \dot{G}_B(\tau_3, \tau_1) \dot{G}_B(\tau_4, \tau_3) - k^\nu_{1}k^\beta_{2} \dot{G}_B(\tau_3, \tau_1) \dot{G}_B(\tau_4, \tau_2)
\nonumber\\
&&- k^\nu_{2} k^\beta_{1} \dot{G}_B(\tau_3, \tau_2) \dot{G}_B(\tau_4, \tau_1) - k^\nu_{2} k^\beta_{3} \dot{G}_B(\tau_3, \tau_2) \dot{G}_B(\tau_4, \tau_3) - k^\nu_{4} k^\beta_{1} \dot{G}_B(\tau_3, \tau_4) \dot{G}_B(\tau_4, \tau_1) - k^\nu_{4} k^\beta_{3} \dot{G}_B(\tau_3, \tau_4) \dot{G}_B(\tau_4, \tau_3)
\nonumber\\
&& - k^\nu_{4} k^\beta_{2} \dot{G}_B(\tau_3, \tau_4) \dot{G}_B(\tau_4, \tau_2) \Big] \frac{1}{8}\epsilon^{\mu\rho\kappa\eta}\epsilon^{\alpha\lambda}_{\ \ \ \kappa\eta} k_{1\rho} k_{2\lambda}
- (\mu\leftrightarrow\nu)\Big) \langle e^{i\sum^4_{i=1}k_i x_i} \rangle
\nonumber
\end{eqnarray}

To cleanly separate out the tensorial structure of the ``external" indices ($\mu,\nu,\alpha,\beta$) in the r.h.s of the above expression, we use the fact that 
\begin{eqnarray}
&&\Big(g^{\delta\nu} \epsilon^{\mu\rho\kappa\zeta}\epsilon^{\alpha\lambda}_{\ \ \ \kappa\zeta} - (\mu\leftrightarrow\nu)\Big)\,,
\end{eqnarray}
can be rewritten as
\begin{eqnarray}
\frac{1}{2}\Big(g^{\delta\nu}\epsilon^{\mu\rho\kappa\zeta}\epsilon^{\alpha\lambda}_{\ \ \ \kappa\zeta} - (\mu\leftrightarrow\nu) \Big)  
&=& (- g^{\delta\nu}g^{\mu\alpha}g^{\rho\lambda} + g^{\delta\nu} g^{\mu\lambda} g^{\rho\alpha}
+ g^{\delta\mu}g^{\nu\alpha}g^{\rho\lambda} - g^{\delta\mu} g^{\nu\lambda} g^{\rho\alpha})
\nonumber\\
&\equiv& \frac{1}{2}\epsilon^{\mu\nu\kappa\eta}\Big( \epsilon^{\alpha\delta}_{\ \ \ \kappa\eta} g^{\rho\lambda} - \epsilon^{\lambda\delta}_{\ \ \ \kappa\eta} g^{\rho\alpha} \Big)\,,
\end{eqnarray}
where we used the identity in Eq.~(\ref{eq:LC1}). We can further simplify this expression such that the external $\mu,\nu$ indices are manifest on the r.h.s:
\begin{eqnarray}
&&\frac{1}{2}\Big(g^{\delta\nu}\epsilon^{\mu\rho\kappa\zeta}\epsilon^{\alpha\lambda}_{\ \ \ \kappa\zeta} - (\mu\leftrightarrow\nu) \Big)  = \frac{1}{2}\epsilon^{\mu\nu}_{\ \ \ \kappa\eta}\Big( g^{\rho\lambda} \epsilon^{\alpha\delta\kappa\eta}  - g^{\rho\alpha} \epsilon^{\lambda\delta\kappa\eta}  \Big)
\nonumber\\
&&= \frac{1}{2}\epsilon^{\mu\nu}_{\ \ \ \kappa\eta}\Big( - g^{\rho\delta } \epsilon^{\kappa\eta\lambda\alpha} - g^{\rho \kappa} \epsilon^{\eta\lambda\alpha\delta} - g^{\rho \eta} \epsilon^{\lambda\alpha\delta\kappa}  \Big)
= \frac{1}{2}\epsilon^{\mu\nu}_{\ \ \ \kappa\eta}\Big( - g^{\rho\delta } \epsilon^{\kappa\eta\lambda\alpha} - 2 g^{\rho \eta} \epsilon^{\lambda\alpha\delta\kappa} \Big)
\end{eqnarray}
This then allows us to write 
\begin{eqnarray}
&&\frac{1}{4\pi^2 T^2}~\mathcal{I}^{\mu\nu\alpha\beta}_{1;(\tau_1,\tau_2,\tau_3,\tau_4)}[k_1, k_3, k_2, k_4] \langle e^{i\sum^4_{i=1}k_i x_i}\rangle
\\
&&= (4\pi T)^{-2} 4~ \Big\{\Big[ \delta^{\beta}_\delta\frac{\partial^2}{\partial \tau_3 \partial \tau_4}G_B(\tau_3, \tau_4) 
+ k_{1\delta}k^\beta_{3} \dot{G}_B(\tau_3, \tau_1) \dot{G}_B(\tau_4, \tau_3) + k_{1\delta}k^\beta_{2} \dot{G}_B(\tau_3, \tau_1) \dot{G}_B(\tau_4, \tau_2)
\nonumber\\
&&+ k_{2\delta} k^\beta_{1} \dot{G}_B(\tau_3, \tau_2) \dot{G}_B(\tau_4, \tau_1) + k_{2\delta} k^\beta_{3} \dot{G}_B(\tau_3, \tau_2) \dot{G}_B(\tau_4, \tau_3) + k_{4\delta} k^\beta_{1} \dot{G}_B(\tau_3, \tau_4) \dot{G}_B(\tau_4, \tau_1) + k_{4\delta} k^\beta_{3} \dot{G}_B(\tau_3, \tau_4) \dot{G}_B(\tau_4, \tau_3)
\nonumber\\
&& + k_{4\delta} k^\beta_{2} \dot{G}_B(\tau_3, \tau_4) \dot{G}_B(\tau_4, \tau_2) \Big] \frac{1}{2} \epsilon^{\mu\nu}_{\ \ \ \kappa\eta}\Big( - k^\delta_{1} \epsilon^{\kappa\eta\lambda\alpha} - 2 k^\eta_{1}\epsilon^{\lambda\alpha\delta\kappa} \Big)  k_{2\lambda}
\Big\} \langle e^{i\sum^4_{i=1}k_i x_i} \rangle\,.
\nonumber
\end{eqnarray}

To further simplify the expression, we will perform a partial integration of the first term in the square bracket. This acts on the plane waves in $\langle \cdots \rangle$. Since the latter satisfy the identity~\cite{Schubert:2001he},
\begin{eqnarray}
&&\langle e^{i\sum^4_{i=1}k_i x_i} \rangle = \exp\Big[
k_{1} \cdot k_{2} G_B(\tau_1, \tau_2) + k_{1} \cdot k_{3} G_B(\tau_1, \tau_3) + k_{1}\cdot k_{4} G_B(\tau_1, \tau_4)\nonumber \\
&& + k_{2} \cdot k_{3} G_B(\tau_2, \tau_3) + k_{2} \cdot k_{4} G_B(\tau_2, \tau_4) + k_{3} \cdot k_{4} G_B(\tau_3, \tau_4) \Big]\,,
\end{eqnarray}
we can express $\mathcal{I}^{\mu\nu\alpha\beta}_{1;(\tau_1,\tau_2,\tau_3,\tau_4)}[k_1, k_3, k_2, k_4] \langle e^{i\sum^4_{i=1}k_i x_i}\rangle$ finally as 
\begin{eqnarray}
&&\frac{1}{4\pi^2 T^2}~\mathcal{I}^{\mu\nu\alpha\beta}_{1;(\tau_1,\tau_2,\tau_3,\tau_4)}[k_1, k_3, k_2, k_4] \langle e^{i\sum^4_{i=1}k_i x_i}\rangle
\nonumber\\
&&= (4\pi T)^{-2} 4~ \Big\{\Big[ \delta^{\beta}_\delta \dot{G}_B(\tau_3, \tau_4)  \Big(
 k_{1}\cdot k_{4} \dot{G}_B(\tau_1, \tau_4) + k_{2} \cdot k_{4} \dot{G}_B(\tau_2, \tau_4) + k_{3} \cdot k_{4} \dot{G}_B(\tau_3, \tau_4) \Big)
\nonumber\\
&&+ k_{1\delta}k^\beta_{3} \dot{G}_B(\tau_3, \tau_1) \dot{G}_B(\tau_4, \tau_3) + k_{1\delta}k^\beta_{2} \dot{G}_B(\tau_3, \tau_1) \dot{G}_B(\tau_4, \tau_2)
\nonumber\\
&&+ k_{2\delta} k^\beta_{1} \dot{G}_B(\tau_3, \tau_2) \dot{G}_B(\tau_4, \tau_1) + k_{2\delta} k^\beta_{3} \dot{G}_B(\tau_3, \tau_2) \dot{G}_B(\tau_4, \tau_3) + k_{4\delta} k^\beta_{1} \dot{G}_B(\tau_3, \tau_4) \dot{G}_B(\tau_4, \tau_1) + k_{4\delta} k^\beta_{3} \dot{G}_B(\tau_3, \tau_4) \dot{G}_B(\tau_4, \tau_3)
\nonumber\\
&& + k_{4\delta} k^\beta_{2} \dot{G}_B(\tau_3, \tau_4) \dot{G}_B(\tau_4, \tau_2) \Big] \frac{1}{2} \epsilon^{\mu\nu}_{\ \ \ \kappa\eta}\Big( - k^\delta_{1} \epsilon^{\kappa\eta\lambda\alpha} - 2 k^\eta_{1}\epsilon^{\lambda\alpha\delta\kappa} \Big)  k_{2\lambda}
\Big\} \langle e^{i\sum^4_{i=1}k_i x_i} \rangle\,.
\nonumber
\end{eqnarray}
One can likewise derive analogous expressions for the other 8 terms in Eq.~(\ref{CtoI}). They all have the common tensorial structure 
$\frac{1}{2} \epsilon^{\mu\nu}_{\ \ \ \kappa\eta}\Big( - k^\delta_{1} \epsilon^{\kappa\eta\lambda\alpha} - 2 k^\eta_{1}\epsilon^{\lambda\alpha\delta\kappa} \Big)  k_{2\lambda}$ and differ only in the expressions inside the square brackets.

In the forward limit, the second term of the tensorial structure yields the non-trivial contribution $\sim  \epsilon^{\mu\nu\eta}_{\ \ \ \ \kappa} q_\eta$. This is identical to that appearing in the the general decomposition of the hadron tensor in Eq.~ (\ref{WA}). In contrast, the first term vanishes due to a Ward identity. As a result, we obtain finally,
\begin{eqnarray}
&&\frac{1}{4\pi^2 T^2}~\mathcal{I}^{\mu\nu\alpha\beta}_{1;(\tau_1,\tau_2,\tau_3,\tau_4)}[k_1, k_3, k_2, k_4] \langle e^{i\sum^4_{i=1}k_i x_i}\rangle
\\
&&= (4\pi T)^{-2} 4~ \Big\{\Big[ \delta^{\beta}_\delta \dot{G}_B(\tau_3, \tau_4)  \Big(
 k_{1}\cdot k_{4} \dot{G}_B(\tau_1, \tau_4) + k_{2} \cdot k_{4} \dot{G}_B(\tau_2, \tau_4) + k_{3} \cdot k_{4} \dot{G}_B(\tau_3, \tau_4) \Big)
\nonumber\\
&&+ k_{1\delta}k^\beta_{3} \dot{G}_B(\tau_3, \tau_1) \dot{G}_B(\tau_4, \tau_3) + k_{1\delta}k^\beta_{2} \dot{G}_B(\tau_3, \tau_1) \dot{G}_B(\tau_4, \tau_2) + k_{2\delta} k^\beta_{1} \dot{G}_B(\tau_3, \tau_2) \dot{G}_B(\tau_4, \tau_1) + k_{2\delta} k^\beta_{3} \dot{G}_B(\tau_3, \tau_2) \dot{G}_B(\tau_4, \tau_3)
\nonumber\\
&& + k_{4\delta} k^\beta_{1} \dot{G}_B(\tau_3, \tau_4) \dot{G}_B(\tau_4, \tau_1) + k_{4\delta} k^\beta_{3} \dot{G}_B(\tau_3, \tau_4) \dot{G}_B(\tau_4, \tau_3)
 + k_{4\delta} k^\beta_{2} \dot{G}_B(\tau_3, \tau_4) \dot{G}_B(\tau_4, \tau_2) \Big] \epsilon^{\mu\nu\eta}_{\ \ \ \ \kappa} k_{1\eta}\epsilon^{\lambda\alpha\delta\kappa} k_{2\lambda}
\Big\} \langle e^{i\sum^4_{i=1}k_i x_i} \rangle\,.
\nonumber
\end{eqnarray}

\section{Coefficients $\mathcal{I}^{\mu\nu\alpha\beta}_{n;(\tau_1,\tau_2,\tau_3,\tau_4)}[k_1, k_3, k_2, k_4]$ in the Bjorken limit $u_1 = u_3$}
In this appendix, we provide explicit expressions for the coefficients $\mathcal{I}^{\mu\nu\alpha\beta}_{n;(\tau_1,\tau_2,\tau_3,\tau_4)}[k_1, k_3, k_2, k_4]$ in the Bjorken limit $u_1 = u_3$. These coefficients were employed in deriving Eq. (\ref{suminBj}). As stated in the main text, this result can be straightforward adapted to the Regge limit.

The complete list results for the coefficients are as follows.
\begin{eqnarray}
&&\mathcal{I}^{\mu\nu\alpha\beta}_{1;(\tau_1,\tau_2,\tau_3,\tau_4)}[k_1, k_3, k_2, k_4]\Big|_{u_1 = u_3} = \Big( \delta^{\beta}_\delta \dot{G}_B(\tau_1, \tau_4)  \big(
 (k_{1} + k_3)\cdot k_{4} \dot{G}_B(\tau_1, \tau_4) + k_{2} \cdot k_{4} \dot{G}_B(\tau_2, \tau_4) \big) \nonumber 
\\
&&- k_{2\delta} (k^\beta_{1} + k^\beta_{3}) \dot{G}_B(\tau_1, \tau_2) \dot{G}_B(\tau_1, \tau_4)  - k_{4\delta} (k^\beta_{1} + k^\beta_{3}) \dot{G}^2_B(\tau_1, \tau_4) - k_{4\delta} k^\beta_{2} \dot{G}_B(\tau_1, \tau_4) \dot{G}_B(\tau_2, \tau_4) \Big) \epsilon^{\mu\nu\eta}_{\ \ \ \ \kappa}   k_{1\eta}\epsilon^{\lambda\alpha\delta\kappa} k_{2\lambda}\,.
\nonumber\\
\end{eqnarray}
\begin{eqnarray}
&&\mathcal{I}^{\mu\nu\alpha\beta}_{2;(\tau_1,\tau_2,\tau_3,\tau_4)}[k_1, k_3, k_2, k_4]\Big|_{u_1 = u_3} = \Big( \delta^{\alpha}_\delta \dot{G}_B(\tau_1, \tau_2) \big( (k_{1} + k_3) \cdot k_{2}  \dot{G}_B(\tau_1, \tau_2) + k_{2} \cdot k_{4} \dot{G}_B(\tau_4, \tau_2) \big)\nonumber
\\
&&- k_{4\delta} (k^\alpha_{1} + k^\alpha_{3}) \dot{G}_B(\tau_1, \tau_2) \dot{G}_B(\tau_1, \tau_4)  - k_{2\delta} (k^\alpha_{1} + k^\alpha_{3}) \dot{G}^2_B(\tau_1, \tau_2) - k_{2\delta} k^\alpha_{4} \dot{G}_B(\tau_1, \tau_2) \dot{G}_B(\tau_4, \tau_2) \Big) \epsilon^{\mu\nu\eta}_{\ \ \ \ \kappa}   k_{1\eta} \epsilon^{\lambda\beta\delta\kappa} k_{4\lambda}\,.
\nonumber\\
\end{eqnarray}
\begin{eqnarray}
&&\mathcal{I}^{\mu\nu\alpha\beta}_{3;(\tau_1,\tau_2,\tau_3,\tau_4)}[k_1, k_3, k_2, k_4]\Big|_{u_1 = u_3} = - \Big(
 \delta^{\beta}_\delta \dot{G}_B(\tau_1, \tau_4) \big( (k_{1} + k_3) \cdot k_{4}  \dot{G}_B(\tau_1, \tau_4) + k_{2} \cdot k_{4} \dot{G}_B(\tau_2, \tau_4) \big)
\nonumber \\
&&- k_{2\delta} (k^\beta_{1} + k^\beta_{3}) \dot{G}_B(\tau_1, \tau_2) \dot{G}_B(\tau_1, \tau_4)  - k_{4\delta} (k^\beta_{1} + k^\beta_{3}) \dot{G}^2_B(\tau_1, \tau_4) - k_{4\delta} k^\beta_{2} \dot{G}_B(\tau_1, \tau_4) \dot{G}_B(\tau_2, \tau_4) \Big) \epsilon^{\mu\nu\eta}_{\ \ \ \ \kappa}  k_{3\eta}\epsilon^{\lambda\alpha\delta\kappa} k_{2\lambda}\,.
\nonumber\\
\end{eqnarray}
\begin{eqnarray}
&&\mathcal{I}^{\mu\nu\alpha\beta}_{4;(\tau_1,\tau_2,\tau_3,\tau_4)}[k_1, k_3, k_2, k_4]\Big|_{u_1 = u_3} =
- \Big( \delta^{\alpha}_\delta \dot{G}_B(\tau_1, \tau_2) \big( (k_{1} + k_3) \cdot k_{2} \dot{G}_B(\tau_1, \tau_2) + k_{2} \cdot k_{4} \dot{G}_B(\tau_4, \tau_2) \big)
\nonumber \\
&&- k_{4\delta} (k^\alpha_{1} + k^\alpha_{3}) \dot{G}_B(\tau_1, \tau_2) \dot{G}_B(\tau_1, \tau_4)  - k_{2\delta} (k^\alpha_{1} + k^\alpha_{3}) \dot{G}^2_B(\tau_1, \tau_2) - k_{2\delta} k^\alpha_{4} \dot{G}_B(\tau_1, \tau_2) \dot{G}_B(\tau_4, \tau_2) \Big) \epsilon^{\mu\nu\eta}_{\ \ \ \ \kappa}   k_{3\eta}\epsilon^{\lambda\beta\delta\kappa}  k_{4\lambda}\,.
\nonumber\\
\end{eqnarray}
\begin{eqnarray}
&&\mathcal{I}^{\mu\nu\alpha\beta}_{5;(\tau_1,\tau_2,\tau_3,\tau_4)}[k_1, k_3, k_2, k_4]\Big|_{u_1 = u_3} = \Big( \dot{G}_B(\tau_1, \tau_2) k_{2\delta}
+ \dot{G}_B(\tau_1, \tau_4) k_{4\delta} \Big)
\epsilon^{\mu\nu}_{\ \ \ \kappa\zeta} \Big( g^{\rho\kappa} \epsilon^{\zeta\beta\delta\alpha} g^{\lambda\eta}
+ g^{\rho\kappa} \epsilon^{\zeta\eta\alpha\delta} g^{\lambda\beta}
\nonumber \\
&&+ g^{\rho\kappa} \epsilon^{\zeta\beta\lambda\delta} g^{\alpha\eta}
 + g^{\rho\kappa} \epsilon^{\zeta\eta\delta\lambda}  g^{\alpha\beta}
 \Big) \Big(\dot{G}(\tau_1, \tau_2) + \dot{G}(\tau_2, \tau_4) + \dot{G}(\tau_4, \tau_1)\Big) ~ k_{1\rho} k_{2\lambda}  k_{4\eta}\,.
  \nonumber\\
\end{eqnarray}
\begin{eqnarray}
&&\mathcal{I}^{\mu\nu\alpha\beta}_{6;(\tau_1,\tau_2,\tau_3,\tau_4)}[k_1, k_3, k_2, k_4]\Big|_{u_1 = u_3} = -\Big( \dot{G}_B(\tau_1, \tau_2) k_{2\delta}
+ \dot{G}_B(\tau_1, \tau_4) k_{4\delta} \Big) \epsilon^{\mu\nu}_{\ \ \ \kappa\zeta} \Big( g^{\rho\kappa} \epsilon^{\zeta\beta\delta\alpha} g^{\lambda\eta}
+ g^{\rho\kappa} \epsilon^{\zeta\eta\alpha\delta} g^{\lambda\beta}
\nonumber \\
 &&+ g^{\rho\kappa} \epsilon^{\zeta\beta\lambda\delta} g^{\alpha\eta}
 + g^{\rho\kappa} \epsilon^{\zeta\eta\delta\lambda} g^{\alpha\beta}
 \Big) \Big(\dot{G}_B(\tau_1, \tau_2) + \dot{G}_B(\tau_2, \tau_4) + \dot{G}_B(\tau_4, \tau_1)\Big) ~ k_{3\rho} k_{2\lambda}  k_{4\eta}\,.
 \nonumber\\
\end{eqnarray}
\begin{eqnarray}
&&\mathcal{I}^{\mu\nu\alpha\beta}_{7;(\tau_1,\tau_2,\tau_3,\tau_4)}[k_1, k_3, k_2, k_4]\Big|_{u_1 = u_3} = \mathcal{I}^{\mu\nu\alpha\beta}_{8;(\tau_1,\tau_2,\tau_3,\tau_4)}[k_1, k_3, k_2, k_4]\Big|_{u_1 = u_3} = 0\,.
\end{eqnarray}
\begin{eqnarray}
&&\mathcal{I}^{\mu\nu\alpha\beta}_{9;(\tau_1,\tau_2,\tau_3,\tau_4)}[k_1, k_3, k_2, k_4]\Big|_{u_1 = u_3} = -2 \epsilon^{\mu\nu\eta\kappa} \epsilon^{\alpha\beta\lambda\sigma} k_{1\eta} k_{3\kappa} k_{2\lambda} k_{4\sigma}\,.
\end{eqnarray}

\section{Calculation of the triangle diagram in the worldline approach}
\label{sec:triangle}
We will compute here the triangle graph in the worldline approach. To do this, we will   extend the worldline representation of the QCD effective action to include an auxialliary axial vector interaction $\slashed{A}_5\gamma_5 $~\cite{Mondragon:1995ab,DHoker:1995aat,Schubert:2001he,Mueller:2017arw,Mueller:2017lzw}:
\begin{eqnarray}
&&\Gamma[A, A_5] = -\frac{1}{2} {\rm Tr_c} \int^\infty_0 \frac{dT}{T} \, \int \mathcal{D}x \int_{AP} \mathcal{D} \psi
\nonumber\\
&&\times\exp\Big\{-\int^T_0 d\tau \Big(\frac{1}{4} \dot{x}^2 + \frac{1}{2}\psi_\mu\dot{\psi}^\mu + ig\dot{x}^\mu A_\mu - ig \psi^\mu \psi^\nu F_{\mu\nu} - 2i\psi_5 \dot{x}^\mu \psi_\mu \psi_\nu A^\nu_5 + i \psi_5 \partial_\mu A^\mu_5 + (D-2)A^2_5\Big)\Big\}\,,
\label{A5action}
\end{eqnarray}
where $\psi_5$ is the Grassmann counterpart of  $\gamma_5$ matrix in the worldline framework.

To compute the triangle graph of the anomaly, since $J_5$ couples to $A_5$, we first take the functional derivative with respect to $A_5$, and then set it equal to zero. Hence, 
\begin{eqnarray}
\langle P^\prime,S| J^\kappa_5 |P,S\rangle = \int d^4y \frac{\partial}{\partial A_{5\kappa}(y)}\Gamma[A, A_5]\Big|_{A_5=0} e^{ily} \equiv \Gamma^\kappa_5[l]\,.
\end{eqnarray}
Then employing the worldline action in Eq.~(\ref{A5action}) including $\Gamma[A, A_5]$, we get
\begin{eqnarray}
\Gamma^\kappa_5[l] &=& \frac{i}{2} {\rm Tr_c} \int^\infty_0 \frac{dT}{T} \, \int \mathcal{D}x \int_{AP} \mathcal{D} \psi
~\int^T_0 d\tau_l~ \psi_5 \Big(i  l^\kappa + 2 \psi^\kappa_l \dot{x}_l \cdot\psi_l \Big) e^{ilx_l}
\nonumber\\
&\times& \exp\Big\{-\int^T_0 d\tau \Big(\frac{1}{4} \dot{x}^2 + \frac{1}{2}\psi_\mu\dot{\psi}^\mu + ig\dot{x}^\mu A_\mu - ig \psi^\mu \psi^\nu F_{\mu\nu} \Big)\Big\}\,,
\label{trianglebefexp}
\end{eqnarray}
where $\tau_l$ is the proper time coordinate of the $A_5$ insertion into the worldline, and $l$ is the incoming momentum.  As usual, we also use the shorthand notation $x_l\equiv x(\tau_l)$, $\psi_l\equiv\psi(\tau_l)$.

\begin{figure}[htb]
 \begin{center}
 \includegraphics[width=120mm]{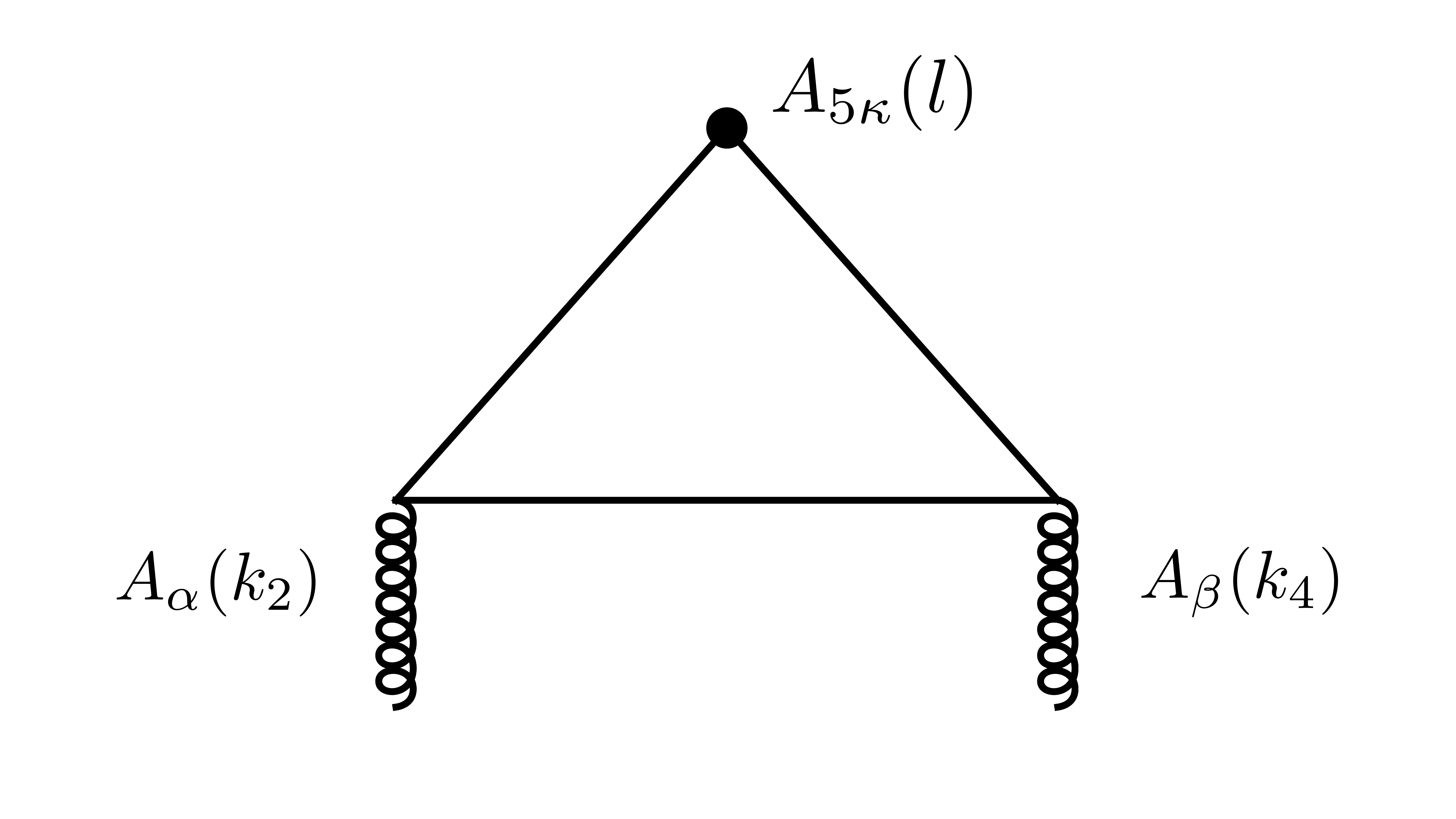}
 \end{center}
 \caption{\label{fig:tringlegraph}The  triangle graph representing the vector-vector-axial vector (VVA) coupling of the chiral anomaly.}
 \end{figure}
 
To compute Fig. \ref{fig:tringlegraph}, we will first expand the phase in Eq.~(\ref{trianglebefexp}) to  second order in the coupling constant:
\begin{eqnarray}
\label{triangleaftexp}
&&\Gamma^\kappa_5[l] = -\frac{ig^2}{2} {\rm Tr_c} \int^\infty_0 \frac{dT}{T} \, \int \mathcal{D}x \int_{AP} \mathcal{D} \psi
~\int^T_0 d\tau_l~ \psi_5 \Big(i  l^\kappa + 2 \psi^\kappa_l \dot{x}_l \cdot\psi_l \Big) e^{ilx_l}\int^T_0 d\tau_2 \int^T_0 d\tau_4 
\nonumber \\
&&\times \Big( \dot{x}^\alpha_2 A_\alpha(x_2) -  2 \psi^\lambda_2 \psi^\alpha_2 \partial_\lambda A_\alpha(x_2) \Big) \Big( \dot{x}^\beta_4 A_\beta(x_4) -  2\psi^\eta_4 \psi^\beta_4 \partial_\eta A_\beta(x_4) \Big)
\exp\Big\{-\int^T_0 d\tau \Big(\frac{1}{4} \dot{x}^2 + \frac{1}{2}\psi_\mu\dot{\psi}^\mu \Big)\Big\}\,.
\nonumber\\
\end{eqnarray}
We can rewrite this equation as 
\begin{eqnarray}
&&\Gamma^\kappa_5[l] = \int \frac{d^4k_2}{(2\pi)^4} \int \frac{d^4k_4}{(2\pi)^4} ~\Gamma^{\kappa\alpha\beta}_5[l,k_2,k_4]~{\rm Tr_c} A_\alpha(k_2)  A_\beta(k_4) \,,
\label{trianglFour}
\end{eqnarray}
where the VVA vertex function,
\begin{eqnarray}
&&\Gamma^{\kappa\alpha\beta}_5[l,k_2,k_4] \equiv -\frac{ig^2}{2}  \int^\infty_0 \frac{dT}{T} \, \int \mathcal{D}x \int_{AP} \mathcal{D} \psi
~\int^T_0 d\tau_l~ \psi_5 \Big(i  l^\kappa + 2 \psi^\kappa_l \dot{x}_l \cdot\psi_l \Big) e^{ilx_l}
\label{trianglFourArg}\\
&&\times\int^T_0 d\tau_2 \int^T_0 d\tau_4 \Big( \dot{x}^\alpha_2  +  2 i \psi^\alpha_2 \psi^\lambda_2 k_{2\lambda} \Big)  e^{ik_2x_2} \Big( \dot{x}^\beta_4 + 2 i \psi^\beta_4 \psi^\eta_4 k_{4\eta} \Big) e^{ik_4 x_4}
\exp\Big\{-\int^T_0 d\tau \Big(\frac{1}{4} \dot{x}^2 + \frac{1}{2}\psi_\mu\dot{\psi}^\mu \Big)\Big\}\,.
\nonumber
\end{eqnarray}
This structure is illustrated in  Fig. \ref{fig:tringlegraph}.

Examining  $\Gamma^{\kappa\alpha\beta}_5[l,k_2,k_4]$, we notice that it has a $\psi_5$ in the argument of the Grassmannian functional integral; this changes the boundary condition  from being antiperiodic (AP) to being periodic (P).  As a result, 
the Grassmann variables in the functional integral acquire a zero mode, which can be 
separated out from the nonzero modes in the action and in the measure as,
\begin{eqnarray}
\psi^\mu(\tau) = \psi^\mu_0 + \xi^\mu(\tau)\,;\ \ \ \int_P \mathcal{D}\psi = \int d^4\psi_0 \int_P \mathcal{D}\xi\,;\ \ \ \int^T_0 d\tau \,\xi(\tau) = 0\,.
\end{eqnarray}
Separating out the zero mode thus, we obtain 
\begin{eqnarray}
&&\Gamma^{\kappa\alpha\beta}_5[l,k_2,k_4] \equiv -\frac{ig^2}{2}  \int^\infty_0 \frac{dT}{T} \, \int \mathcal{D}x \int d^4\psi_0 \int_P \mathcal{D}\xi
~\int^T_0 d\tau_l~ \Big(i  l^\kappa + 2 \psi^\kappa_l \dot{x}_l \cdot\psi_l \Big) e^{ilx_l}
\nonumber\\
&&\times\int^T_0 d\tau_2 \int^T_0 d\tau_4 \Big( \dot{x}^\alpha_2  +  2 i \psi^\alpha_2 \psi^\lambda_2 k_{2\lambda} \Big)  e^{ik_2x_2} \Big( \dot{x}^\beta_4 + 2 i \psi^\beta_4 \psi^\eta_4 k_{4\eta} \Big) e^{ik_4x_4}
\exp\Big\{-\int^T_0 d\tau \Big(\frac{1}{4} \dot{x}^2 + \frac{1}{2}\psi_\mu\dot{\psi}^\mu \Big)\Big\}\Big|_{\psi = \psi_0+\xi}\,.
\nonumber\\
\end{eqnarray}

The evaluation of the functional integrals over $x$ and $\xi$, as well as the integral over zero mode $\psi_0$, is straightforward. In particular, we use the identities, 
\begin{eqnarray}
\int d^4\psi_0~ \psi^\mu_0\psi^\nu_0\psi^\rho_0\psi^\sigma_0 = \epsilon^{\mu\nu\rho\sigma}\,,
\end{eqnarray}
and
\begin{eqnarray}
\int_P \mathcal{D}\xi~ \xi^\mu(\tau_1)\xi^\nu(\tau_2)~\exp\Big\{-\int^T_0 d\tau \frac{1}{2}\xi_\mu\dot{\xi}^\mu \Big\} = g^{\mu\nu}\frac{1}{2}\dot{G}_B(\tau_1, \tau_2)\,,
\end{eqnarray}
where the first derivative of the bosonic worldline propagator is
\begin{eqnarray}
\dot{G}_B(\tau_1, \tau_2) \equiv \frac{\partial}{\partial \tau_1}G_B(\tau_1, \tau_2) = {\rm sign}(\tau_1 - \tau_2) - 2\frac{\tau_1 - \tau_2}{T}\,.
\end{eqnarray}
The details of the calculation of the functional integral over bosonic worldline trajectories $x$ can be found in \cite{Tarasov:2019rfp}.

Evaluating the integrals, we obtain after lengthy algebraric manipulations,
\begin{eqnarray}
&&\Gamma^{\kappa\alpha\beta}_5[l, k_2, k_4] = 2 \int^\infty_0 \frac{dT}{T}(4\pi T)^{-2} \int^T_0 d\tau_l \int^T_0 d\tau_2 \int^T_0 d\tau_4
\nonumber\\
&&\times \Big[ \Big\{ - \dot{G}^2_B(\tau_l, \tau_4) + \dot{G}_B(\tau_l, \tau_4) \dot{G}_B(\tau_2, \tau_4) + \dot{G}_B(\tau_l, \tau_2)\dot{G}_B(\tau_l, \tau_4)  - \dot{G}_B(\tau_l, \tau_2)\dot{G}_B(\tau_2, \tau_4) \Big\} k_2 \cdot k_4 \epsilon^{\kappa\alpha\beta\sigma} k_{2\sigma}  
\nonumber\\
&&+ \Big\{ \dot{G}^2_B(\tau_l, \tau_2) + \dot{G}_B(\tau_l, \tau_2) \dot{G}_B(\tau_2, \tau_4) - \dot{G}_B(\tau_l, \tau_4)\dot{G}_B(\tau_l, \tau_2) - \dot{G}_B(\tau_l, \tau_4) \dot{G}_B(\tau_2, \tau_4) \Big\} k_2\cdot k_4 \epsilon^{\kappa\alpha\beta\sigma} k_{4\sigma}
\nonumber\\
&&+ \Big\{ - \dot{G}^2_B(\tau_l, \tau_4) + \dot{G}_B(\tau_l, \tau_4) \dot{G}_B(\tau_2, \tau_4) + \dot{G}_B(\tau_l, \tau_2) \dot{G}_B(\tau_l, \tau_4) - \dot{G}_B(\tau_l, \tau_2)\dot{G}_B(\tau_2, \tau_4)  \Big\} \epsilon^{\kappa\alpha\sigma\lambda} k^\beta_2 k_{2\sigma} k_{4\lambda}
\nonumber\\
&&+ \Big\{ \dot{G}^2_B(\tau_l, \tau_2) + \dot{G}_B(\tau_l, \tau_2) \dot{G}_B(\tau_2, \tau_4) - \dot{G}_B(\tau_l, \tau_4) \dot{G}_B(\tau_l, \tau_2) - \dot{G}_B(\tau_l, \tau_4) \dot{G}_B(\tau_2, \tau_4) \Big\}\epsilon^{\kappa\beta \sigma\lambda} k^\alpha_4  k_{2\sigma} k_{4\lambda}
\nonumber\\
&&+ \Big\{ - 1 + \dot{G}^2_B(\tau_l, \tau_2) \Big\} \epsilon^{\alpha\beta\sigma\lambda} k^\kappa_2 k_{2\sigma} k_{4\lambda}
+ \Big\{ - 1 + \dot{G}^2_B(\tau_l, \tau_4) \Big\} \epsilon^{\alpha\beta\sigma\lambda} k^\kappa_4 k_{2\sigma} k_{4\lambda} \Big]
\nonumber\\
&&\times \exp\Big[ - k_2 \cdot k_4 G_B(\tau_l, \tau_2) - k_2 \cdot k_4 G_B(\tau_l, \tau_4) + k_2\cdot k_4 G_B(\tau_2, \tau_4) \Big] (2\pi)^4\delta^4(l + k_2 + k_4)\,.
\end{eqnarray}
As in the main text, we use the on mass-shell condition $k^2_2 = k^2_4 = 0$ for the background gluons.

Now using the identity,
\begin{eqnarray}
\epsilon^{\alpha\beta\sigma\lambda} k^\kappa_2 k_{2\sigma} k_{4\lambda} = - k_2 \cdot k_{4} \epsilon^{\kappa\alpha\beta\sigma} k_{2\sigma} -  \epsilon^{\kappa\alpha\sigma\lambda} k^\beta_2 k_{2\sigma} k_{4\lambda} - \epsilon^{\beta\sigma\lambda\kappa} k^\alpha_2 k_{2\sigma} k_{4\lambda} - k^2_2 \epsilon^{\lambda\kappa\alpha\beta} k_{4\lambda} \,,
\end{eqnarray}
that we employed in the main text, and a similar identity for  $\epsilon^{\alpha\beta\sigma\lambda} k^\kappa_4 k_{2\sigma} k_{4\lambda}$, we obtain,
\begin{eqnarray}
&&\Gamma^{\kappa\alpha\beta}_5[l, k_2, k_4] = 2 \int^\infty_0 \frac{dT}{T}(4\pi T)^{-2} \int^T_0 d\tau_l \int^T_0 d\tau_2 \int^T_0 d\tau_4
\nonumber\\
&&\times \Big( \Big[ 1 - \dot{G}^2_B(\tau_l, \tau_4) + \dot{G}_B(\tau_l, \tau_4) \dot{G}_B(\tau_2, \tau_4) 
- \dot{G}_B(\tau_l, \tau_2) \big( \dot{G}_B(\tau_l, \tau_2) + \dot{G}_B(\tau_2, \tau_4) + \dot{G}_B(\tau_4, \tau_l) \big)  \Big] 
\nonumber\\
&&\times ( k_2 \cdot k_4 \epsilon^{\kappa\alpha\beta\sigma} k_{2\sigma}  + \epsilon^{\kappa\alpha\sigma\lambda} k^\beta_2 k_{2\sigma} k_{4\lambda} )
\nonumber\\
&&+ \Big[ - 1 + \dot{G}^2_B(\tau_l, \tau_2) - \dot{G}_B(\tau_l, \tau_2) \dot{G}_B(\tau_4, \tau_2) + \dot{G}_B(\tau_l, \tau_4) \big( \dot{G}_B(\tau_l, \tau_4) + \dot{G}_B(\tau_4, \tau_2) + \dot{G}_B(\tau_2, \tau_l) \big)\Big] 
\nonumber\\
&&\times ( k_2\cdot k_4 \epsilon^{\kappa\alpha\beta\sigma} k_{4\sigma} + \epsilon^{\kappa\beta \sigma\lambda} k^\alpha_4  k_{2\sigma} k_{4\lambda} ) \Big)
\nonumber\\
&&\times \exp\Big[ - k_2 \cdot k_4 G_B(\tau_l, \tau_2) - k_2 \cdot k_4 G_B(\tau_l, \tau_4) + k_2\cdot k_4 G_B(\tau_2, \tau_4) \Big] (2\pi)^4\delta^4(l + k_2 + k_4) \,,
\end{eqnarray}
where we took into account the constrains $k^\alpha_2 A_\alpha(k_2) = 0$ and $k^\beta_4 A_\beta(k_4) = 0$ from gauge invariance. Further simplifying this result, and using the identity,
\begin{eqnarray}
1 - \dot{G}^2_B(\tau_i, \tau_j) = \frac{4}{T}G_B(\tau_i, \tau_j)\,,
\end{eqnarray}
we rewrite our result as
\begin{eqnarray}
&&\Gamma^{\kappa\alpha\beta}_5[l, k_2, k_4] =  \int^\infty_0 \frac{dT}{T}(4\pi T)^{-2} \int^T_0 d\tau_l \int^T_0 d\tau_2 \int^T_0 d\tau_4~ \frac{4}{T}\Big[ G_B(\tau_l, \tau_2) - G_B(\tau_2, \tau_4) + G_B(\tau_l, \tau_4) \Big] \epsilon^{\alpha\beta\sigma\lambda} k_{2\sigma} k_{4\lambda}
\nonumber\\
&&\times ( - k^\kappa_{2} - k^\kappa_{4} ) \exp\Big[ - k_2\cdot k_4 (G_B(\tau_l, \tau_2) - G_B(\tau_2, \tau_4) + G_B(\tau_l, \tau_4)) \Big] (2\pi)^4 \delta^4(l + k_2 + k_4)
\end{eqnarray}

Introducing the  variable $u \equiv \tau/T$, with $u\in [0, 1]$, we can integrate 
over the worldline period $T$. By doing so, one finds that the numerator and denominator of the expressions containing the Green's functions cancel each other,  yielding
\begin{eqnarray}
\label{eq:VVA1}
&&\Gamma^{\kappa\alpha\beta}_5[l, k_2, k_4] =  \frac{ 1 }{2\pi^2}\frac{k^\kappa_{2} + k^\kappa_{4}}{ (k_2 + k_4)^2 } ~ \epsilon^{\alpha\sigma\beta\lambda} k_{2\sigma} k_{4\lambda}  (2\pi)^4 \delta^4(l + k_2 + k_4)\,.
\end{eqnarray}
Note that this result can be expressed in the form stated in Eq.~(6.49) of Ref.~\cite{Schubert:2001he}, thereby providing a nice consistency check of our derivation.

Substituting this VVA vertex function back into Eq. (\ref{trianglFour}), we  obtain our final result,
\begin{eqnarray}
&&\Gamma^\kappa_5[l] = \frac{ 1 }{4\pi^2} \frac{l^\kappa}{ l^2 } \int \frac{d^4k_2}{(2\pi)^4} \int \frac{d^4k_4}{(2\pi)^4} ~ {\rm Tr_c} F_{\alpha\beta}(k_2)  \tilde{F}^{\alpha\beta}(k_4)~(2\pi)^4 \delta^4(l + k_2 + k_4)\,.
\end{eqnarray}

\bibliography{wlines}

\begin{thebibliography}{121}%
\makeatletter
\providecommand \@ifxundefined [1]{%
 \@ifx{#1\undefined}
}%
\providecommand \@ifnum [1]{%
 \ifnum #1\expandafter \@firstoftwo
 \else \expandafter \@secondoftwo
 \fi
}%
\providecommand \@ifx [1]{%
 \ifx #1\expandafter \@firstoftwo
 \else \expandafter \@secondoftwo
 \fi
}%
\providecommand \natexlab [1]{#1}%
\providecommand \enquote  [1]{``#1''}%
\providecommand \bibnamefont  [1]{#1}%
\providecommand \bibfnamefont [1]{#1}%
\providecommand \citenamefont [1]{#1}%
\providecommand \href@noop [0]{\@secondoftwo}%
\providecommand \href [0]{\begingroup \@sanitize@url \@href}%
\providecommand \@href[1]{\@@startlink{#1}\@@href}%
\providecommand \@@href[1]{\endgroup#1\@@endlink}%
\providecommand \@sanitize@url [0]{\catcode `\\12\catcode `\$12\catcode
  `\&12\catcode `\#12\catcode `\^12\catcode `\_12\catcode `\%12\relax}%
\providecommand \@@startlink[1]{}%
\providecommand \@@endlink[0]{}%
\providecommand \url  [0]{\begingroup\@sanitize@url \@url }%
\providecommand \@url [1]{\endgroup\@href {#1}{\urlprefix }}%
\providecommand \urlprefix  [0]{URL }%
\providecommand \Eprint [0]{\href }%
\providecommand \doibase [0]{http://dx.doi.org/}%
\providecommand \selectlanguage [0]{\@gobble}%
\providecommand \bibinfo  [0]{\@secondoftwo}%
\providecommand \bibfield  [0]{\@secondoftwo}%
\providecommand \translation [1]{[#1]}%
\providecommand \BibitemOpen [0]{}%
\providecommand \bibitemStop [0]{}%
\providecommand \bibitemNoStop [0]{.\EOS\space}%
\providecommand \EOS [0]{\spacefactor3000\relax}%
\providecommand \BibitemShut  [1]{\csname bibitem#1\endcsname}%
\let\auto@bib@innerbib\@empty
%</preamble>
\bibitem [{\citenamefont {Kodaira}(1980)}]{Kodaira:1979pa}%
  \BibitemOpen
  \bibfield  {author} {\bibinfo {author} {\bibfnamefont {J.}~\bibnamefont
  {Kodaira}},\ }\href {\doibase 10.1016/0550-3213(80)90310-7} {\bibfield
  {journal} {\bibinfo  {journal} {Nucl. Phys. B}\ }\textbf {\bibinfo {volume}
  {165}},\ \bibinfo {pages} {129} (\bibinfo {year} {1980})}\BibitemShut
  {NoStop}%
\bibitem [{\citenamefont {Altarelli}\ and\ \citenamefont
  {Ross}(1988)}]{Altarelli:1988nr}%
  \BibitemOpen
  \bibfield  {author} {\bibinfo {author} {\bibfnamefont {G.}~\bibnamefont
  {Altarelli}}\ and\ \bibinfo {author} {\bibfnamefont {G.~G.}\ \bibnamefont
  {Ross}},\ }\href {\doibase 10.1016/0370-2693(88)91335-4} {\bibfield
  {journal} {\bibinfo  {journal} {Phys. Lett.}\ }\textbf {\bibinfo {volume}
  {B212}},\ \bibinfo {pages} {391} (\bibinfo {year} {1988})}\BibitemShut
  {NoStop}%
%%CITATION = PHLTA,B212,391;%%
\bibitem [{\citenamefont {Carlitz}\ \emph {et~al.}(1988)\citenamefont
  {Carlitz}, \citenamefont {Collins},\ and\ \citenamefont
  {Mueller}}]{Carlitz:1988ab}%
  \BibitemOpen
  \bibfield  {author} {\bibinfo {author} {\bibfnamefont {R.~D.}\ \bibnamefont
  {Carlitz}}, \bibinfo {author} {\bibfnamefont {J.~C.}\ \bibnamefont
  {Collins}}, \ and\ \bibinfo {author} {\bibfnamefont {A.~H.}\ \bibnamefont
  {Mueller}},\ }\href {\doibase 10.1016/0370-2693(88)91474-8} {\bibfield
  {journal} {\bibinfo  {journal} {Phys. Lett.}\ }\textbf {\bibinfo {volume}
  {B214}},\ \bibinfo {pages} {229} (\bibinfo {year} {1988})}\BibitemShut
  {NoStop}%
%%CITATION = PHLTA,B214,229;%%
\bibitem [{\citenamefont {Jaffe}\ and\ \citenamefont
  {Manohar}(1990)}]{Jaffe:1989jz}%
  \BibitemOpen
  \bibfield  {author} {\bibinfo {author} {\bibfnamefont {R.~L.}\ \bibnamefont
  {Jaffe}}\ and\ \bibinfo {author} {\bibfnamefont {A.}~\bibnamefont
  {Manohar}},\ }\href {\doibase 10.1016/0550-3213(90)90506-9} {\bibfield
  {journal} {\bibinfo  {journal} {Nucl. Phys.}\ }\textbf {\bibinfo {volume}
  {B337}},\ \bibinfo {pages} {509} (\bibinfo {year} {1990})}\BibitemShut
  {NoStop}%
%%CITATION = NUPHA,B337,509;%%
\bibitem [{\citenamefont {Adler}(1969)}]{Adler:1969gk}%
  \BibitemOpen
  \bibfield  {author} {\bibinfo {author} {\bibfnamefont {S.~L.}\ \bibnamefont
  {Adler}},\ }\href {\doibase 10.1103/PhysRev.177.2426} {\bibfield  {journal}
  {\bibinfo  {journal} {Phys. Rev.}\ }\textbf {\bibinfo {volume} {177}},\
  \bibinfo {pages} {2426} (\bibinfo {year} {1969})}\BibitemShut {NoStop}%
\bibitem [{\citenamefont {Adler}\ and\ \citenamefont
  {Bardeen}(1969)}]{Adler:1969er}%
  \BibitemOpen
  \bibfield  {author} {\bibinfo {author} {\bibfnamefont {S.~L.}\ \bibnamefont
  {Adler}}\ and\ \bibinfo {author} {\bibfnamefont {W.~A.}\ \bibnamefont
  {Bardeen}},\ }\href {\doibase 10.1103/PhysRev.182.1517} {\bibfield  {journal}
  {\bibinfo  {journal} {Phys. Rev.}\ }\textbf {\bibinfo {volume} {182}},\
  \bibinfo {pages} {1517} (\bibinfo {year} {1969})}\BibitemShut {NoStop}%
\bibitem [{\citenamefont {Bell}\ and\ \citenamefont
  {Jackiw}(1969)}]{Bell:1969ts}%
  \BibitemOpen
  \bibfield  {author} {\bibinfo {author} {\bibfnamefont {J.}~\bibnamefont
  {Bell}}\ and\ \bibinfo {author} {\bibfnamefont {R.}~\bibnamefont {Jackiw}},\
  }\href {\doibase 10.1007/BF02823296} {\bibfield  {journal} {\bibinfo
  {journal} {Nuovo Cim. A}\ }\textbf {\bibinfo {volume} {60}},\ \bibinfo
  {pages} {47} (\bibinfo {year} {1969})}\BibitemShut {NoStop}%
\bibitem [{\citenamefont {Kogut}\ and\ \citenamefont
  {Susskind}(1975)}]{Kogut:1974kt}%
  \BibitemOpen
  \bibfield  {author} {\bibinfo {author} {\bibfnamefont {J.~B.}\ \bibnamefont
  {Kogut}}\ and\ \bibinfo {author} {\bibfnamefont {L.}~\bibnamefont
  {Susskind}},\ }\href {\doibase 10.1103/PhysRevD.11.3594} {\bibfield
  {journal} {\bibinfo  {journal} {Phys. Rev. D}\ }\textbf {\bibinfo {volume}
  {11}},\ \bibinfo {pages} {3594} (\bibinfo {year} {1975})}\BibitemShut
  {NoStop}%
\bibitem [{\citenamefont {Polyakov}(1987)}]{Polyakov:1987ez}%
  \BibitemOpen
  \bibfield  {author} {\bibinfo {author} {\bibfnamefont {A.~M.}\ \bibnamefont
  {Polyakov}},\ }\href@noop {} {\emph {\bibinfo {title} {{Gauge Fields and
  Strings}}}},\ Vol.~\bibinfo {volume} {3}\ (\bibinfo {year}
  {1987})\BibitemShut {NoStop}%
\bibitem [{\citenamefont {Alvarez-Gaume}\ and\ \citenamefont
  {Witten}(1984)}]{AlvarezGaume:1983ig}%
  \BibitemOpen
  \bibfield  {author} {\bibinfo {author} {\bibfnamefont {L.}~\bibnamefont
  {Alvarez-Gaume}}\ and\ \bibinfo {author} {\bibfnamefont {E.}~\bibnamefont
  {Witten}},\ }\href {\doibase 10.1016/0550-3213(84)90066-X} {\bibfield
  {journal} {\bibinfo  {journal} {Nucl. Phys. B}\ }\textbf {\bibinfo {volume}
  {234}},\ \bibinfo {pages} {269} (\bibinfo {year} {1984})}\BibitemShut
  {NoStop}%
\bibitem [{\citenamefont {Strassler}(1992)}]{Strassler:1992zr}%
  \BibitemOpen
  \bibfield  {author} {\bibinfo {author} {\bibfnamefont {M.~J.}\ \bibnamefont
  {Strassler}},\ }\href {\doibase 10.1016/0550-3213(92)90098-V} {\bibfield
  {journal} {\bibinfo  {journal} {Nucl. Phys.}\ }\textbf {\bibinfo {volume}
  {B385}},\ \bibinfo {pages} {145} (\bibinfo {year} {1992})},\ \Eprint
  {http://arxiv.org/abs/hep-ph/9205205} {arXiv:hep-ph/9205205 [hep-ph]}
  \BibitemShut {NoStop}%
%%CITATION = HEP-PH/9205205;%%
\bibitem [{\citenamefont {D'Hoker}\ and\ \citenamefont
  {Gagne}(1996{\natexlab{a}})}]{DHoker:1995aat}%
  \BibitemOpen
  \bibfield  {author} {\bibinfo {author} {\bibfnamefont {E.}~\bibnamefont
  {D'Hoker}}\ and\ \bibinfo {author} {\bibfnamefont {D.~G.}\ \bibnamefont
  {Gagne}},\ }\href {\doibase 10.1016/0550-3213(96)00125-3} {\bibfield
  {journal} {\bibinfo  {journal} {Nucl. Phys.}\ }\textbf {\bibinfo {volume}
  {B467}},\ \bibinfo {pages} {272} (\bibinfo {year} {1996}{\natexlab{a}})},\
  \Eprint {http://arxiv.org/abs/hep-th/9508131} {arXiv:hep-th/9508131 [hep-th]}
  \BibitemShut {NoStop}%
%%CITATION = HEP-TH/9508131;%%
\bibitem [{\citenamefont {D'Hoker}\ and\ \citenamefont
  {Gagne}(1996{\natexlab{b}})}]{DHoker:1995uyv}%
  \BibitemOpen
  \bibfield  {author} {\bibinfo {author} {\bibfnamefont {E.}~\bibnamefont
  {D'Hoker}}\ and\ \bibinfo {author} {\bibfnamefont {D.~G.}\ \bibnamefont
  {Gagne}},\ }\href {\doibase 10.1016/0550-3213(96)00126-5} {\bibfield
  {journal} {\bibinfo  {journal} {Nucl. Phys.}\ }\textbf {\bibinfo {volume}
  {B467}},\ \bibinfo {pages} {297} (\bibinfo {year} {1996}{\natexlab{b}})},\
  \Eprint {http://arxiv.org/abs/hep-th/9512080} {arXiv:hep-th/9512080 [hep-th]}
  \BibitemShut {NoStop}%
%%CITATION = HEP-TH/9512080;%%
\bibitem [{\citenamefont {Haack}\ and\ \citenamefont
  {Schmidt}(1999)}]{Haack:1998uy}%
  \BibitemOpen
  \bibfield  {author} {\bibinfo {author} {\bibfnamefont {M.}~\bibnamefont
  {Haack}}\ and\ \bibinfo {author} {\bibfnamefont {M.~G.}\ \bibnamefont
  {Schmidt}},\ }\href {\doibase 10.1007/s100529800982} {\bibfield  {journal}
  {\bibinfo  {journal} {Eur. Phys. J. C}\ }\textbf {\bibinfo {volume} {7}},\
  \bibinfo {pages} {149} (\bibinfo {year} {1999})},\ \Eprint
  {http://arxiv.org/abs/hep-th/9806138} {arXiv:hep-th/9806138} \BibitemShut
  {NoStop}%
\bibitem [{\citenamefont {McKeon}\ and\ \citenamefont
  {Schubert}(1998)}]{McKeon:1998et}%
  \BibitemOpen
  \bibfield  {author} {\bibinfo {author} {\bibfnamefont {D.}~\bibnamefont
  {McKeon}}\ and\ \bibinfo {author} {\bibfnamefont {C.}~\bibnamefont
  {Schubert}},\ }\href {\doibase 10.1016/S0370-2693(98)01074-0} {\bibfield
  {journal} {\bibinfo  {journal} {Phys. Lett. B}\ }\textbf {\bibinfo {volume}
  {440}},\ \bibinfo {pages} {101} (\bibinfo {year} {1998})},\ \Eprint
  {http://arxiv.org/abs/hep-th/9807072} {arXiv:hep-th/9807072} \BibitemShut
  {NoStop}%
\bibitem [{\citenamefont {Schubert}(2001)}]{Schubert:2001he}%
  \BibitemOpen
  \bibfield  {author} {\bibinfo {author} {\bibfnamefont {C.}~\bibnamefont
  {Schubert}},\ }\href {\doibase 10.1016/S0370-1573(01)00013-8} {\bibfield
  {journal} {\bibinfo  {journal} {Phys. Rept.}\ }\textbf {\bibinfo {volume}
  {355}},\ \bibinfo {pages} {73} (\bibinfo {year} {2001})},\ \Eprint
  {http://arxiv.org/abs/hep-th/0101036} {arXiv:hep-th/0101036 [hep-th]}
  \BibitemShut {NoStop}%
%%CITATION = HEP-TH/0101036;%%
\bibitem [{\citenamefont {Bastianelli}\ \emph {et~al.}(2004)\citenamefont
  {Bastianelli}, \citenamefont {Corradini},\ and\ \citenamefont
  {Zirotti}}]{Bastianelli:2003bg}%
  \BibitemOpen
  \bibfield  {author} {\bibinfo {author} {\bibfnamefont {F.}~\bibnamefont
  {Bastianelli}}, \bibinfo {author} {\bibfnamefont {O.}~\bibnamefont
  {Corradini}}, \ and\ \bibinfo {author} {\bibfnamefont {A.}~\bibnamefont
  {Zirotti}},\ }\href {\doibase 10.1088/1126-6708/2004/01/023} {\bibfield
  {journal} {\bibinfo  {journal} {JHEP}\ }\textbf {\bibinfo {volume} {01}},\
  \bibinfo {pages} {023} (\bibinfo {year} {2004})},\ \Eprint
  {http://arxiv.org/abs/hep-th/0312064} {arXiv:hep-th/0312064} \BibitemShut
  {NoStop}%
\bibitem [{\citenamefont {Jaffe}(1996)}]{Jaffe:1996zw}%
  \BibitemOpen
  \bibfield  {author} {\bibinfo {author} {\bibfnamefont {R.~L.}\ \bibnamefont
  {Jaffe}},\ }in\ \href@noop {} {\emph {\bibinfo {booktitle} {{Ettore Majorana
  International School of Nucleon Structure: 1st Course: The Spin Structure of
  the Nucleon}}}}\ (\bibinfo {year} {1996})\ pp.\ \bibinfo {pages} {42--129},\
  \Eprint {http://arxiv.org/abs/hep-ph/9602236} {arXiv:hep-ph/9602236}
  \BibitemShut {NoStop}%
\bibitem [{\citenamefont {Bass}(1998)}]{Bass:1997zz}%
  \BibitemOpen
  \bibfield  {author} {\bibinfo {author} {\bibfnamefont {S.}~\bibnamefont
  {Bass}},\ }\href {\doibase 10.1142/S0217732398000851} {\bibfield  {journal}
  {\bibinfo  {journal} {Mod. Phys. Lett. A}\ }\textbf {\bibinfo {volume}
  {13}},\ \bibinfo {pages} {791} (\bibinfo {year} {1998})},\ \Eprint
  {http://arxiv.org/abs/hep-ph/9712507} {arXiv:hep-ph/9712507} \BibitemShut
  {NoStop}%
\bibitem [{\citenamefont {Bass}(2005)}]{Bass:2004xa}%
  \BibitemOpen
  \bibfield  {author} {\bibinfo {author} {\bibfnamefont {S.~D.}\ \bibnamefont
  {Bass}},\ }\href {\doibase 10.1103/RevModPhys.77.1257} {\bibfield  {journal}
  {\bibinfo  {journal} {Rev. Mod. Phys.}\ }\textbf {\bibinfo {volume} {77}},\
  \bibinfo {pages} {1257} (\bibinfo {year} {2005})},\ \Eprint
  {http://arxiv.org/abs/hep-ph/0411005} {arXiv:hep-ph/0411005} \BibitemShut
  {NoStop}%
\bibitem [{\citenamefont {Wakamatsu}(2019)}]{Wakamatsu:2019ain}%
  \BibitemOpen
  \bibfield  {author} {\bibinfo {author} {\bibfnamefont {M.}~\bibnamefont
  {Wakamatsu}},\ }\href {\doibase 10.1140/epja/i2019-12800-9} {\bibfield
  {journal} {\bibinfo  {journal} {Eur. Phys. J. A}\ }\textbf {\bibinfo {volume}
  {55}},\ \bibinfo {pages} {123} (\bibinfo {year} {2019})},\ \Eprint
  {http://arxiv.org/abs/1905.03412} {arXiv:1905.03412 [hep-ph]} \BibitemShut
  {NoStop}%
\bibitem [{\citenamefont {Accardi}\ \emph {et~al.}(2016)\citenamefont {Accardi}
  \emph {et~al.}}]{Accardi:2012qut}%
  \BibitemOpen
  \bibfield  {author} {\bibinfo {author} {\bibfnamefont {A.}~\bibnamefont
  {Accardi}} \emph {et~al.},\ }\href {\doibase 10.1140/epja/i2016-16268-9}
  {\bibfield  {journal} {\bibinfo  {journal} {Eur. Phys. J.}\ }\textbf
  {\bibinfo {volume} {A52}},\ \bibinfo {pages} {268} (\bibinfo {year}
  {2016})},\ \Eprint {http://arxiv.org/abs/1212.1701} {arXiv:1212.1701
  [nucl-ex]} \BibitemShut {NoStop}%
%%CITATION = ARXIV:1212.1701;%%
\bibitem [{\citenamefont {Aschenauer}\ \emph {et~al.}(2019)\citenamefont
  {Aschenauer}, \citenamefont {Fazio}, \citenamefont {Lee}, \citenamefont
  {Mantysaari}, \citenamefont {Page}, \citenamefont {Schenke}, \citenamefont
  {Ullrich}, \citenamefont {Venugopalan},\ and\ \citenamefont
  {Zurita}}]{Aschenauer:2017jsk}%
  \BibitemOpen
  \bibfield  {author} {\bibinfo {author} {\bibfnamefont {E.~C.}\ \bibnamefont
  {Aschenauer}}, \bibinfo {author} {\bibfnamefont {S.}~\bibnamefont {Fazio}},
  \bibinfo {author} {\bibfnamefont {J.~H.}\ \bibnamefont {Lee}}, \bibinfo
  {author} {\bibfnamefont {H.}~\bibnamefont {Mantysaari}}, \bibinfo {author}
  {\bibfnamefont {B.~S.}\ \bibnamefont {Page}}, \bibinfo {author}
  {\bibfnamefont {B.}~\bibnamefont {Schenke}}, \bibinfo {author} {\bibfnamefont
  {T.}~\bibnamefont {Ullrich}}, \bibinfo {author} {\bibfnamefont
  {R.}~\bibnamefont {Venugopalan}}, \ and\ \bibinfo {author} {\bibfnamefont
  {P.}~\bibnamefont {Zurita}},\ }\href {\doibase 10.1088/1361-6633/aaf216}
  {\bibfield  {journal} {\bibinfo  {journal} {Rept. Prog. Phys.}\ }\textbf
  {\bibinfo {volume} {82}},\ \bibinfo {pages} {024301} (\bibinfo {year}
  {2019})},\ \Eprint {http://arxiv.org/abs/1708.01527} {arXiv:1708.01527
  [nucl-ex]} \BibitemShut {NoStop}%
%%CITATION = ARXIV:1708.01527;%%
\bibitem [{\citenamefont {Aschenauer}\ \emph {et~al.}(2020)\citenamefont
  {Aschenauer}, \citenamefont {Borsa}, \citenamefont {Lucero}, \citenamefont
  {Nunes},\ and\ \citenamefont {Sassot}}]{Aschenauer:2020pdk}%
  \BibitemOpen
  \bibfield  {author} {\bibinfo {author} {\bibfnamefont {E.~C.}\ \bibnamefont
  {Aschenauer}}, \bibinfo {author} {\bibfnamefont {I.}~\bibnamefont {Borsa}},
  \bibinfo {author} {\bibfnamefont {G.}~\bibnamefont {Lucero}}, \bibinfo
  {author} {\bibfnamefont {A.~S.}\ \bibnamefont {Nunes}}, \ and\ \bibinfo
  {author} {\bibfnamefont {R.}~\bibnamefont {Sassot}},\ }\href@noop {} {\
  (\bibinfo {year} {2020})},\ \Eprint {http://arxiv.org/abs/2007.08300}
  {arXiv:2007.08300 [hep-ph]} \BibitemShut {NoStop}%
\bibitem [{\citenamefont {Shore}\ and\ \citenamefont
  {Veneziano}(1990)}]{Shore:1990zu}%
  \BibitemOpen
  \bibfield  {author} {\bibinfo {author} {\bibfnamefont {G.}~\bibnamefont
  {Shore}}\ and\ \bibinfo {author} {\bibfnamefont {G.}~\bibnamefont
  {Veneziano}},\ }\href {\doibase 10.1016/0370-2693(90)90272-8} {\bibfield
  {journal} {\bibinfo  {journal} {Phys. Lett. B}\ }\textbf {\bibinfo {volume}
  {244}},\ \bibinfo {pages} {75} (\bibinfo {year} {1990})}\BibitemShut
  {NoStop}%
\bibitem [{\citenamefont {Shore}\ and\ \citenamefont
  {Veneziano}(1992)}]{Shore:1991dv}%
  \BibitemOpen
  \bibfield  {author} {\bibinfo {author} {\bibfnamefont {G.}~\bibnamefont
  {Shore}}\ and\ \bibinfo {author} {\bibfnamefont {G.}~\bibnamefont
  {Veneziano}},\ }\href {\doibase 10.1016/0550-3213(92)90639-S} {\bibfield
  {journal} {\bibinfo  {journal} {Nucl. Phys. B}\ }\textbf {\bibinfo {volume}
  {381}},\ \bibinfo {pages} {23} (\bibinfo {year} {1992})}\BibitemShut
  {NoStop}%
\bibitem [{\citenamefont {Narison}\ \emph {et~al.}(1995)\citenamefont
  {Narison}, \citenamefont {Shore},\ and\ \citenamefont
  {Veneziano}}]{Narison:1994hv}%
  \BibitemOpen
  \bibfield  {author} {\bibinfo {author} {\bibfnamefont {S.}~\bibnamefont
  {Narison}}, \bibinfo {author} {\bibfnamefont {G.}~\bibnamefont {Shore}}, \
  and\ \bibinfo {author} {\bibfnamefont {G.}~\bibnamefont {Veneziano}},\ }\href
  {\doibase 10.1016/0550-3213(94)00329-D} {\bibfield  {journal} {\bibinfo
  {journal} {Nucl. Phys. B}\ }\textbf {\bibinfo {volume} {433}},\ \bibinfo
  {pages} {209} (\bibinfo {year} {1995})},\ \Eprint
  {http://arxiv.org/abs/hep-ph/9404277} {arXiv:hep-ph/9404277} \BibitemShut
  {NoStop}%
\bibitem [{\citenamefont {Narison}\ \emph {et~al.}(1999)\citenamefont
  {Narison}, \citenamefont {Shore},\ and\ \citenamefont
  {Veneziano}}]{Narison:1998aq}%
  \BibitemOpen
  \bibfield  {author} {\bibinfo {author} {\bibfnamefont {S.}~\bibnamefont
  {Narison}}, \bibinfo {author} {\bibfnamefont {G.}~\bibnamefont {Shore}}, \
  and\ \bibinfo {author} {\bibfnamefont {G.}~\bibnamefont {Veneziano}},\ }\href
  {\doibase 10.1016/S0550-3213(99)00061-9} {\bibfield  {journal} {\bibinfo
  {journal} {Nucl. Phys. B}\ }\textbf {\bibinfo {volume} {546}},\ \bibinfo
  {pages} {235} (\bibinfo {year} {1999})},\ \Eprint
  {http://arxiv.org/abs/hep-ph/9812333} {arXiv:hep-ph/9812333} \BibitemShut
  {NoStop}%
\bibitem [{\citenamefont {Shore}(2008)}]{Shore:2007yn}%
  \BibitemOpen
  \bibfield  {author} {\bibinfo {author} {\bibfnamefont {G.}~\bibnamefont
  {Shore}},\ }\enquote {\bibinfo {title} {{The U(1)(A) Anomaly and QCD
  Phenomenology}},}\ \ (\bibinfo {year} {2008})\ pp.\ \bibinfo {pages}
  {235--288},\ \Eprint {http://arxiv.org/abs/hep-ph/0701171}
  {arXiv:hep-ph/0701171} \BibitemShut {NoStop}%
\bibitem [{\citenamefont {Wess}\ and\ \citenamefont
  {Zumino}(1971)}]{Wess:1971yu}%
  \BibitemOpen
  \bibfield  {author} {\bibinfo {author} {\bibfnamefont {J.}~\bibnamefont
  {Wess}}\ and\ \bibinfo {author} {\bibfnamefont {B.}~\bibnamefont {Zumino}},\
  }\href {\doibase 10.1016/0370-2693(71)90582-X} {\bibfield  {journal}
  {\bibinfo  {journal} {Phys. Lett. B}\ }\textbf {\bibinfo {volume} {37}},\
  \bibinfo {pages} {95} (\bibinfo {year} {1971})}\BibitemShut {NoStop}%
\bibitem [{\citenamefont {Kirschner}\ and\ \citenamefont
  {Lipatov}(1983)}]{Kirschner:1983di}%
  \BibitemOpen
  \bibfield  {author} {\bibinfo {author} {\bibfnamefont {R.}~\bibnamefont
  {Kirschner}}\ and\ \bibinfo {author} {\bibfnamefont {L.}~\bibnamefont
  {Lipatov}},\ }\href {\doibase 10.1016/0550-3213(83)90178-5} {\bibfield
  {journal} {\bibinfo  {journal} {Nucl. Phys. B}\ }\textbf {\bibinfo {volume}
  {213}},\ \bibinfo {pages} {122} (\bibinfo {year} {1983})}\BibitemShut
  {NoStop}%
\bibitem [{\citenamefont {Bartels}\ \emph
  {et~al.}(1996{\natexlab{a}})\citenamefont {Bartels}, \citenamefont
  {Ermolaev},\ and\ \citenamefont {Ryskin}}]{Bartels:1995iu}%
  \BibitemOpen
  \bibfield  {author} {\bibinfo {author} {\bibfnamefont {J.}~\bibnamefont
  {Bartels}}, \bibinfo {author} {\bibfnamefont {B.}~\bibnamefont {Ermolaev}}, \
  and\ \bibinfo {author} {\bibfnamefont {M.}~\bibnamefont {Ryskin}},\
  }\href@noop {} {\bibfield  {journal} {\bibinfo  {journal} {Z. Phys. C}\
  }\textbf {\bibinfo {volume} {70}},\ \bibinfo {pages} {273} (\bibinfo {year}
  {1996}{\natexlab{a}})},\ \Eprint {http://arxiv.org/abs/hep-ph/9507271}
  {arXiv:hep-ph/9507271} \BibitemShut {NoStop}%
\bibitem [{\citenamefont {Bartels}\ \emph
  {et~al.}(1996{\natexlab{b}})\citenamefont {Bartels}, \citenamefont
  {Ermolaev},\ and\ \citenamefont {Ryskin}}]{Bartels:1996wc}%
  \BibitemOpen
  \bibfield  {author} {\bibinfo {author} {\bibfnamefont {J.}~\bibnamefont
  {Bartels}}, \bibinfo {author} {\bibfnamefont {B.~I.}\ \bibnamefont
  {Ermolaev}}, \ and\ \bibinfo {author} {\bibfnamefont {M.~G.}\ \bibnamefont
  {Ryskin}},\ }\href {\doibase 10.1007/s002880050285, 10.1007/BF02909194}
  {\bibfield  {journal} {\bibinfo  {journal} {Z. Phys.}\ }\textbf {\bibinfo
  {volume} {C72}},\ \bibinfo {pages} {627} (\bibinfo {year}
  {1996}{\natexlab{b}})},\ \Eprint {http://arxiv.org/abs/hep-ph/9603204}
  {arXiv:hep-ph/9603204 [hep-ph]} \BibitemShut {NoStop}%
%%CITATION = HEP-PH/9603204;%%
\bibitem [{\citenamefont {Kovchegov}\ \emph {et~al.}(2016)\citenamefont
  {Kovchegov}, \citenamefont {Pitonyak},\ and\ \citenamefont
  {Sievert}}]{Kovchegov:2015pbl}%
  \BibitemOpen
  \bibfield  {author} {\bibinfo {author} {\bibfnamefont {Y.~V.}\ \bibnamefont
  {Kovchegov}}, \bibinfo {author} {\bibfnamefont {D.}~\bibnamefont {Pitonyak}},
  \ and\ \bibinfo {author} {\bibfnamefont {M.~D.}\ \bibnamefont {Sievert}},\
  }\href {\doibase 10.1007/JHEP01(2016)072, 10.1007/JHEP10(2016)148} {\bibfield
   {journal} {\bibinfo  {journal} {JHEP}\ }\textbf {\bibinfo {volume} {01}},\
  \bibinfo {pages} {072} (\bibinfo {year} {2016})},\ \bibinfo {note} {[Erratum:
  JHEP10,148(2016)]},\ \Eprint {http://arxiv.org/abs/1511.06737}
  {arXiv:1511.06737 [hep-ph]} \BibitemShut {NoStop}%
%%CITATION = ARXIV:1511.06737;%%
\bibitem [{\citenamefont {Kovchegov}\ \emph
  {et~al.}(2017{\natexlab{a}})\citenamefont {Kovchegov}, \citenamefont
  {Pitonyak},\ and\ \citenamefont {Sievert}}]{Kovchegov:2016weo}%
  \BibitemOpen
  \bibfield  {author} {\bibinfo {author} {\bibfnamefont {Y.~V.}\ \bibnamefont
  {Kovchegov}}, \bibinfo {author} {\bibfnamefont {D.}~\bibnamefont {Pitonyak}},
  \ and\ \bibinfo {author} {\bibfnamefont {M.~D.}\ \bibnamefont {Sievert}},\
  }\href {\doibase 10.1103/PhysRevLett.118.052001} {\bibfield  {journal}
  {\bibinfo  {journal} {Phys. Rev. Lett.}\ }\textbf {\bibinfo {volume} {118}},\
  \bibinfo {pages} {052001} (\bibinfo {year} {2017}{\natexlab{a}})},\ \Eprint
  {http://arxiv.org/abs/1610.06188} {arXiv:1610.06188 [hep-ph]} \BibitemShut
  {NoStop}%
%%CITATION = ARXIV:1610.06188;%%
\bibitem [{\citenamefont {Kovchegov}\ \emph
  {et~al.}(2017{\natexlab{b}})\citenamefont {Kovchegov}, \citenamefont
  {Pitonyak},\ and\ \citenamefont {Sievert}}]{Kovchegov:2017jxc}%
  \BibitemOpen
  \bibfield  {author} {\bibinfo {author} {\bibfnamefont {Y.~V.}\ \bibnamefont
  {Kovchegov}}, \bibinfo {author} {\bibfnamefont {D.}~\bibnamefont {Pitonyak}},
  \ and\ \bibinfo {author} {\bibfnamefont {M.~D.}\ \bibnamefont {Sievert}},\
  }\href {\doibase 10.1016/j.physletb.2017.06.032} {\bibfield  {journal}
  {\bibinfo  {journal} {Phys. Lett. B}\ }\textbf {\bibinfo {volume} {772}},\
  \bibinfo {pages} {136} (\bibinfo {year} {2017}{\natexlab{b}})},\ \Eprint
  {http://arxiv.org/abs/1703.05809} {arXiv:1703.05809 [hep-ph]} \BibitemShut
  {NoStop}%
\bibitem [{\citenamefont {Chirilli}(2019)}]{Chirilli:2018kkw}%
  \BibitemOpen
  \bibfield  {author} {\bibinfo {author} {\bibfnamefont {G.~A.}\ \bibnamefont
  {Chirilli}},\ }\href {\doibase 10.1007/JHEP01(2019)118} {\bibfield  {journal}
  {\bibinfo  {journal} {JHEP}\ }\textbf {\bibinfo {volume} {01}},\ \bibinfo
  {pages} {118} (\bibinfo {year} {2019})},\ \Eprint
  {http://arxiv.org/abs/1807.11435} {arXiv:1807.11435 [hep-ph]} \BibitemShut
  {NoStop}%
\bibitem [{\citenamefont {Boussarie}\ \emph {et~al.}(2019)\citenamefont
  {Boussarie}, \citenamefont {Hatta},\ and\ \citenamefont
  {Yuan}}]{Boussarie:2019icw}%
  \BibitemOpen
  \bibfield  {author} {\bibinfo {author} {\bibfnamefont {R.}~\bibnamefont
  {Boussarie}}, \bibinfo {author} {\bibfnamefont {Y.}~\bibnamefont {Hatta}}, \
  and\ \bibinfo {author} {\bibfnamefont {F.}~\bibnamefont {Yuan}},\ }\href
  {\doibase 10.1016/j.physletb.2019.134817} {\bibfield  {journal} {\bibinfo
  {journal} {Phys. Lett. B}\ }\textbf {\bibinfo {volume} {797}},\ \bibinfo
  {pages} {134817} (\bibinfo {year} {2019})},\ \Eprint
  {http://arxiv.org/abs/1904.02693} {arXiv:1904.02693 [hep-ph]} \BibitemShut
  {NoStop}%
\bibitem [{\citenamefont {Cougoulic}\ and\ \citenamefont
  {Kovchegov}(2019)}]{Cougoulic:2019aja}%
  \BibitemOpen
  \bibfield  {author} {\bibinfo {author} {\bibfnamefont {F.}~\bibnamefont
  {Cougoulic}}\ and\ \bibinfo {author} {\bibfnamefont {Y.~V.}\ \bibnamefont
  {Kovchegov}},\ }\href {\doibase 10.1103/PhysRevD.100.114020} {\bibfield
  {journal} {\bibinfo  {journal} {Phys. Rev. D}\ }\textbf {\bibinfo {volume}
  {100}},\ \bibinfo {pages} {114020} (\bibinfo {year} {2019})},\ \Eprint
  {http://arxiv.org/abs/1910.04268} {arXiv:1910.04268 [hep-ph]} \BibitemShut
  {NoStop}%
\bibitem [{\citenamefont {Kovchegov}\ and\ \citenamefont
  {Tawabutr}(2020)}]{Kovchegov:2020hgb}%
  \BibitemOpen
  \bibfield  {author} {\bibinfo {author} {\bibfnamefont {Y.~V.}\ \bibnamefont
  {Kovchegov}}\ and\ \bibinfo {author} {\bibfnamefont {Y.}~\bibnamefont
  {Tawabutr}},\ }\href@noop {} {\  (\bibinfo {year} {2020})},\ \Eprint
  {http://arxiv.org/abs/2005.07285} {arXiv:2005.07285 [hep-ph]} \BibitemShut
  {NoStop}%
\bibitem [{\citenamefont {Cougoulic}\ and\ \citenamefont
  {Kovchegov}(2020)}]{Cougoulic:2020tbc}%
  \BibitemOpen
  \bibfield  {author} {\bibinfo {author} {\bibfnamefont {F.}~\bibnamefont
  {Cougoulic}}\ and\ \bibinfo {author} {\bibfnamefont {Y.~V.}\ \bibnamefont
  {Kovchegov}},\ }\href@noop {} {\  (\bibinfo {year} {2020})},\ \Eprint
  {http://arxiv.org/abs/2005.14688} {arXiv:2005.14688 [hep-ph]} \BibitemShut
  {NoStop}%
\bibitem [{\citenamefont {Blumlein}(2013)}]{Blumlein:2012bf}%
  \BibitemOpen
  \bibfield  {author} {\bibinfo {author} {\bibfnamefont {J.}~\bibnamefont
  {Blumlein}},\ }\href {\doibase 10.1016/j.ppnp.2012.09.006} {\bibfield
  {journal} {\bibinfo  {journal} {Prog. Part. Nucl. Phys.}\ }\textbf {\bibinfo
  {volume} {69}},\ \bibinfo {pages} {28} (\bibinfo {year} {2013})},\ \Eprint
  {http://arxiv.org/abs/1208.6087} {arXiv:1208.6087 [hep-ph]} \BibitemShut
  {NoStop}%
\bibitem [{\citenamefont {Anselmino}\ \emph {et~al.}(1995)\citenamefont
  {Anselmino}, \citenamefont {Efremov},\ and\ \citenamefont
  {Leader}}]{Anselmino:1994gn}%
  \BibitemOpen
  \bibfield  {author} {\bibinfo {author} {\bibfnamefont {M.}~\bibnamefont
  {Anselmino}}, \bibinfo {author} {\bibfnamefont {A.}~\bibnamefont {Efremov}},
  \ and\ \bibinfo {author} {\bibfnamefont {E.}~\bibnamefont {Leader}},\ }\href
  {\doibase 10.1016/0370-1573(95)00011-5} {\bibfield  {journal} {\bibinfo
  {journal} {Phys. Rept.}\ }\textbf {\bibinfo {volume} {261}},\ \bibinfo
  {pages} {1} (\bibinfo {year} {1995})},\ \bibinfo {note} {[Erratum: Phys.Rept.
  281, 399--400 (1997)]},\ \Eprint {http://arxiv.org/abs/hep-ph/9501369}
  {arXiv:hep-ph/9501369} \BibitemShut {NoStop}%
\bibitem [{\citenamefont {Leader}(2011)}]{Leader:2001gr}%
  \BibitemOpen
  \bibfield  {author} {\bibinfo {author} {\bibfnamefont {E.}~\bibnamefont
  {Leader}},\ }\href@noop {} {\bibfield  {journal} {\bibinfo  {journal} {Camb.
  Monogr. Part. Phys. Nucl. Phys. Cosmol.}\ }\textbf {\bibinfo {volume} {15}},\
  \bibinfo {pages} {pp.1} (\bibinfo {year} {2011})}\BibitemShut {NoStop}%
%%CITATION = CMPCE,15,pp.1;%%
\bibitem [{\citenamefont {Ellis}\ and\ \citenamefont
  {Jaffe}(1974)}]{Ellis:1973kp}%
  \BibitemOpen
  \bibfield  {author} {\bibinfo {author} {\bibfnamefont {J.~R.}\ \bibnamefont
  {Ellis}}\ and\ \bibinfo {author} {\bibfnamefont {R.~L.}\ \bibnamefont
  {Jaffe}},\ }\href {\doibase 10.1103/PhysRevD.9.1444} {\bibfield  {journal}
  {\bibinfo  {journal} {Phys. Rev. D}\ }\textbf {\bibinfo {volume} {9}},\
  \bibinfo {pages} {1444} (\bibinfo {year} {1974})},\ \bibinfo {note}
  {[Erratum: Phys.Rev.D 10, 1669 (1974)]}\BibitemShut {NoStop}%
\bibitem [{\citenamefont {Ashman}\ \emph {et~al.}(1989)\citenamefont {Ashman}
  \emph {et~al.}}]{Ashman:1989ig}%
  \BibitemOpen
  \bibfield  {author} {\bibinfo {author} {\bibfnamefont {J.}~\bibnamefont
  {Ashman}} \emph {et~al.} (\bibinfo {collaboration} {European Muon}),\ }\href
  {\doibase 10.1016/0550-3213(89)90089-8} {\bibfield  {journal} {\bibinfo
  {journal} {Nucl. Phys. B}\ }\textbf {\bibinfo {volume} {328}},\ \bibinfo
  {pages} {1} (\bibinfo {year} {1989})}\BibitemShut {NoStop}%
\bibitem [{\citenamefont {Ashman}\ \emph {et~al.}(1988)\citenamefont {Ashman}
  \emph {et~al.}}]{Ashman:1987hv}%
  \BibitemOpen
  \bibfield  {author} {\bibinfo {author} {\bibfnamefont {J.}~\bibnamefont
  {Ashman}} \emph {et~al.} (\bibinfo {collaboration} {European Muon}),\ }\href
  {\doibase 10.1016/0370-2693(88)91523-7} {\bibfield  {journal} {\bibinfo
  {journal} {Phys. Lett. B}\ }\textbf {\bibinfo {volume} {206}},\ \bibinfo
  {pages} {364} (\bibinfo {year} {1988})}\BibitemShut {NoStop}%
\bibitem [{\citenamefont {Alekseev}\ \emph {et~al.}(2010)\citenamefont
  {Alekseev} \emph {et~al.}}]{Alekseev:2010ub}%
  \BibitemOpen
  \bibfield  {author} {\bibinfo {author} {\bibfnamefont {M.}~\bibnamefont
  {Alekseev}} \emph {et~al.} (\bibinfo {collaboration} {COMPASS}),\ }\href
  {\doibase 10.1016/j.physletb.2010.08.034} {\bibfield  {journal} {\bibinfo
  {journal} {Phys. Lett. B}\ }\textbf {\bibinfo {volume} {693}},\ \bibinfo
  {pages} {227} (\bibinfo {year} {2010})},\ \Eprint
  {http://arxiv.org/abs/1007.4061} {arXiv:1007.4061 [hep-ex]} \BibitemShut
  {NoStop}%
\bibitem [{\citenamefont {Airapetian}\ \emph {et~al.}(2007)\citenamefont
  {Airapetian} \emph {et~al.}}]{Airapetian:2007mh}%
  \BibitemOpen
  \bibfield  {author} {\bibinfo {author} {\bibfnamefont {A.}~\bibnamefont
  {Airapetian}} \emph {et~al.} (\bibinfo {collaboration} {HERMES}),\ }\href
  {\doibase 10.1103/PhysRevD.75.012007} {\bibfield  {journal} {\bibinfo
  {journal} {Phys. Rev. D}\ }\textbf {\bibinfo {volume} {75}},\ \bibinfo
  {pages} {012007} (\bibinfo {year} {2007})},\ \Eprint
  {http://arxiv.org/abs/hep-ex/0609039} {arXiv:hep-ex/0609039} \BibitemShut
  {NoStop}%
\bibitem [{\citenamefont {Aidala}\ \emph {et~al.}(2013)\citenamefont {Aidala},
  \citenamefont {Bass}, \citenamefont {Hasch},\ and\ \citenamefont
  {Mallot}}]{Aidala:2012mv}%
  \BibitemOpen
  \bibfield  {author} {\bibinfo {author} {\bibfnamefont {C.~A.}\ \bibnamefont
  {Aidala}}, \bibinfo {author} {\bibfnamefont {S.~D.}\ \bibnamefont {Bass}},
  \bibinfo {author} {\bibfnamefont {D.}~\bibnamefont {Hasch}}, \ and\ \bibinfo
  {author} {\bibfnamefont {G.~K.}\ \bibnamefont {Mallot}},\ }\href {\doibase
  10.1103/RevModPhys.85.655} {\bibfield  {journal} {\bibinfo  {journal} {Rev.
  Mod. Phys.}\ }\textbf {\bibinfo {volume} {85}},\ \bibinfo {pages} {655}
  (\bibinfo {year} {2013})},\ \Eprint {http://arxiv.org/abs/1209.2803}
  {arXiv:1209.2803 [hep-ph]} \BibitemShut {NoStop}%
\bibitem [{\citenamefont {Deur}\ \emph {et~al.}(2019)\citenamefont {Deur},
  \citenamefont {Brodsky},\ and\ \citenamefont {De~Téramond}}]{Deur:2018roz}%
  \BibitemOpen
  \bibfield  {author} {\bibinfo {author} {\bibfnamefont {A.}~\bibnamefont
  {Deur}}, \bibinfo {author} {\bibfnamefont {S.~J.}\ \bibnamefont {Brodsky}}, \
  and\ \bibinfo {author} {\bibfnamefont {G.~F.}\ \bibnamefont {De~Téramond}},\
  }\href {\doibase 10.1088/1361-6633/ab0b8f} {\bibfield  {journal} {\bibinfo
  {journal} {Rept. Prog. Phys.}\ }\textbf {\bibinfo {volume} {82}} (\bibinfo
  {year} {2019}),\ 10.1088/1361-6633/ab0b8f},\ \Eprint
  {http://arxiv.org/abs/1807.05250} {arXiv:1807.05250 [hep-ph]} \BibitemShut
  {NoStop}%
\bibitem [{\citenamefont {Kuhn}\ \emph {et~al.}(2009)\citenamefont {Kuhn},
  \citenamefont {Chen},\ and\ \citenamefont {Leader}}]{Kuhn:2008sy}%
  \BibitemOpen
  \bibfield  {author} {\bibinfo {author} {\bibfnamefont {S.}~\bibnamefont
  {Kuhn}}, \bibinfo {author} {\bibfnamefont {J.-P.}\ \bibnamefont {Chen}}, \
  and\ \bibinfo {author} {\bibfnamefont {E.}~\bibnamefont {Leader}},\ }\href
  {\doibase 10.1016/j.ppnp.2009.02.001} {\bibfield  {journal} {\bibinfo
  {journal} {Prog. Part. Nucl. Phys.}\ }\textbf {\bibinfo {volume} {63}},\
  \bibinfo {pages} {1} (\bibinfo {year} {2009})},\ \Eprint
  {http://arxiv.org/abs/0812.3535} {arXiv:0812.3535 [hep-ph]} \BibitemShut
  {NoStop}%
\bibitem [{\citenamefont {Lampe}\ and\ \citenamefont
  {Reya}(2000)}]{Lampe:1998eu}%
  \BibitemOpen
  \bibfield  {author} {\bibinfo {author} {\bibfnamefont {B.}~\bibnamefont
  {Lampe}}\ and\ \bibinfo {author} {\bibfnamefont {E.}~\bibnamefont {Reya}},\
  }\href {\doibase 10.1016/S0370-1573(99)00100-3} {\bibfield  {journal}
  {\bibinfo  {journal} {Phys. Rept.}\ }\textbf {\bibinfo {volume} {332}},\
  \bibinfo {pages} {1} (\bibinfo {year} {2000})},\ \Eprint
  {http://arxiv.org/abs/hep-ph/9810270} {arXiv:hep-ph/9810270} \BibitemShut
  {NoStop}%
\bibitem [{\citenamefont {Efremov}\ \emph {et~al.}(1990)\citenamefont
  {Efremov}, \citenamefont {Soffer},\ and\ \citenamefont
  {Teryaev}}]{Efremov:1989sn}%
  \BibitemOpen
  \bibfield  {author} {\bibinfo {author} {\bibfnamefont {A.}~\bibnamefont
  {Efremov}}, \bibinfo {author} {\bibfnamefont {J.}~\bibnamefont {Soffer}}, \
  and\ \bibinfo {author} {\bibfnamefont {O.}~\bibnamefont {Teryaev}},\ }\href
  {\doibase 10.1016/0550-3213(90)90239-A} {\bibfield  {journal} {\bibinfo
  {journal} {Nucl. Phys. B}\ }\textbf {\bibinfo {volume} {346}},\ \bibinfo
  {pages} {97} (\bibinfo {year} {1990})}\BibitemShut {NoStop}%
\bibitem [{\citenamefont {Witten}(1979)}]{Witten:1979vv}%
  \BibitemOpen
  \bibfield  {author} {\bibinfo {author} {\bibfnamefont {E.}~\bibnamefont
  {Witten}},\ }\href {\doibase 10.1016/0550-3213(79)90031-2} {\bibfield
  {journal} {\bibinfo  {journal} {Nucl. Phys. B}\ }\textbf {\bibinfo {volume}
  {156}},\ \bibinfo {pages} {269} (\bibinfo {year} {1979})}\BibitemShut
  {NoStop}%
\bibitem [{\citenamefont {Veneziano}(1979)}]{Veneziano:1979ec}%
  \BibitemOpen
  \bibfield  {author} {\bibinfo {author} {\bibfnamefont {G.}~\bibnamefont
  {Veneziano}},\ }\href {\doibase 10.1016/0550-3213(79)90332-8} {\bibfield
  {journal} {\bibinfo  {journal} {Nucl. Phys. B}\ }\textbf {\bibinfo {volume}
  {159}},\ \bibinfo {pages} {213} (\bibinfo {year} {1979})}\BibitemShut
  {NoStop}%
\bibitem [{\citenamefont {Veneziano}(1989)}]{Veneziano:1989ei}%
  \BibitemOpen
  \bibfield  {author} {\bibinfo {author} {\bibfnamefont {G.}~\bibnamefont
  {Veneziano}},\ }\href {\doibase 10.1142/S0217732389001830} {\bibfield
  {journal} {\bibinfo  {journal} {Mod. Phys. Lett. A}\ }\textbf {\bibinfo
  {volume} {4}},\ \bibinfo {pages} {1605} (\bibinfo {year} {1989})}\BibitemShut
  {NoStop}%
\bibitem [{\citenamefont {Hatsuda}(1990)}]{Hatsuda:1989bi}%
  \BibitemOpen
  \bibfield  {author} {\bibinfo {author} {\bibfnamefont {T.}~\bibnamefont
  {Hatsuda}},\ }\href {\doibase 10.1016/0550-3213(90)90148-7} {\bibfield
  {journal} {\bibinfo  {journal} {Nucl. Phys. B}\ }\textbf {\bibinfo {volume}
  {329}},\ \bibinfo {pages} {376} (\bibinfo {year} {1990})}\BibitemShut
  {NoStop}%
\bibitem [{\citenamefont {Diakonov}\ and\ \citenamefont
  {Eides}(1981)}]{Diakonov:1981nv}%
  \BibitemOpen
  \bibfield  {author} {\bibinfo {author} {\bibfnamefont {D.}~\bibnamefont
  {Diakonov}}\ and\ \bibinfo {author} {\bibfnamefont {M.~I.}\ \bibnamefont
  {Eides}},\ }\href@noop {} {\bibfield  {journal} {\bibinfo  {journal} {Sov.
  Phys. JETP}\ }\textbf {\bibinfo {volume} {54}},\ \bibinfo {pages} {232}
  (\bibinfo {year} {1981})}\BibitemShut {NoStop}%
\bibitem [{\citenamefont {Ji}(1990)}]{Ji:1990fj}%
  \BibitemOpen
  \bibfield  {author} {\bibinfo {author} {\bibfnamefont {X.-D.}\ \bibnamefont
  {Ji}},\ }\href {\doibase 10.1103/PhysRevLett.65.408} {\bibfield  {journal}
  {\bibinfo  {journal} {Phys. Rev. Lett.}\ }\textbf {\bibinfo {volume} {65}},\
  \bibinfo {pages} {408} (\bibinfo {year} {1990})}\BibitemShut {NoStop}%
\bibitem [{\citenamefont {Efremov}\ \emph {et~al.}(1991)\citenamefont
  {Efremov}, \citenamefont {Soffer},\ and\ \citenamefont
  {Tornqvist}}]{Efremov:1990nj}%
  \BibitemOpen
  \bibfield  {author} {\bibinfo {author} {\bibfnamefont {A.~V.}\ \bibnamefont
  {Efremov}}, \bibinfo {author} {\bibfnamefont {J.}~\bibnamefont {Soffer}}, \
  and\ \bibinfo {author} {\bibfnamefont {N.~A.}\ \bibnamefont {Tornqvist}},\
  }\href {\doibase 10.1103/PhysRevD.44.1369} {\bibfield  {journal} {\bibinfo
  {journal} {Phys. Rev. D}\ }\textbf {\bibinfo {volume} {44}},\ \bibinfo
  {pages} {1369} (\bibinfo {year} {1991})}\BibitemShut {NoStop}%
\bibitem [{\citenamefont {Dvali}(2005)}]{Dvali:2005an}%
  \BibitemOpen
  \bibfield  {author} {\bibinfo {author} {\bibfnamefont {G.}~\bibnamefont
  {Dvali}},\ }\href@noop {} {\  (\bibinfo {year} {2005})},\ \Eprint
  {http://arxiv.org/abs/hep-th/0507215} {arXiv:hep-th/0507215} \BibitemShut
  {NoStop}%
\bibitem [{\citenamefont {Dvali}\ \emph {et~al.}(2006)\citenamefont {Dvali},
  \citenamefont {Jackiw},\ and\ \citenamefont {Pi}}]{Dvali:2005ws}%
  \BibitemOpen
  \bibfield  {author} {\bibinfo {author} {\bibfnamefont {G.}~\bibnamefont
  {Dvali}}, \bibinfo {author} {\bibfnamefont {R.}~\bibnamefont {Jackiw}}, \
  and\ \bibinfo {author} {\bibfnamefont {S.-Y.}\ \bibnamefont {Pi}},\ }\href
  {\doibase 10.1103/PhysRevLett.96.081602} {\bibfield  {journal} {\bibinfo
  {journal} {Phys. Rev. Lett.}\ }\textbf {\bibinfo {volume} {96}},\ \bibinfo
  {pages} {081602} (\bibinfo {year} {2006})},\ \Eprint
  {http://arxiv.org/abs/hep-th/0511175} {arXiv:hep-th/0511175} \BibitemShut
  {NoStop}%
\bibitem [{\citenamefont {Kharzeev}\ and\ \citenamefont
  {Levin}(2015)}]{Kharzeev:2015xsa}%
  \BibitemOpen
  \bibfield  {author} {\bibinfo {author} {\bibfnamefont {D.~E.}\ \bibnamefont
  {Kharzeev}}\ and\ \bibinfo {author} {\bibfnamefont {E.~M.}\ \bibnamefont
  {Levin}},\ }\href {\doibase 10.1103/PhysRevLett.114.242001} {\bibfield
  {journal} {\bibinfo  {journal} {Phys. Rev. Lett.}\ }\textbf {\bibinfo
  {volume} {114}},\ \bibinfo {pages} {242001} (\bibinfo {year} {2015})},\
  \Eprint {http://arxiv.org/abs/1501.04622} {arXiv:1501.04622 [hep-ph]}
  \BibitemShut {NoStop}%
\bibitem [{\citenamefont {Witten}(1983)}]{Witten:1983tw}%
  \BibitemOpen
  \bibfield  {author} {\bibinfo {author} {\bibfnamefont {E.}~\bibnamefont
  {Witten}},\ }\href {\doibase 10.1016/0550-3213(83)90063-9} {\bibfield
  {journal} {\bibinfo  {journal} {Nucl. Phys. B}\ }\textbf {\bibinfo {volume}
  {223}},\ \bibinfo {pages} {422} (\bibinfo {year} {1983})}\BibitemShut
  {NoStop}%
\bibitem [{\citenamefont {Leutwyler}(1996)}]{Leutwyler:1996sa}%
  \BibitemOpen
  \bibfield  {author} {\bibinfo {author} {\bibfnamefont {H.}~\bibnamefont
  {Leutwyler}},\ }\href {\doibase 10.1016/0370-2693(96)85876-X} {\bibfield
  {journal} {\bibinfo  {journal} {Phys. Lett. B}\ }\textbf {\bibinfo {volume}
  {374}},\ \bibinfo {pages} {163} (\bibinfo {year} {1996})},\ \Eprint
  {http://arxiv.org/abs/hep-ph/9601234} {arXiv:hep-ph/9601234} \BibitemShut
  {NoStop}%
\bibitem [{\citenamefont {Manohar}(1991)}]{Manohar:1990jx}%
  \BibitemOpen
  \bibfield  {author} {\bibinfo {author} {\bibfnamefont {A.~V.}\ \bibnamefont
  {Manohar}},\ }\href {\doibase 10.1103/PhysRevLett.66.289} {\bibfield
  {journal} {\bibinfo  {journal} {Phys. Rev. Lett.}\ }\textbf {\bibinfo
  {volume} {66}},\ \bibinfo {pages} {289} (\bibinfo {year} {1991})}\BibitemShut
  {NoStop}%
\bibitem [{\citenamefont {Bodwin}\ and\ \citenamefont
  {Qiu}(1991)}]{Bodwin:1990fk}%
  \BibitemOpen
  \bibfield  {author} {\bibinfo {author} {\bibfnamefont {G.~T.}\ \bibnamefont
  {Bodwin}}\ and\ \bibinfo {author} {\bibfnamefont {J.-w.}\ \bibnamefont
  {Qiu}},\ }\href {\doibase 10.1063/1.40493} {\bibfield  {journal} {\bibinfo
  {journal} {AIP Conf. Proc.}\ }\textbf {\bibinfo {volume} {223}},\ \bibinfo
  {pages} {285} (\bibinfo {year} {1991})}\BibitemShut {NoStop}%
\bibitem [{\citenamefont {Vogelsang}(1991)}]{Vogelsang:1990ug}%
  \BibitemOpen
  \bibfield  {author} {\bibinfo {author} {\bibfnamefont {W.}~\bibnamefont
  {Vogelsang}},\ }\href {\doibase 10.1007/BF01474080} {\bibfield  {journal}
  {\bibinfo  {journal} {Z. Phys. C}\ }\textbf {\bibinfo {volume} {50}},\
  \bibinfo {pages} {275} (\bibinfo {year} {1991})}\BibitemShut {NoStop}%
\bibitem [{\citenamefont {'t~Hooft}(1976)}]{tHooft:1976snw}%
  \BibitemOpen
  \bibfield  {author} {\bibinfo {author} {\bibfnamefont {G.}~\bibnamefont
  {'t~Hooft}},\ }\href {\doibase 10.1103/PhysRevD.14.3432} {\bibfield
  {journal} {\bibinfo  {journal} {Phys. Rev. D}\ }\textbf {\bibinfo {volume}
  {14}},\ \bibinfo {pages} {3432} (\bibinfo {year} {1976})},\ \bibinfo {note}
  {[Erratum: Phys.Rev.D 18, 2199 (1978)]}\BibitemShut {NoStop}%
\bibitem [{\citenamefont {'t~Hooft}(1986)}]{tHooft:1986ooh}%
  \BibitemOpen
  \bibfield  {author} {\bibinfo {author} {\bibfnamefont {G.}~\bibnamefont
  {'t~Hooft}},\ }\href {\doibase 10.1016/0370-1573(86)90117-1} {\bibfield
  {journal} {\bibinfo  {journal} {Phys. Rept.}\ }\textbf {\bibinfo {volume}
  {142}},\ \bibinfo {pages} {357} (\bibinfo {year} {1986})}\BibitemShut
  {NoStop}%
\bibitem [{\citenamefont {Forte}(1990)}]{Forte:1989qq}%
  \BibitemOpen
  \bibfield  {author} {\bibinfo {author} {\bibfnamefont {S.}~\bibnamefont
  {Forte}},\ }\href {\doibase 10.1016/0550-3213(90)90015-6} {\bibfield
  {journal} {\bibinfo  {journal} {Nucl. Phys. B}\ }\textbf {\bibinfo {volume}
  {331}},\ \bibinfo {pages} {1} (\bibinfo {year} {1990})}\BibitemShut {NoStop}%
\bibitem [{\citenamefont {Forte}\ and\ \citenamefont
  {Shuryak}(1991)}]{Forte:1990xb}%
  \BibitemOpen
  \bibfield  {author} {\bibinfo {author} {\bibfnamefont {S.}~\bibnamefont
  {Forte}}\ and\ \bibinfo {author} {\bibfnamefont {E.~V.}\ \bibnamefont
  {Shuryak}},\ }\href {\doibase 10.1016/0550-3213(91)90462-7} {\bibfield
  {journal} {\bibinfo  {journal} {Nucl. Phys. B}\ }\textbf {\bibinfo {volume}
  {357}},\ \bibinfo {pages} {153} (\bibinfo {year} {1991})}\BibitemShut
  {NoStop}%
\bibitem [{\citenamefont {Dorokhov}\ \emph {et~al.}(1993)\citenamefont
  {Dorokhov}, \citenamefont {Kochelev},\ and\ \citenamefont
  {Zubov}}]{Dorokhov:1993ym}%
  \BibitemOpen
  \bibfield  {author} {\bibinfo {author} {\bibfnamefont {A.}~\bibnamefont
  {Dorokhov}}, \bibinfo {author} {\bibfnamefont {N.}~\bibnamefont {Kochelev}},
  \ and\ \bibinfo {author} {\bibfnamefont {Y.}~\bibnamefont {Zubov}},\ }\href
  {\doibase 10.1142/S0217751X93000242} {\bibfield  {journal} {\bibinfo
  {journal} {Int. J. Mod. Phys. A}\ }\textbf {\bibinfo {volume} {8}},\ \bibinfo
  {pages} {603} (\bibinfo {year} {1993})}\BibitemShut {NoStop}%
\bibitem [{\citenamefont {Qian}\ and\ \citenamefont
  {Zahed}(2016)}]{Qian:2015wyq}%
  \BibitemOpen
  \bibfield  {author} {\bibinfo {author} {\bibfnamefont {Y.}~\bibnamefont
  {Qian}}\ and\ \bibinfo {author} {\bibfnamefont {I.}~\bibnamefont {Zahed}},\
  }\href {\doibase 10.1016/j.aop.2016.09.002} {\bibfield  {journal} {\bibinfo
  {journal} {Annals Phys.}\ }\textbf {\bibinfo {volume} {374}},\ \bibinfo
  {pages} {314} (\bibinfo {year} {2016})},\ \Eprint
  {http://arxiv.org/abs/1512.08172} {arXiv:1512.08172 [hep-ph]} \BibitemShut
  {NoStop}%
\bibitem [{\citenamefont {Preskill}(1991)}]{Preskill:1990fr}%
  \BibitemOpen
  \bibfield  {author} {\bibinfo {author} {\bibfnamefont {J.}~\bibnamefont
  {Preskill}},\ }\href {\doibase 10.1016/0003-4916(91)90046-B} {\bibfield
  {journal} {\bibinfo  {journal} {Annals Phys.}\ }\textbf {\bibinfo {volume}
  {210}},\ \bibinfo {pages} {323} (\bibinfo {year} {1991})}\BibitemShut
  {NoStop}%
\bibitem [{\citenamefont {Leutwyler}(1998)}]{Leutwyler:1997yr}%
  \BibitemOpen
  \bibfield  {author} {\bibinfo {author} {\bibfnamefont {H.}~\bibnamefont
  {Leutwyler}},\ }\href {\doibase 10.1016/S0920-5632(97)01065-7} {\bibfield
  {journal} {\bibinfo  {journal} {Nucl. Phys. B Proc. Suppl.}\ }\textbf
  {\bibinfo {volume} {64}},\ \bibinfo {pages} {223} (\bibinfo {year} {1998})},\
  \Eprint {http://arxiv.org/abs/hep-ph/9709408} {arXiv:hep-ph/9709408}
  \BibitemShut {NoStop}%
\bibitem [{\citenamefont {Herrera-Siklody}\ \emph {et~al.}(1997)\citenamefont
  {Herrera-Siklody}, \citenamefont {Latorre}, \citenamefont {Pascual},\ and\
  \citenamefont {Taron}}]{HerreraSiklody:1996pm}%
  \BibitemOpen
  \bibfield  {author} {\bibinfo {author} {\bibfnamefont {P.}~\bibnamefont
  {Herrera-Siklody}}, \bibinfo {author} {\bibfnamefont {J.}~\bibnamefont
  {Latorre}}, \bibinfo {author} {\bibfnamefont {P.}~\bibnamefont {Pascual}}, \
  and\ \bibinfo {author} {\bibfnamefont {J.}~\bibnamefont {Taron}},\ }\href
  {\doibase 10.1016/S0550-3213(97)00260-5} {\bibfield  {journal} {\bibinfo
  {journal} {Nucl. Phys. B}\ }\textbf {\bibinfo {volume} {497}},\ \bibinfo
  {pages} {345} (\bibinfo {year} {1997})},\ \Eprint
  {http://arxiv.org/abs/hep-ph/9610549} {arXiv:hep-ph/9610549} \BibitemShut
  {NoStop}%
\bibitem [{\citenamefont {Liang}\ \emph {et~al.}(2018)\citenamefont {Liang},
  \citenamefont {Yang}, \citenamefont {Draper}, \citenamefont {Gong},\ and\
  \citenamefont {Liu}}]{Liang:2018pis}%
  \BibitemOpen
  \bibfield  {author} {\bibinfo {author} {\bibfnamefont {J.}~\bibnamefont
  {Liang}}, \bibinfo {author} {\bibfnamefont {Y.-B.}\ \bibnamefont {Yang}},
  \bibinfo {author} {\bibfnamefont {T.}~\bibnamefont {Draper}}, \bibinfo
  {author} {\bibfnamefont {M.}~\bibnamefont {Gong}}, \ and\ \bibinfo {author}
  {\bibfnamefont {K.-F.}\ \bibnamefont {Liu}},\ }\href {\doibase
  10.1103/PhysRevD.98.074505} {\bibfield  {journal} {\bibinfo  {journal} {Phys.
  Rev. D}\ }\textbf {\bibinfo {volume} {98}},\ \bibinfo {pages} {074505}
  (\bibinfo {year} {2018})},\ \Eprint {http://arxiv.org/abs/1806.08366}
  {arXiv:1806.08366 [hep-ph]} \BibitemShut {NoStop}%
\bibitem [{\citenamefont {Tarasov}\ and\ \citenamefont
  {Venugopalan}(2019)}]{Tarasov:2019rfp}%
  \BibitemOpen
  \bibfield  {author} {\bibinfo {author} {\bibfnamefont {A.}~\bibnamefont
  {Tarasov}}\ and\ \bibinfo {author} {\bibfnamefont {R.}~\bibnamefont
  {Venugopalan}},\ }\href {\doibase 10.1103/PhysRevD.100.054007} {\bibfield
  {journal} {\bibinfo  {journal} {Phys. Rev. D}\ }\textbf {\bibinfo {volume}
  {100}},\ \bibinfo {pages} {054007} (\bibinfo {year} {2019})},\ \Eprint
  {http://arxiv.org/abs/1903.11624} {arXiv:1903.11624 [hep-ph]} \BibitemShut
  {NoStop}%
\bibitem [{\citenamefont {Fliegner}\ \emph {et~al.}(1997)\citenamefont
  {Fliegner}, \citenamefont {Reuter}, \citenamefont {Schmidt},\ and\
  \citenamefont {Schubert}}]{Fliegner:1997ra}%
  \BibitemOpen
  \bibfield  {author} {\bibinfo {author} {\bibfnamefont {D.}~\bibnamefont
  {Fliegner}}, \bibinfo {author} {\bibfnamefont {M.}~\bibnamefont {Reuter}},
  \bibinfo {author} {\bibfnamefont {M.}~\bibnamefont {Schmidt}}, \ and\
  \bibinfo {author} {\bibfnamefont {C.}~\bibnamefont {Schubert}},\ }\href
  {\doibase 10.1007/BF02634170} {\bibfield  {journal} {\bibinfo  {journal}
  {Theor. Math. Phys.}\ }\textbf {\bibinfo {volume} {113}},\ \bibinfo {pages}
  {1442} (\bibinfo {year} {1997})},\ \Eprint
  {http://arxiv.org/abs/hep-th/9704194} {arXiv:hep-th/9704194} \BibitemShut
  {NoStop}%
\bibitem [{\citenamefont {Pawlowski}\ \emph {et~al.}(2009)\citenamefont
  {Pawlowski}, \citenamefont {Schmidt},\ and\ \citenamefont
  {Zhang}}]{Pawlowski:2008xh}%
  \BibitemOpen
  \bibfield  {author} {\bibinfo {author} {\bibfnamefont {J.~M.}\ \bibnamefont
  {Pawlowski}}, \bibinfo {author} {\bibfnamefont {M.~G.}\ \bibnamefont
  {Schmidt}}, \ and\ \bibinfo {author} {\bibfnamefont {J.-H.}\ \bibnamefont
  {Zhang}},\ }\href {\doibase 10.1016/j.physletb.2009.05.005} {\bibfield
  {journal} {\bibinfo  {journal} {Phys. Lett. B}\ }\textbf {\bibinfo {volume}
  {677}},\ \bibinfo {pages} {100} (\bibinfo {year} {2009})},\ \Eprint
  {http://arxiv.org/abs/0812.0008} {arXiv:0812.0008 [hep-th]} \BibitemShut
  {NoStop}%
\bibitem [{\citenamefont {Magnea}\ \emph {et~al.}(2013)\citenamefont {Magnea},
  \citenamefont {Playle}, \citenamefont {Russo},\ and\ \citenamefont
  {Sciuto}}]{Magnea:2013lna}%
  \BibitemOpen
  \bibfield  {author} {\bibinfo {author} {\bibfnamefont {L.}~\bibnamefont
  {Magnea}}, \bibinfo {author} {\bibfnamefont {S.}~\bibnamefont {Playle}},
  \bibinfo {author} {\bibfnamefont {R.}~\bibnamefont {Russo}}, \ and\ \bibinfo
  {author} {\bibfnamefont {S.}~\bibnamefont {Sciuto}},\ }\href {\doibase
  10.1007/JHEP09(2013)081} {\bibfield  {journal} {\bibinfo  {journal} {JHEP}\
  }\textbf {\bibinfo {volume} {09}},\ \bibinfo {pages} {081} (\bibinfo {year}
  {2013})},\ \Eprint {http://arxiv.org/abs/1305.6631} {arXiv:1305.6631
  [hep-th]} \BibitemShut {NoStop}%
\bibitem [{\citenamefont {Mueller}\ and\ \citenamefont
  {Venugopalan}(2017)}]{Mueller:2017arw}%
  \BibitemOpen
  \bibfield  {author} {\bibinfo {author} {\bibfnamefont {N.}~\bibnamefont
  {Mueller}}\ and\ \bibinfo {author} {\bibfnamefont {R.}~\bibnamefont
  {Venugopalan}},\ }\href {\doibase 10.1103/PhysRevD.96.016023} {\bibfield
  {journal} {\bibinfo  {journal} {Phys. Rev.}\ }\textbf {\bibinfo {volume}
  {D96}},\ \bibinfo {pages} {016023} (\bibinfo {year} {2017})},\ \Eprint
  {http://arxiv.org/abs/1702.01233} {arXiv:1702.01233 [hep-ph]} \BibitemShut
  {NoStop}%
%%CITATION = ARXIV:1702.01233;%%
\bibitem [{\citenamefont {Alvarez-Gaume}\ and\ \citenamefont
  {Vazquez-Mozo}(2006)}]{AlvarezGaume:2005qb}%
  \BibitemOpen
  \bibfield  {author} {\bibinfo {author} {\bibfnamefont {L.}~\bibnamefont
  {Alvarez-Gaume}}\ and\ \bibinfo {author} {\bibfnamefont {M.~A.}\ \bibnamefont
  {Vazquez-Mozo}},\ }in\ \href {\doibase 10.5170/CERN-2006-015.1} {\emph
  {\bibinfo {booktitle} {{3rd CERN-CLAF School of High Energy Physics}}}}\
  (\bibinfo {year} {2006})\ pp.\ \bibinfo {pages} {1--80},\ \Eprint
  {http://arxiv.org/abs/hep-th/0510040} {arXiv:hep-th/0510040} \BibitemShut
  {NoStop}%
\bibitem [{\citenamefont {Bilal}(2008)}]{Bilal:2008qx}%
  \BibitemOpen
  \bibfield  {author} {\bibinfo {author} {\bibfnamefont {A.}~\bibnamefont
  {Bilal}},\ }\href@noop {} {\  (\bibinfo {year} {2008})},\ \Eprint
  {http://arxiv.org/abs/0802.0634} {arXiv:0802.0634 [hep-th]} \BibitemShut
  {NoStop}%
\bibitem [{\citenamefont {Bern}\ and\ \citenamefont
  {Kosower}(1991)}]{Bern:1990ux}%
  \BibitemOpen
  \bibfield  {author} {\bibinfo {author} {\bibfnamefont {Z.}~\bibnamefont
  {Bern}}\ and\ \bibinfo {author} {\bibfnamefont {D.~A.}\ \bibnamefont
  {Kosower}},\ }\href {\doibase 10.1016/0550-3213(91)90567-H} {\bibfield
  {journal} {\bibinfo  {journal} {Nucl. Phys. B}\ }\textbf {\bibinfo {volume}
  {362}},\ \bibinfo {pages} {389} (\bibinfo {year} {1991})}\BibitemShut
  {NoStop}%
\bibitem [{\citenamefont {Bern}\ and\ \citenamefont
  {Kosower}(1992)}]{Bern:1991aq}%
  \BibitemOpen
  \bibfield  {author} {\bibinfo {author} {\bibfnamefont {Z.}~\bibnamefont
  {Bern}}\ and\ \bibinfo {author} {\bibfnamefont {D.~A.}\ \bibnamefont
  {Kosower}},\ }\href {\doibase 10.1016/0550-3213(92)90134-W} {\bibfield
  {journal} {\bibinfo  {journal} {Nucl. Phys.}\ }\textbf {\bibinfo {volume}
  {B379}},\ \bibinfo {pages} {451} (\bibinfo {year} {1992})}\BibitemShut
  {NoStop}%
%%CITATION = NUPHA,B379,451;%%
\bibitem [{\citenamefont {Bern}(1992)}]{Bern:1992ad}%
  \BibitemOpen
  \bibfield  {author} {\bibinfo {author} {\bibfnamefont {Z.}~\bibnamefont
  {Bern}},\ }in\ \href@noop {} {\emph {\bibinfo {booktitle} {{Theoretical
  Advanced Study Institute (TASI 92): From Black Holes and Strings to
  Particles}}}}\ (\bibinfo {year} {1992})\ pp.\ \bibinfo {pages} {0471--536},\
  \Eprint {http://arxiv.org/abs/hep-ph/9304249} {arXiv:hep-ph/9304249}
  \BibitemShut {NoStop}%
\bibitem [{\citenamefont {Rosenberg}(1963)}]{Rosenberg:1962pp}%
  \BibitemOpen
  \bibfield  {author} {\bibinfo {author} {\bibfnamefont {L.}~\bibnamefont
  {Rosenberg}},\ }\href {\doibase 10.1103/PhysRev.129.2786} {\bibfield
  {journal} {\bibinfo  {journal} {Phys. Rev.}\ }\textbf {\bibinfo {volume}
  {129}},\ \bibinfo {pages} {2786} (\bibinfo {year} {1963})}\BibitemShut
  {NoStop}%
\bibitem [{\citenamefont {Armillis}\ \emph {et~al.}(2009)\citenamefont
  {Armillis}, \citenamefont {Coriano}, \citenamefont {Delle~Rose},\ and\
  \citenamefont {Guzzi}}]{Armillis:2009sm}%
  \BibitemOpen
  \bibfield  {author} {\bibinfo {author} {\bibfnamefont {R.}~\bibnamefont
  {Armillis}}, \bibinfo {author} {\bibfnamefont {C.}~\bibnamefont {Coriano}},
  \bibinfo {author} {\bibfnamefont {L.}~\bibnamefont {Delle~Rose}}, \ and\
  \bibinfo {author} {\bibfnamefont {M.}~\bibnamefont {Guzzi}},\ }\href
  {\doibase 10.1088/1126-6708/2009/12/029} {\bibfield  {journal} {\bibinfo
  {journal} {JHEP}\ }\textbf {\bibinfo {volume} {12}},\ \bibinfo {pages} {029}
  (\bibinfo {year} {2009})},\ \Eprint {http://arxiv.org/abs/0905.0865}
  {arXiv:0905.0865 [hep-ph]} \BibitemShut {NoStop}%
\bibitem [{\citenamefont {Usyukina}\ and\ \citenamefont
  {Davydychev}(1993)}]{Usyukina:1992jd}%
  \BibitemOpen
  \bibfield  {author} {\bibinfo {author} {\bibfnamefont {N.}~\bibnamefont
  {Usyukina}}\ and\ \bibinfo {author} {\bibfnamefont {A.~I.}\ \bibnamefont
  {Davydychev}},\ }\href {\doibase 10.1016/0370-2693(93)91834-A} {\bibfield
  {journal} {\bibinfo  {journal} {Phys. Lett. B}\ }\textbf {\bibinfo {volume}
  {298}},\ \bibinfo {pages} {363} (\bibinfo {year} {1993})}\BibitemShut
  {NoStop}%
\bibitem [{\citenamefont {'t~Hooft}\ and\ \citenamefont
  {Veltman}(1979)}]{tHooft:1978jhc}%
  \BibitemOpen
  \bibfield  {author} {\bibinfo {author} {\bibfnamefont {G.}~\bibnamefont
  {'t~Hooft}}\ and\ \bibinfo {author} {\bibfnamefont {M.}~\bibnamefont
  {Veltman}},\ }\href {\doibase 10.1016/0550-3213(79)90605-9} {\bibfield
  {journal} {\bibinfo  {journal} {Nucl. Phys. B}\ }\textbf {\bibinfo {volume}
  {153}},\ \bibinfo {pages} {365} (\bibinfo {year} {1979})}\BibitemShut
  {NoStop}%
\bibitem [{\citenamefont {Fujikawa}(1979)}]{Fujikawa:1979ay}%
  \BibitemOpen
  \bibfield  {author} {\bibinfo {author} {\bibfnamefont {K.}~\bibnamefont
  {Fujikawa}},\ }\href {\doibase 10.1103/PhysRevLett.42.1195} {\bibfield
  {journal} {\bibinfo  {journal} {Phys. Rev. Lett.}\ }\textbf {\bibinfo
  {volume} {42}},\ \bibinfo {pages} {1195} (\bibinfo {year}
  {1979})}\BibitemShut {NoStop}%
\bibitem [{\citenamefont {Mueller}\ and\ \citenamefont
  {Venugopalan}(2019)}]{Mueller:2019gjj}%
  \BibitemOpen
  \bibfield  {author} {\bibinfo {author} {\bibfnamefont {N.}~\bibnamefont
  {Mueller}}\ and\ \bibinfo {author} {\bibfnamefont {R.}~\bibnamefont
  {Venugopalan}},\ }\href@noop {} {\  (\bibinfo {year} {2019})},\ \Eprint
  {http://arxiv.org/abs/1901.10492} {arXiv:1901.10492 [hep-th]} \BibitemShut
  {NoStop}%
%%CITATION = ARXIV:1901.10492;%%
\bibitem [{\citenamefont {Kodaira}\ and\ \citenamefont
  {Tanaka}(1999)}]{Kodaira:1998jn}%
  \BibitemOpen
  \bibfield  {author} {\bibinfo {author} {\bibfnamefont {J.}~\bibnamefont
  {Kodaira}}\ and\ \bibinfo {author} {\bibfnamefont {K.}~\bibnamefont
  {Tanaka}},\ }\href {\doibase 10.1143/PTP.101.191} {\bibfield  {journal}
  {\bibinfo  {journal} {Prog. Theor. Phys.}\ }\textbf {\bibinfo {volume}
  {101}},\ \bibinfo {pages} {191} (\bibinfo {year} {1999})},\ \Eprint
  {http://arxiv.org/abs/hep-ph/9812449} {arXiv:hep-ph/9812449} \BibitemShut
  {NoStop}%
\bibitem [{\citenamefont {Zijlstra}\ and\ \citenamefont {van
  Neerven}(1994)}]{Zijlstra:1993sh}%
  \BibitemOpen
  \bibfield  {author} {\bibinfo {author} {\bibfnamefont {E.}~\bibnamefont
  {Zijlstra}}\ and\ \bibinfo {author} {\bibfnamefont {W.}~\bibnamefont {van
  Neerven}},\ }\href {\doibase 10.1016/0550-3213(94)90538-X} {\bibfield
  {journal} {\bibinfo  {journal} {Nucl. Phys. B}\ }\textbf {\bibinfo {volume}
  {417}},\ \bibinfo {pages} {61} (\bibinfo {year} {1994})},\ \bibinfo {note}
  {[Erratum: Nucl.Phys.B 426, 245 (1994), Erratum: Nucl.Phys.B 773, 105--106
  (2007), Erratum: Nucl.Phys.B 501, 599--599 (1997)]}\BibitemShut {NoStop}%
\bibitem [{\citenamefont {Ball}\ \emph {et~al.}(1995)\citenamefont {Ball},
  \citenamefont {Forte},\ and\ \citenamefont {Ridolfi}}]{Ball:1995ye}%
  \BibitemOpen
  \bibfield  {author} {\bibinfo {author} {\bibfnamefont {R.~D.}\ \bibnamefont
  {Ball}}, \bibinfo {author} {\bibfnamefont {S.}~\bibnamefont {Forte}}, \ and\
  \bibinfo {author} {\bibfnamefont {G.}~\bibnamefont {Ridolfi}},\ }\href
  {\doibase 10.1016/0550-3213(95)00178-U} {\bibfield  {journal} {\bibinfo
  {journal} {Nucl. Phys. B}\ }\textbf {\bibinfo {volume} {444}},\ \bibinfo
  {pages} {287} (\bibinfo {year} {1995})},\ \bibinfo {note} {[Erratum:
  Nucl.Phys.B 449, 680--680 (1995)]},\ \Eprint
  {http://arxiv.org/abs/hep-ph/9502340} {arXiv:hep-ph/9502340} \BibitemShut
  {NoStop}%
\bibitem [{\citenamefont {Blumlein}\ and\ \citenamefont
  {Vogt}(1996)}]{Blumlein:1996hb}%
  \BibitemOpen
  \bibfield  {author} {\bibinfo {author} {\bibfnamefont {J.}~\bibnamefont
  {Blumlein}}\ and\ \bibinfo {author} {\bibfnamefont {A.}~\bibnamefont
  {Vogt}},\ }\href {\doibase 10.1016/0370-2693(96)00958-6} {\bibfield
  {journal} {\bibinfo  {journal} {Phys. Lett. B}\ }\textbf {\bibinfo {volume}
  {386}},\ \bibinfo {pages} {350} (\bibinfo {year} {1996})},\ \Eprint
  {http://arxiv.org/abs/hep-ph/9606254} {arXiv:hep-ph/9606254} \BibitemShut
  {NoStop}%
\bibitem [{\citenamefont {Ahmed}\ and\ \citenamefont
  {Ross}(1976)}]{Ahmed:1976ee}%
  \BibitemOpen
  \bibfield  {author} {\bibinfo {author} {\bibfnamefont {M.}~\bibnamefont
  {Ahmed}}\ and\ \bibinfo {author} {\bibfnamefont {G.~G.}\ \bibnamefont
  {Ross}},\ }\href {\doibase 10.1016/0550-3213(76)90328-X} {\bibfield
  {journal} {\bibinfo  {journal} {Nucl. Phys. B}\ }\textbf {\bibinfo {volume}
  {111}},\ \bibinfo {pages} {441} (\bibinfo {year} {1976})}\BibitemShut
  {NoStop}%
\bibitem [{\citenamefont {Altarelli}\ and\ \citenamefont
  {Parisi}(1977)}]{Altarelli:1977zs}%
  \BibitemOpen
  \bibfield  {author} {\bibinfo {author} {\bibfnamefont {G.}~\bibnamefont
  {Altarelli}}\ and\ \bibinfo {author} {\bibfnamefont {G.}~\bibnamefont
  {Parisi}},\ }\href {\doibase 10.1016/0550-3213(77)90384-4} {\bibfield
  {journal} {\bibinfo  {journal} {Nucl. Phys. B}\ }\textbf {\bibinfo {volume}
  {126}},\ \bibinfo {pages} {298} (\bibinfo {year} {1977})}\BibitemShut
  {NoStop}%
\bibitem [{\citenamefont {Mertig}\ and\ \citenamefont {van
  Neerven}(1996)}]{Mertig:1995ny}%
  \BibitemOpen
  \bibfield  {author} {\bibinfo {author} {\bibfnamefont {R.}~\bibnamefont
  {Mertig}}\ and\ \bibinfo {author} {\bibfnamefont {W.}~\bibnamefont {van
  Neerven}},\ }\href {\doibase 10.1007/s002880050138} {\bibfield  {journal}
  {\bibinfo  {journal} {Z. Phys. C}\ }\textbf {\bibinfo {volume} {70}},\
  \bibinfo {pages} {637} (\bibinfo {year} {1996})},\ \Eprint
  {http://arxiv.org/abs/hep-ph/9506451} {arXiv:hep-ph/9506451} \BibitemShut
  {NoStop}%
\bibitem [{\citenamefont {Vogelsang}(1996{\natexlab{a}})}]{Vogelsang:1995vh}%
  \BibitemOpen
  \bibfield  {author} {\bibinfo {author} {\bibfnamefont {W.}~\bibnamefont
  {Vogelsang}},\ }\href {\doibase 10.1103/PhysRevD.54.2023} {\bibfield
  {journal} {\bibinfo  {journal} {Phys. Rev. D}\ }\textbf {\bibinfo {volume}
  {54}},\ \bibinfo {pages} {2023} (\bibinfo {year} {1996}{\natexlab{a}})},\
  \Eprint {http://arxiv.org/abs/hep-ph/9512218} {arXiv:hep-ph/9512218}
  \BibitemShut {NoStop}%
\bibitem [{\citenamefont {Vogelsang}(1996{\natexlab{b}})}]{Vogelsang:1996im}%
  \BibitemOpen
  \bibfield  {author} {\bibinfo {author} {\bibfnamefont {W.}~\bibnamefont
  {Vogelsang}},\ }\href {\doibase 10.1016/0550-3213(96)00306-9} {\bibfield
  {journal} {\bibinfo  {journal} {Nucl. Phys. B}\ }\textbf {\bibinfo {volume}
  {475}},\ \bibinfo {pages} {47} (\bibinfo {year} {1996}{\natexlab{b}})},\
  \Eprint {http://arxiv.org/abs/hep-ph/9603366} {arXiv:hep-ph/9603366}
  \BibitemShut {NoStop}%
\bibitem [{\citenamefont {Moch}\ \emph {et~al.}(2014)\citenamefont {Moch},
  \citenamefont {Vermaseren},\ and\ \citenamefont {Vogt}}]{Moch:2014sna}%
  \BibitemOpen
  \bibfield  {author} {\bibinfo {author} {\bibfnamefont {S.}~\bibnamefont
  {Moch}}, \bibinfo {author} {\bibfnamefont {J.}~\bibnamefont {Vermaseren}}, \
  and\ \bibinfo {author} {\bibfnamefont {A.}~\bibnamefont {Vogt}},\ }\href
  {\doibase 10.1016/j.nuclphysb.2014.10.016} {\bibfield  {journal} {\bibinfo
  {journal} {Nucl. Phys. B}\ }\textbf {\bibinfo {volume} {889}},\ \bibinfo
  {pages} {351} (\bibinfo {year} {2014})},\ \Eprint
  {http://arxiv.org/abs/1409.5131} {arXiv:1409.5131 [hep-ph]} \BibitemShut
  {NoStop}%
\bibitem [{\citenamefont {Moch}\ \emph {et~al.}(2015)\citenamefont {Moch},
  \citenamefont {Vermaseren},\ and\ \citenamefont {Vogt}}]{Moch:2015usa}%
  \BibitemOpen
  \bibfield  {author} {\bibinfo {author} {\bibfnamefont {S.}~\bibnamefont
  {Moch}}, \bibinfo {author} {\bibfnamefont {J.}~\bibnamefont {Vermaseren}}, \
  and\ \bibinfo {author} {\bibfnamefont {A.}~\bibnamefont {Vogt}},\ }\href
  {\doibase 10.1016/j.physletb.2015.07.027} {\bibfield  {journal} {\bibinfo
  {journal} {Phys. Lett. B}\ }\textbf {\bibinfo {volume} {748}},\ \bibinfo
  {pages} {432} (\bibinfo {year} {2015})},\ \Eprint
  {http://arxiv.org/abs/1506.04517} {arXiv:1506.04517 [hep-ph]} \BibitemShut
  {NoStop}%
\bibitem [{\citenamefont {Altarelli}\ and\ \citenamefont
  {Lampe}(1990)}]{Altarelli:1990jp}%
  \BibitemOpen
  \bibfield  {author} {\bibinfo {author} {\bibfnamefont {G.}~\bibnamefont
  {Altarelli}}\ and\ \bibinfo {author} {\bibfnamefont {B.}~\bibnamefont
  {Lampe}},\ }\href {\doibase 10.1007/BF01552357} {\bibfield  {journal}
  {\bibinfo  {journal} {Z. Phys. C}\ }\textbf {\bibinfo {volume} {47}},\
  \bibinfo {pages} {315} (\bibinfo {year} {1990})}\BibitemShut {NoStop}%
\bibitem [{\citenamefont {de~Florian}\ and\ \citenamefont
  {Vogelsang}(2019)}]{deFlorian:2019egz}%
  \BibitemOpen
  \bibfield  {author} {\bibinfo {author} {\bibfnamefont {D.}~\bibnamefont
  {de~Florian}}\ and\ \bibinfo {author} {\bibfnamefont {W.}~\bibnamefont
  {Vogelsang}},\ }\href {\doibase 10.1103/PhysRevD.99.054001} {\bibfield
  {journal} {\bibinfo  {journal} {Phys. Rev. D}\ }\textbf {\bibinfo {volume}
  {99}},\ \bibinfo {pages} {054001} (\bibinfo {year} {2019})},\ \Eprint
  {http://arxiv.org/abs/1902.04636} {arXiv:1902.04636 [hep-ph]} \BibitemShut
  {NoStop}%
\bibitem [{\citenamefont {Gelis}\ \emph {et~al.}(2010)\citenamefont {Gelis},
  \citenamefont {Iancu}, \citenamefont {Jalilian-Marian},\ and\ \citenamefont
  {Venugopalan}}]{Gelis:2010nm}%
  \BibitemOpen
  \bibfield  {author} {\bibinfo {author} {\bibfnamefont {F.}~\bibnamefont
  {Gelis}}, \bibinfo {author} {\bibfnamefont {E.}~\bibnamefont {Iancu}},
  \bibinfo {author} {\bibfnamefont {J.}~\bibnamefont {Jalilian-Marian}}, \ and\
  \bibinfo {author} {\bibfnamefont {R.}~\bibnamefont {Venugopalan}},\ }\href
  {\doibase 10.1146/annurev.nucl.010909.083629} {\bibfield  {journal} {\bibinfo
   {journal} {Ann. Rev. Nucl. Part. Sci.}\ }\textbf {\bibinfo {volume} {60}},\
  \bibinfo {pages} {463} (\bibinfo {year} {2010})},\ \Eprint
  {http://arxiv.org/abs/1002.0333} {arXiv:1002.0333 [hep-ph]} \BibitemShut
  {NoStop}%
%%CITATION = ARXIV:1002.0333;%%
\bibitem [{\citenamefont {Kovchegov}\ and\ \citenamefont
  {Levin}(2012)}]{Kovchegov:2012mbw}%
  \BibitemOpen
  \bibfield  {author} {\bibinfo {author} {\bibfnamefont {Y.~V.}\ \bibnamefont
  {Kovchegov}}\ and\ \bibinfo {author} {\bibfnamefont {E.}~\bibnamefont
  {Levin}},\ }\href {http://www.cambridge.org/de/knowledge/isbn/item6803159}
  {\emph {\bibinfo {title} {{Quantum chromodynamics at high energy}}}},\
  Vol.~\bibinfo {volume} {33}\ (\bibinfo  {publisher} {Cambridge University
  Press},\ \bibinfo {year} {2012})\BibitemShut {NoStop}%
%%CITATION = CMPCE,33,;%%
\bibitem [{\citenamefont {Mueller}(1997)}]{Mueller:1996hm}%
  \BibitemOpen
  \bibfield  {author} {\bibinfo {author} {\bibfnamefont {A.~H.}\ \bibnamefont
  {Mueller}},\ }\href {\doibase 10.1016/S0370-2693(97)00116-0} {\bibfield
  {journal} {\bibinfo  {journal} {Phys. Lett. B}\ }\textbf {\bibinfo {volume}
  {396}},\ \bibinfo {pages} {251} (\bibinfo {year} {1997})},\ \Eprint
  {http://arxiv.org/abs/hep-ph/9612251} {arXiv:hep-ph/9612251} \BibitemShut
  {NoStop}%
\bibitem [{\citenamefont {McLerran}\ and\ \citenamefont
  {Venugopalan}(1994{\natexlab{a}})}]{McLerran:1993ni}%
  \BibitemOpen
  \bibfield  {author} {\bibinfo {author} {\bibfnamefont {L.~D.}\ \bibnamefont
  {McLerran}}\ and\ \bibinfo {author} {\bibfnamefont {R.}~\bibnamefont
  {Venugopalan}},\ }\href {\doibase 10.1103/PhysRevD.49.2233} {\bibfield
  {journal} {\bibinfo  {journal} {Phys. Rev.}\ }\textbf {\bibinfo {volume}
  {D49}},\ \bibinfo {pages} {2233} (\bibinfo {year} {1994}{\natexlab{a}})},\
  \Eprint {http://arxiv.org/abs/hep-ph/9309289} {arXiv:hep-ph/9309289 [hep-ph]}
  \BibitemShut {NoStop}%
%%CITATION = HEP-PH/9309289;%%
\bibitem [{\citenamefont {McLerran}\ and\ \citenamefont
  {Venugopalan}(1994{\natexlab{b}})}]{McLerran:1993ka}%
  \BibitemOpen
  \bibfield  {author} {\bibinfo {author} {\bibfnamefont {L.~D.}\ \bibnamefont
  {McLerran}}\ and\ \bibinfo {author} {\bibfnamefont {R.}~\bibnamefont
  {Venugopalan}},\ }\href {\doibase 10.1103/PhysRevD.49.3352} {\bibfield
  {journal} {\bibinfo  {journal} {Phys. Rev.}\ }\textbf {\bibinfo {volume}
  {D49}},\ \bibinfo {pages} {3352} (\bibinfo {year} {1994}{\natexlab{b}})},\
  \Eprint {http://arxiv.org/abs/hep-ph/9311205} {arXiv:hep-ph/9311205 [hep-ph]}
  \BibitemShut {NoStop}%
%%CITATION = HEP-PH/9311205;%%
\bibitem [{\citenamefont {McLerran}\ and\ \citenamefont
  {Venugopalan}(1994{\natexlab{c}})}]{McLerran:1994vd}%
  \BibitemOpen
  \bibfield  {author} {\bibinfo {author} {\bibfnamefont {L.~D.}\ \bibnamefont
  {McLerran}}\ and\ \bibinfo {author} {\bibfnamefont {R.}~\bibnamefont
  {Venugopalan}},\ }\href {\doibase 10.1103/PhysRevD.50.2225} {\bibfield
  {journal} {\bibinfo  {journal} {Phys. Rev.}\ }\textbf {\bibinfo {volume}
  {D50}},\ \bibinfo {pages} {2225} (\bibinfo {year} {1994}{\natexlab{c}})},\
  \Eprint {http://arxiv.org/abs/hep-ph/9402335} {arXiv:hep-ph/9402335 [hep-ph]}
  \BibitemShut {NoStop}%
%%CITATION = HEP-PH/9402335;%%
\bibitem [{\citenamefont {Schäfer}\ and\ \citenamefont
  {Shuryak}(1998)}]{Schafer:1996wv}%
  \BibitemOpen
  \bibfield  {author} {\bibinfo {author} {\bibfnamefont {T.}~\bibnamefont
  {Schäfer}}\ and\ \bibinfo {author} {\bibfnamefont {E.~V.}\ \bibnamefont
  {Shuryak}},\ }\href {\doibase 10.1103/RevModPhys.70.323} {\bibfield
  {journal} {\bibinfo  {journal} {Rev. Mod. Phys.}\ }\textbf {\bibinfo {volume}
  {70}},\ \bibinfo {pages} {323} (\bibinfo {year} {1998})},\ \Eprint
  {http://arxiv.org/abs/hep-ph/9610451} {arXiv:hep-ph/9610451} \BibitemShut
  {NoStop}%
\bibitem [{\citenamefont {Kuzmin}\ \emph {et~al.}(1985)\citenamefont {Kuzmin},
  \citenamefont {Rubakov},\ and\ \citenamefont {Shaposhnikov}}]{Kuzmin:1985mm}%
  \BibitemOpen
  \bibfield  {author} {\bibinfo {author} {\bibfnamefont {V.}~\bibnamefont
  {Kuzmin}}, \bibinfo {author} {\bibfnamefont {V.}~\bibnamefont {Rubakov}}, \
  and\ \bibinfo {author} {\bibfnamefont {M.}~\bibnamefont {Shaposhnikov}},\
  }\href {\doibase 10.1016/0370-2693(85)91028-7} {\bibfield  {journal}
  {\bibinfo  {journal} {Phys. Lett. B}\ }\textbf {\bibinfo {volume} {155}},\
  \bibinfo {pages} {36} (\bibinfo {year} {1985})}\BibitemShut {NoStop}%
\bibitem [{\citenamefont {McLerran}\ \emph {et~al.}(1991)\citenamefont
  {McLerran}, \citenamefont {Mottola},\ and\ \citenamefont
  {Shaposhnikov}}]{McLerran:1990de}%
  \BibitemOpen
  \bibfield  {author} {\bibinfo {author} {\bibfnamefont {L.~D.}\ \bibnamefont
  {McLerran}}, \bibinfo {author} {\bibfnamefont {E.}~\bibnamefont {Mottola}}, \
  and\ \bibinfo {author} {\bibfnamefont {M.~E.}\ \bibnamefont {Shaposhnikov}},\
  }\href {\doibase 10.1103/PhysRevD.43.2027} {\bibfield  {journal} {\bibinfo
  {journal} {Phys. Rev. D}\ }\textbf {\bibinfo {volume} {43}},\ \bibinfo
  {pages} {2027} (\bibinfo {year} {1991})}\BibitemShut {NoStop}%
\bibitem [{\citenamefont {Moore}\ and\ \citenamefont
  {Tassler}(2011)}]{Moore:2010jd}%
  \BibitemOpen
  \bibfield  {author} {\bibinfo {author} {\bibfnamefont {G.~D.}\ \bibnamefont
  {Moore}}\ and\ \bibinfo {author} {\bibfnamefont {M.}~\bibnamefont
  {Tassler}},\ }\href {\doibase 10.1007/JHEP02(2011)105} {\bibfield  {journal}
  {\bibinfo  {journal} {JHEP}\ }\textbf {\bibinfo {volume} {02}},\ \bibinfo
  {pages} {105} (\bibinfo {year} {2011})},\ \Eprint
  {http://arxiv.org/abs/1011.1167} {arXiv:1011.1167 [hep-ph]} \BibitemShut
  {NoStop}%
\bibitem [{\citenamefont {Mace}\ \emph {et~al.}(2016)\citenamefont {Mace},
  \citenamefont {Schlichting},\ and\ \citenamefont
  {Venugopalan}}]{Mace:2016svc}%
  \BibitemOpen
  \bibfield  {author} {\bibinfo {author} {\bibfnamefont {M.}~\bibnamefont
  {Mace}}, \bibinfo {author} {\bibfnamefont {S.}~\bibnamefont {Schlichting}}, \
  and\ \bibinfo {author} {\bibfnamefont {R.}~\bibnamefont {Venugopalan}},\
  }\href {\doibase 10.1103/PhysRevD.93.074036} {\bibfield  {journal} {\bibinfo
  {journal} {Phys. Rev. D}\ }\textbf {\bibinfo {volume} {93}},\ \bibinfo
  {pages} {074036} (\bibinfo {year} {2016})},\ \Eprint
  {http://arxiv.org/abs/1601.07342} {arXiv:1601.07342 [hep-ph]} \BibitemShut
  {NoStop}%
\bibitem [{\citenamefont {Mondragon}\ \emph {et~al.}(1996)\citenamefont
  {Mondragon}, \citenamefont {Nellen}, \citenamefont {Schmidt},\ and\
  \citenamefont {Schubert}}]{Mondragon:1995ab}%
  \BibitemOpen
  \bibfield  {author} {\bibinfo {author} {\bibfnamefont {M.}~\bibnamefont
  {Mondragon}}, \bibinfo {author} {\bibfnamefont {L.}~\bibnamefont {Nellen}},
  \bibinfo {author} {\bibfnamefont {M.~G.}\ \bibnamefont {Schmidt}}, \ and\
  \bibinfo {author} {\bibfnamefont {C.}~\bibnamefont {Schubert}},\ }\href
  {\doibase 10.1016/0370-2693(95)01392-X} {\bibfield  {journal} {\bibinfo
  {journal} {Phys. Lett. B}\ }\textbf {\bibinfo {volume} {366}},\ \bibinfo
  {pages} {212} (\bibinfo {year} {1996})},\ \Eprint
  {http://arxiv.org/abs/hep-th/9510036} {arXiv:hep-th/9510036} \BibitemShut
  {NoStop}%
\bibitem [{\citenamefont {Mueller}\ and\ \citenamefont
  {Venugopalan}(2018)}]{Mueller:2017lzw}%
  \BibitemOpen
  \bibfield  {author} {\bibinfo {author} {\bibfnamefont {N.}~\bibnamefont
  {Mueller}}\ and\ \bibinfo {author} {\bibfnamefont {R.}~\bibnamefont
  {Venugopalan}},\ }\href {\doibase 10.1103/PhysRevD.97.051901} {\bibfield
  {journal} {\bibinfo  {journal} {Phys. Rev.}\ }\textbf {\bibinfo {volume}
  {D97}},\ \bibinfo {pages} {051901} (\bibinfo {year} {2018})},\ \Eprint
  {http://arxiv.org/abs/1701.03331} {arXiv:1701.03331 [hep-ph]} \BibitemShut
  {NoStop}%
%%CITATION = ARXIV:1701.03331;%%
\end{thebibliography}%

\end{document}